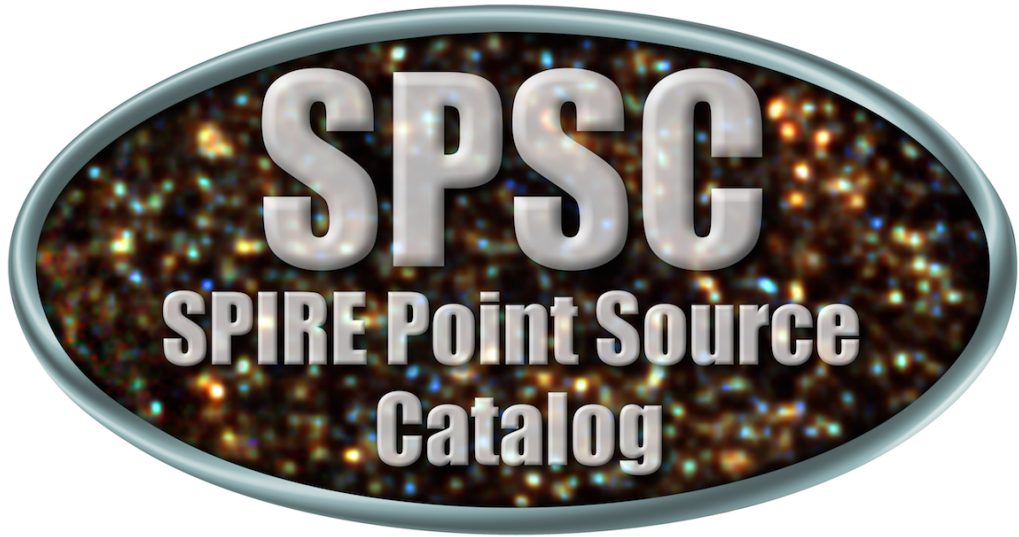

# SPIRE Point Source Catalog Explanatory Supplement

Bernhard Schulz[1], Gábor Marton[2], Ivan Valtchanov[3], Ana María Pérez García[3], Sándor Pintér[4], Phil Appleton[1], Csaba Kiss[2], Tanya Lim[3,6], Nanyao Lu[1,8,9], Andreas Papageorgiou[5], Chris Pearson[6], John Rector[1], Miguel Sánchez Portal[3,7], David Shupe[1], Viktor L. Tóth[4], Schuyler Van Dyk[1], Erika Varga-Verebélyi[2], Kevin Xu[1]

1) Caltech/IPAC, Pasadena, USA
2) Konkoly Observatory, Budapest, Hungary
3) ESAC-ESA, Villanueva de la Cañada, Madrid, Spain
4) Eötvös Loránd University, Budapest, Hungary
5) Cardiff University, Cardiff, UK
6) Rutherford Appleton Labs, STFC, Chilton, UK
7) Joint ALMA Observatory & European Southern Observatory, Santiago, Chile
8) National Astronomical Observatories CAS, Beijing, China
9) South American Center for Astronomy CAS, Santiago, Chile

# Change Record

| Date | Changed |
|---|---|
| 23-Jan-2017 | First version |
| 03-Feb-2017 | Updated references and HTML links; added caveats on source multiplicity and number counts in conclusions; corrected number of sources with astrom_flag set; added new sections Cautionary Notes and Catalog Products, updated text on cross-identification table. |
| 11-May-2017 | Adjusted SPIRE Sky coverage to 8%; added OTKA grant number to acknowledgement. |
| | |

# Contents

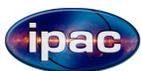

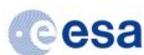
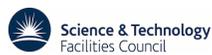

# 1 Introduction

The Spectral and Photometric Imaging Receiver (SPIRE) (Griffin et al. 2010) was launched as one of the scientific instruments on board of the space observatory Herschel on May 14th, 2009 (Pilbratt et al. 2010). It operated for almost four years, until April 29th, 2013, when the liquid Helium coolant boiled off. The SPIRE photometer opened up an entirely new window in the Submillimeter domain for large scale mapping, that up to then was very difficult to observe. Predecessor facilities at these wavelengths include SCUBA (1999), SWAS (Melnick et al. 2000), ODIN (Nordh et al. 2003), and BLAST (Devlin et al. 2004). Without the limitations of Earth's atmosphere, and broad band filters centered at 250µm, 350µm, and 500µm, SPIRE covered about 8% of the entire sky in three scan mapping modes. This wavelength range finally allowed for much better estimates of dust masses from cold spectral energy distributions than prior FIR facilities like IRAS (Neugebauer et al. 1984), ISO (Kessler at al. 1996), Spitzer (Werner et al. 2004), or AKARI (Murakami et al. 2007), as it covered the spectral region beyond the infrared emission peak and encompassed the peak for high redshift galaxies.

Although Herschel carried the largest primary telescope mirror of a space facility of this kind with a physical diameter of 3.5 m, the SPIRE bolometer arrays were so sensitive that the recorded maps were confusion limited after only a few scans. Even in a virtually "empty" region of the sky the maps are practically saturated with background galaxies that are seen through the thermal emission of their dust content in the interstellar medium. A quick back-of-the-envelope calculation, using the SPIRE-based number counts (e.g. Oliver et al. 2010), yields a number of a few million objects that can be found in the SPIRE scan map data, depending on recoverable depth.

There are already several catalogs that were produced by individual Herschel science projects, such as HerMES and Herschel-ATLAS. Yet, we estimate that the objects of only a fraction of these maps will ever be systematically extracted and published by the science teams that originally proposed the observations. The thirty Herschel key programs have the highest probability of a systematic exploitation of their data, but even they only cover about 55% of all the SPIRE scan map area. Furthermore, the SPIRE instrument performed its standard photometric observations in an optically very stable configuration, only moving the telescope across the sky, with variations in its configuration parameters limited to scan speed and sampling rate. This and the scarcity of features in the data that require special processing steps made this dataset very attractive for producing an expert reduced catalog of point sources that is being described in this document.

The Catalog was extracted from a total of 6878 unmodified SPIRE scan map observations, consisting of a serendipitous mix of program science observations with sky coverage to varying depths and calibration observations performed in standard configurations, as they are found as final legacy version in the Herschel Science Archive (HSA). SPIRE maps become confusion limited after a few repeated scans, reducing the overall variation in depth. The photometry was obtained by a systematic and homogeneous source extraction procedure, followed by a rigorous quality check that emphasized reliability over completeness. Having to exclude regions affected by strong Galactic emission, mostly in the Galactic Plane, that pushed the limits of the four source extraction methods that were used, this catalog is aimed primarily at the extragalactic community. The result can serve as a pathfinder for ALMA and other Submillimeter and Far-Infrared facilities.

With a major part of the authors having worked as part of the SPIRE Instrument Control Centre (ICC), we made use of this considerable pool of expertise and detailed knowledge of the instrument and its data processing pipeline. We initially extracted close to 10 million source candidates that, in the interest of reliability, were eventually downselected to 1,693,718 records, splitting into 950688, 524734, 218296 objects for the 250µm, 350µm, and 500µm bands, respectively. Application of the same four different photometric methods to every source, delivered highly accurate photometry for point sources and reasonable, albeit somewhat less accurate flux estimates for extended sources. The catalog comes with well characterized environments, reliability, completeness, and accuracies, that single programs typically cannot provide.

## Cautionary Notes

Although this work is a big step forward in facilitating the archival exploitation of the SPIRE photometric dataset, there are a number of important attributes that every user must be aware of when using this catalog for scientific work. Some of these issues were found in the completed product only after considerable verification efforts and we hope that some can be corrected in a possible future version.

### Completeness

This catalog is not 100% complete, even at the highest flux levels. The source detection is optimized for point sources and missed sources that were too extended like nearby galaxies. We also found that our algorithms didn't do very well on top of strongly structured backgrounds, especially those in the galactic plane, so entire tiles of sky (Q3C tiles) were eliminated where the median structure noise surpassed a certain threshold.

In order to improve reliability, some rigorous filtering was used, that is detailed in the main part of this document. Thus we need to point out that the absence of a source at a given position, is no guarantee for its actual absence at that wavelength in the respective SPIRE map. As we also don't provide rejected source lists. The only way of verification in that case, is the inspection of the original archival SPIRE maps at that position.

At the low flux levels the completeness, according to simulations, drops below 90% for fluxes smaller than 50 mJy in a clean field, and at higher fluxes for more complex backgrounds. The background confusion noise, that never drops below the extragalactic component, represents the fundamental limitation, while the number of scans and the scan speed are secondary factors that only matter appreciably in Fast Scan and Parallel modes.

Completeness was also affected by a software error that left differently calibrated, so-called Serendipity Slew Data originating from telescope slews in the timeline data, leading to non-convergence and failures of parts of the photometry extraction. Only a low percentage of all catalog objects were lost this way, but the effect is visible in many maps.

## Homogeneity

Herschel, executing a multitude of observing programs with different goals, left a sometimes quite arbitrary looking coverage of the sky. Even though the source extraction procedure was homogeneously applied to all sources detected, and the three scan map types are very similar in the way they are executed, their differences in scan speed, sampling rate, scan direction, and repetition factors added further to the inhomogeneous coverage of the sky. The effect on the dynamic range of the noise levels across the covered sky was fortunately lowered by the aforementioned extragalactic confusion limit. Nevertheless, these factors must be well understood before embarking on any statistical studies using this material.

## Cross Wavelength Matching

The source detection and photometric extraction made no use of priors detected at different wavelengths, nor did it attempt a simultaneous extraction at all three SPIRE filter bands. Each of the three bands underwent an independent source detection. Observers need to be aware that two or more catalog objects at one wavelength can correspond to just one apparent point- or slightly extended source at a longer wavelength, especially when close to the confusion limit.

## Reliability

Although we achieved a very high degree of reliability as indicated by the statistics of number of expected source detections (nmap) versus the number of actual detections (ndet), and visual inspection of several hundred catalog positions in actual maps, there are a small number of objects that result from high energy radiation impacts in the bolometers or electronics. Comparison of the four photometer values should help to weed out these objects that evaded the deglitching procedures of the processing pipeline. High Timeline Fitter fluxes, that are contrasting substantially smaller Sussextractor and DAOPHOT fluxes, are good indicators for an undetected glitch. Low coverage and proximity to a map edge are additional risk factors. On the other hand, a point source without peculiarities and consistent photometry in all four values can be used with great confidence.

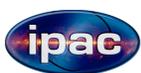 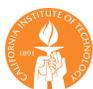 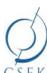 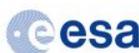 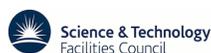 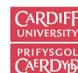

## Photometric Accuracy

In such a case, where all four photometry values are consistent and the point source flag is set, the Timeline Fitter value (TML) is the most accurate flux estimate down to flux levels of ~30 mJy. Relative photometric accuracies of 2% at fluxes greater 50 mJy are achieved in clean fields, which was also verified with Neptune, the primary flux calibrator of SPIRE. Below 30 mJy, background confusion and instrument noise start to take away the benefits of the method, and Daophot and Sussextractor perform similarly. Besides checking for consistency of the 4 methods, the individual uncertainties of the methods, as well as the total uncertainty derived from the local structure noise, need to be verified.

For the slightly extended sources that were accepted into this catalog, the Timeline Fitter 2 (TM2) value provides the best guess for an extended flux. Sussextractor and TML are point source flux methods only, that will underestimate extended source fluxes. The Daophot method will measure a higher flux and is a good indicator for source extension, but due to the small aperture used, it will systematically underestimate extended source fluxes. It should be understood that these fluxes are subject to greater uncertainty, not only at low fluxes due to more free parameters, but also due to the implicit assumption of a Gaussian shape which may or may not be true.

## Shape Parameters

The TM2 run provides also FWHM values, which are used to distinguish point- and extended sources. However, these flags are indicative only. Their statistical uncertainty depends strongly on flux and at low fluxes the range of values that could still be a point source is quite large.

There is also a small number of sources with too small diameters in one direction and too large diameters in the other. These objects appear in the middle of the earlier mentioned Serendipity Slew Trails with typically FWHM2 being too small for a point source at that wavelength, and FWHM1 much larger than that, usually with the position angle oriented in scan direction. These will also show inconsistent photometry with ratios of TML versus DAOPHOT fluxes of greater than 7 and should be excluded.

## Positional Accuracy

The positional accuracy of the objects is usually very good and consistent with the published pointing performance of the observatory of better than 2" (1-sigma). This was verified, where possible, using WISE 22μm catalog sources. Nevertheless, 108 observations were identified with pointing discrepancies of more than 5". Although this is a small number of maps compared to a total of 6878, about 8% of the catalog's objects come from maps that either are one of those maps, or are from combined maps, where at least one of those is a member. In most cases the effect is negligible, but in principle the potential impact could result in side-by-side source doublets and some non-existent catalog objects. All of these objects were flagged. A correction would only be feasible in a second version of the catalog.

### Source Multiplicities

Sources are considered indistinguishable when closer together than the FWHM of a point source at the respective wavelength and will be extracted as one object. This important instrumental limitation must be taken into account when comparing with other catalogs that contain objects with smaller distances between sources at the same wavelength, either because they have a higher spatial resolution, or they employed some other way of flux separation. Some examples we encountered can be found in the Appendix.

## Catalog Products

This Explanatory Supplement is part of the first public version of the SPIRE Point Source Catalog, consisting of an additional set of three catalog tables, one for each of the three filter bands, and a cross-identification table. The four data files are distributed as CSV tables that are easily imported into databases, spreadsheets, or other data processing software. Each line in the catalog tables has a unique identifier and corresponds to a specific location in the sky at one of the three wavelengths.

The source detection and characterization is performed independently in each of the three bands. Source confusion and different spatial resolution at different wavelengths are major factors. We explicitly excluded band-merging from this effort, as it often involves multiple sources that merge into one at longer wavelengths, where disentangling the flux contributions also requires an understanding of the actual physical nature of the objects in question, which is clearly beyond the scope of this work. Consequently, entries for the same object at different wavelengths do not always have the same coordinate based identifiers.

The cross-identification table contains only two columns, the Herschel observation identifier (OBSID), and the catalog identifier (SPSCID). For a given observation in the HSA, this table lists all catalog sources that contain contributions from source detections in this particular observation. Note that it is not a tool to find all catalog sources in the sky area covered by a certain observation, as there could for instance be another overlapping deeper observation, that contains more detections than the one in question, generating additional catalog objects that are not seen in the first map, and thus not recorded as related to it in the cross-identification table.

In the following we describe how the catalog was built, starting with a description of the SPIRE data products, and continuing with a description of the various extraction algorithms, the noise estimation, filtering and object consolidation, to the derivation of additional quality indicators. The next part gives a detailed account of the contents of the columns found in the catalog tables, followed by a part describing a number of validation efforts that help understand the strengths but also limitations of the data presented here. For the sake of completeness, most of the original validation reports generated by team members, are available in an Annex.

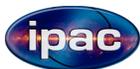

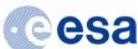
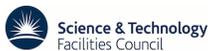

# 2 Source Catalog Construction

The SPIRE Point Source Source Catalog (SPSC) was generated in two major phases: 1) Point source extraction, and 2) Consolidation phase, after the originally planned re-processing of the entire dataset could be dropped in favor of using the official final Legacy Dataset (SPG Version 14.1) available in the [HSA](#) at ESA. The first phase consisted of an initial source detection stage and an outlier rejection stage, both based on FITS maps, followed by a positional and photometric evaluation that goes back to the original signal timelines of the individual detectors. The resulting source candidate tables were fed into a relational database which is the centerpiece of the consolidation phase, where source candidate detections are classified and, based on their positions, consolidated into a list of objects in the sky. This phase ended with establishing fluxes, uncertainties, and a number of characterizing parameters and flags. In the following we will describe the different steps, from the initial data sets through the two major phases, until we arrive at the final source table, and the accompanying cross reference table that links catalog objects to the observations found in the HSA.

## SPIRE Photometer Data Products

The telemetry that was downloaded from the Herschel spacecraft after every observational day was processed by a pipeline software that we refer to as Standard Product Generation (SPG), which is part of the Herschel Common Software System (HCSS; Ott 2010). The intricate details of the data and data reduction are described in much greater detail in the [SPIRE Data Reduction Guide](#). However, for convenience we give here a simplified overview over the photometer scan map data of SPIRE that forms the basis of the SPSC.

### Level 0 and 0.5 Products

First, a Level 0 data product is created that effectively re-arranges the telemetry data into objects that can be manipulated and stored by the software. In a second processing step the digital numbers from the telemetry are turned into engineering units, i.e., voltages, temperatures and a variety of flags. These products are called Level 0.5.

The data is organized in blocks (Building Blocks) that change for the different phases of an observation. For instance there are separate blocks for when the spacecraft slews to the starting point of the observation, for internal calibration, for each scan across the field, and for each positional shift before the next scan leg is started. The building blocks contain tables of science and housekeeping data, where the science data are recorded at the highest data rates and digital resolution. Each row in a table corresponds to a sample in time, where a number of parameters are measured simultaneously. Such a data sample is also called a readout.

## Level 1 Timelines

The processing step from Level 0.5 to Level 1 is for SPIRE photometer data the most important and complex one. Here the voltages of the detector signals are turned into calibrated point source fluxes. They are defined in such a way that, if a detector scans centrally across a point source, the difference between the background level and the peak of the beam profile is equal to the integrated flux density of that source in the respective filter band, expressed in units of $10^{-3}$ Jansky [mJy]. The data remains to be organized in timelines, i.e., tables that state for each sample time and each detector, a flux and a series of qualifying flags. In addition each sample for each detector is assigned a position on the sky. At this point the number of building blocks have been reduced to the essential ones, and the data blocks along each scan line were recombined into one block per scan across the observed field on the sky. In this way, the information of the collection of Level 1 building blocks (scans), is sufficient to reconstruct a map of that region.

## Level 2 and 2.5 Maps

The map reconstruction is performed by the so-called Destriper, a software module that iteratively removes the arbitrary and varying offsets between the scans across the map, using the inherent redundancy of the scans that cross in many places, and the condition that in the same place different scans should show the same flux value. The map grid varies with wavelength and has pixel sizes of 6", 10", 14" for the filter bands at 250µm (PSW), 350µm (PMW), and 500µm (PLW), respectively. In the following we will repeatedly quote triplets of values, that, as a convention, will apply to these filter wavelengths in the same order respectively. The three letter acronyms associated with the filter bands can be found in the existing literature repeatedly and are mentioned here for completeness.

Each map has three data planes (i.e., extensions): The first is the Flux Map ('image' extension) that contains the averages of all detector readouts, where the position falls into the respective pixel square projected onto the sky. The Error Map ('error' extension) contains the respective standard deviation of the mean, and the Coverage Map ('coverage' extension) records the number of readouts that contributed to the respective map pixel.

The same data are delivered in three different map reconstructions and calibrations: i) Point Source Maps that are calibrated in [Jy/beam]; ii) Extended source maps, where the relative photometric gains of the detectors are flat-fielded w.r.t. their respective integrated beam profiles rather than their peaks. These are calibrated in [MJy/sr] and their zero offset is derived via cross-calibration with Planck-HFI; iii) In case of Solar System Objects (SSOs) that were tracked during the observation, an additional map is reconstructed with pointings relative to the SSO centric reference system. For the SPSC only the first two types of maps were used. Fluxes in the SPSC are always in [Jy/beam] and correspond to the calibration of the Point Source Maps.

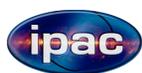 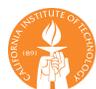 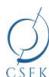 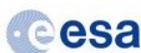 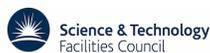 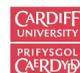

The maps are organized into either Level 2 or Level 2.5 products, which, from a user's point of view, look effectively the same. The difference is that Level 2 maps are produced from just one observation, while a Level 2.5 map results from destriping the Level 1 timelines of an entire group of observations that cover the same area. This approach produces maps with higher Signal to Noise Ratios (SNR) and also benefits catalog construction, as there will be fewer independent source candidates representing the same object in the sky.

## Point Source Extraction

The extraction of point sources from a map starts with a detection step that establishes the coordinates of all sources, followed by a photometric evaluation. The procedure we adopted eventually after many trials involved 4 steps.

After tests of known extractors with injection of artificial Gaussian sources into real SPIRE Level 1 timelines and subsequent map reconstruction, we selected Sussextractor to act as point source detector and Timeline Fitter (TML) to derive accurate photometry. Sussextractor was selected because of its good and quick detection performance, while the TML remained to provide the best possible photometric accuracy of all tested candidates at the cost of a substantially longer processing time. Both modules had the additional advantage of being implemented in the Herschel software, which facilitated the realisation of the project.

Taking into account that, in case of spurious source detections due to instrumental artifacts, a lot of time is wasted during the TML run, an additional aperture photometry run with Daophot was added after the Sussextractor run. It was followed by a discrimination step based on the Daophot Roundness and Sharpness parameters, that prevented running of the TML at all, and elimination of the source candidate, if the source didn't meet certain limits.

The following will first give an account of the tests conducted to find the best source extraction algorithms for our purposes. We will then give a detailed description of the individual point source extraction steps.

## Testing Point Source Extractors

The procedure that finds and evaluates the sources for the catalog is of central importance. We used the data of the GOODS-N maps from the HerMES project (KPGT_soliver_1, PI: Seb Oliver), which consists of 38 SPIRE observations of a Cirrus-free field in the sky, centered at RA=189.246475 deg, Dec=62.24355556 deg. We injected 30 artificial Gaussian shaped sources at brightness levels between 20 and 60 mJy into the Level 1 timelines. The flux levels were 20, 25, 30, 35, 40, 50 and 60 mJy to provide a better sampling at fainter flux levels. The full-width half-maximum (FWHM) values were adjusted corresponding to the beam profiles at the three respective wavelengths. The large number of maps allowed for variance in the magnitude of the instrument noise for the tests. Then, maps were reconstructed from the modified Level 1 products.

In a first step to examine their source detection performance, these maps were used as input for the three algorithms: Sussextractor (Savage & Oliver 2007), Starfinder (Diolaiti 2000), and Getsources (Men'shchikov et al. 2012). All three source detectors worked very well at recovering the injected sources, as illustrated in Figure 1.1, which shows the ratio of the numbers of recovered and injected sources versus flux for the three filter bands at 250μm (PSW), 350μm (PMW), and 500μm (PLW).

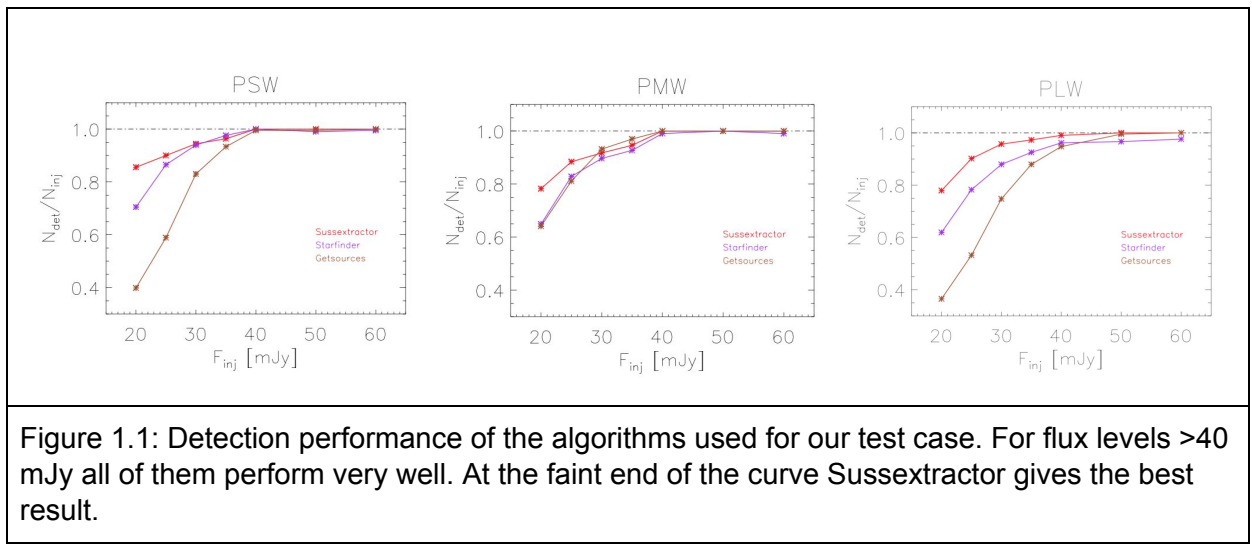

Figure 1.1: Detection performance of the algorithms used for our test case. For flux levels >40 mJy all of them perform very well. At the faint end of the curve Sussextractor gives the best result.

Eventually Sussextractor was chosen for its speed and slightly better performance at low fluxes, but also for the fact that an implementation already existed in the HCSS, that was used also for the standard product generation and data analysis. Sussextractor was developed by the team in University of Sussex for the AKARI/FIS All-Sky Survey, which covered a wavelength range similar to the Herschel ranges. The software applies Bayesian statistics in the source detection process.

Starfinder and Getsources are both very sophisticated algorithms with many parameters to tune in order to achieve the best possible performance. However, this also means that they cannot be used for general purposes, when a large variety of fields needs to be examined. Also, their speed and computational resource requirements cannot be compared to Sussextractor.

To test the second quality, the ability to recover accurate photometry, three additional algorithms were evaluated, that also were available as HCSS implementations: Simultaneous Extractor (Roseboom et al. 2010), Daophot (Landsman 1995), and Timeline Fitter (SPIRE Data Reduction Guide). The ratio of measured versus injected source flux is shown for all three wavelength bands against injected flux in Figure 1.2. This test was carried out by using the same dataset as before for the detection performance test.

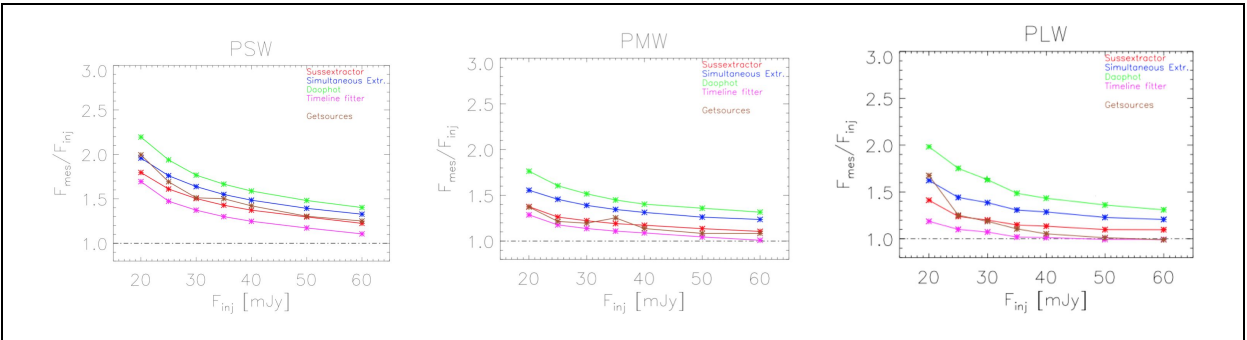

Figure 1.2: Photometric accuracy of the tested algorithms. Timeline Fitter provides the best accuracy for all bands and all flux levels.

In this test series the Timeline Fitter recovered fluxes almost perfectly for flux levels above ~50 mJy, while all others showed a varying degree of discrepancy. This finding reproduced earlier ones by the SPIRE ICC that had conducted tests as well (Pearson et al., 2014). The curves in Figure 1.2 also show a feature called "flux boosting", meaning that faint sources are detected if they are overlapping with brighter sources, resulting in a much higher brightness value than the injected. This effect appears in the diagrams as higher inaccuracy.

## Extraction Procedure Overview

The point source extraction was performed on the final legacy data of SPIRE, processed by SPG version 14.1. A graphic overview of the processing steps is shown in Figure 1.3.

The initial source detection by Sussextractor uses the three point source calibrated maps that are available for the wavelengths 250µm, 350µm, and 500µm. Sussextractor produces a list of coordinates, fluxes and other parameters of which the coordinates are sent to Daophot, an aperture photometry method that is also implemented within the HCSS. The rationale behind using this rather different aperture photometry method is to provide a photometric sanity check, provide additional information in case of a slightly extended source, but mostly, to weed out spurious detections due to residual artifacts in the maps that are often caused by energetic particle hits from Herschel's space environment.

Spurious sources are eliminated from the catalog in a filter stage by detecting extreme values in the Daophot generated parameters Roundness, Sharpness, Flux, and Quality. This early cleaning saves time during the following iterative fitting routines, that would needlessly run to the iteration maximum without ever converging.

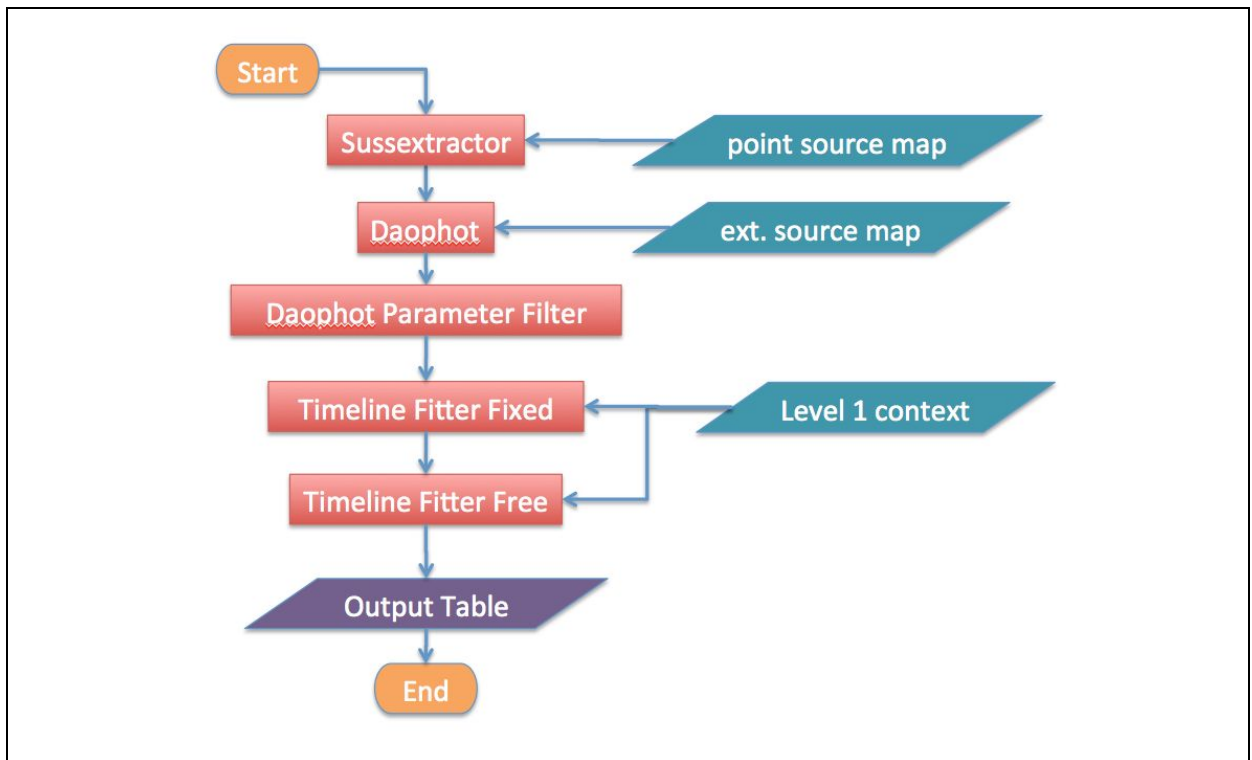

Figure 1.3: Overview flow diagram of the source extraction procedure.

The cleaned catalog is now fed to a first Timeline Fitter stage (TML), that fits a circular Gaussian beam profile to the detector timelines. The size is fixed and depends on the filter band. The values are given in Table 1.1. This method provides refined positions, fluxes and quality parameters that are the most accurate for point sources.

Due to the high sensitivity of SPIRE and the limited telescope size, the confusion limit is reached very quickly, and many sources show wider beam profiles due to multiplicity. Other extended sources are found as galaxies at intermediate distances, Galactic Cirrus knots, etc. To allow another distinction apart from the Daophot aperture photometry, a second Timeline Fitter run is performed, this time allowing for an elliptical Gaussian with free FWHM and a tilted background (TM2).

All parameters from the four different photometric methods are collected in a source table that is ingested into a relational database for further statistical analysis, filtering and consolidation of duplicate detections at the same sky positions. A somewhat more detailed flow diagram of the table generation is shown in Figure 1.4.

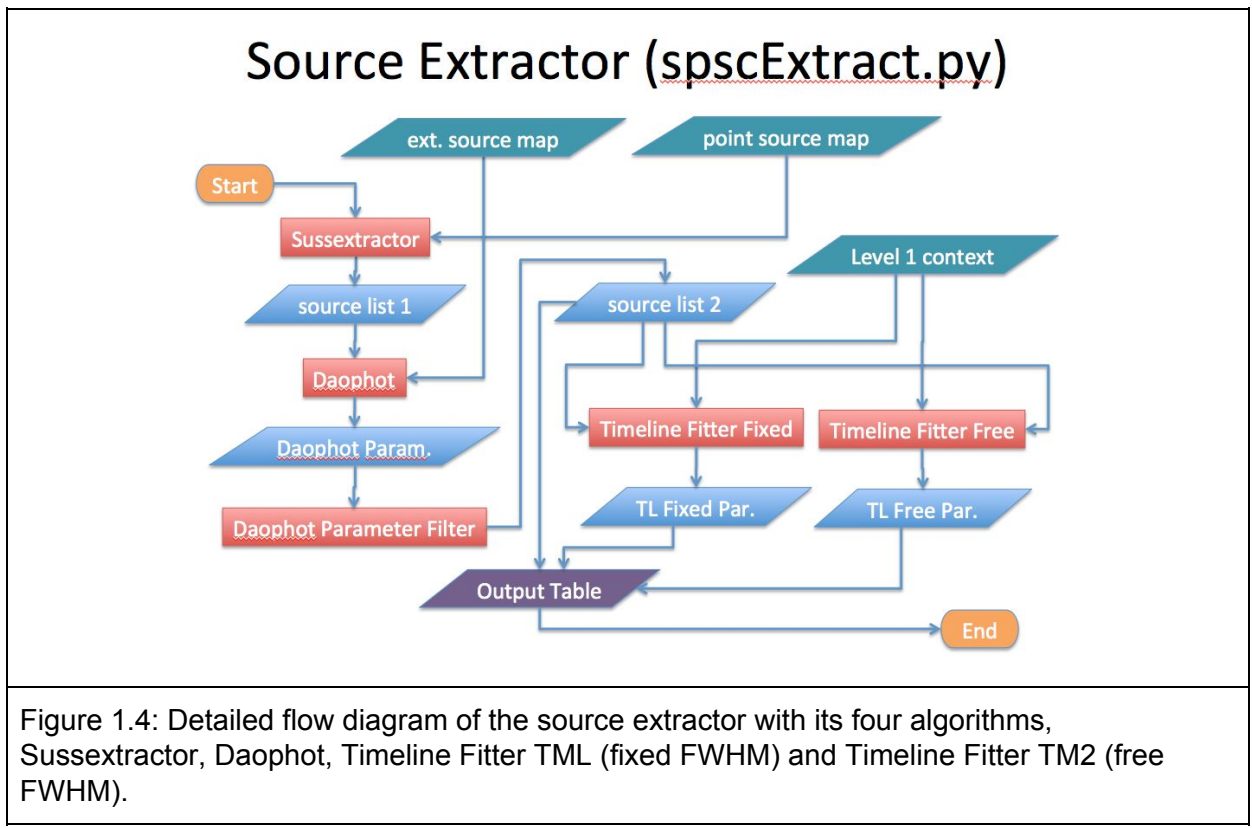

Figure 1.4: Detailed flow diagram of the source extractor with its four algorithms, Sussextractor, Daophot, Timeline Fitter TML (fixed FWHM) and Timeline Fitter TM2 (free FWHM).

## Handling Single and Combined Maps

The four source extraction methods require three different data products. Sussextractor uses point source calibrated maps, Daophot takes extended source calibrated maps, and both Timeline Fitter runs require Level 1 timeline products. All these products are readily available in a standard output product as a small or large map downloaded from the Herschel Science

Archive. However, for a combined Level 2.5 map the Level 1 timelines must be collected from all of the constituent observations and combined. When combining the timelines of different maps, small relative signal offsets between the observations need to be applied, that are stored in the Diagnostic Product of the Level 2.5 map. A flow diagram of the process is shown in Figure 1.5. In addition to the source table, a Run Table provided details about each extraction run for bookkeeping, as well as three PNG images showing the maps and overplotted source detections for quality control purposes.

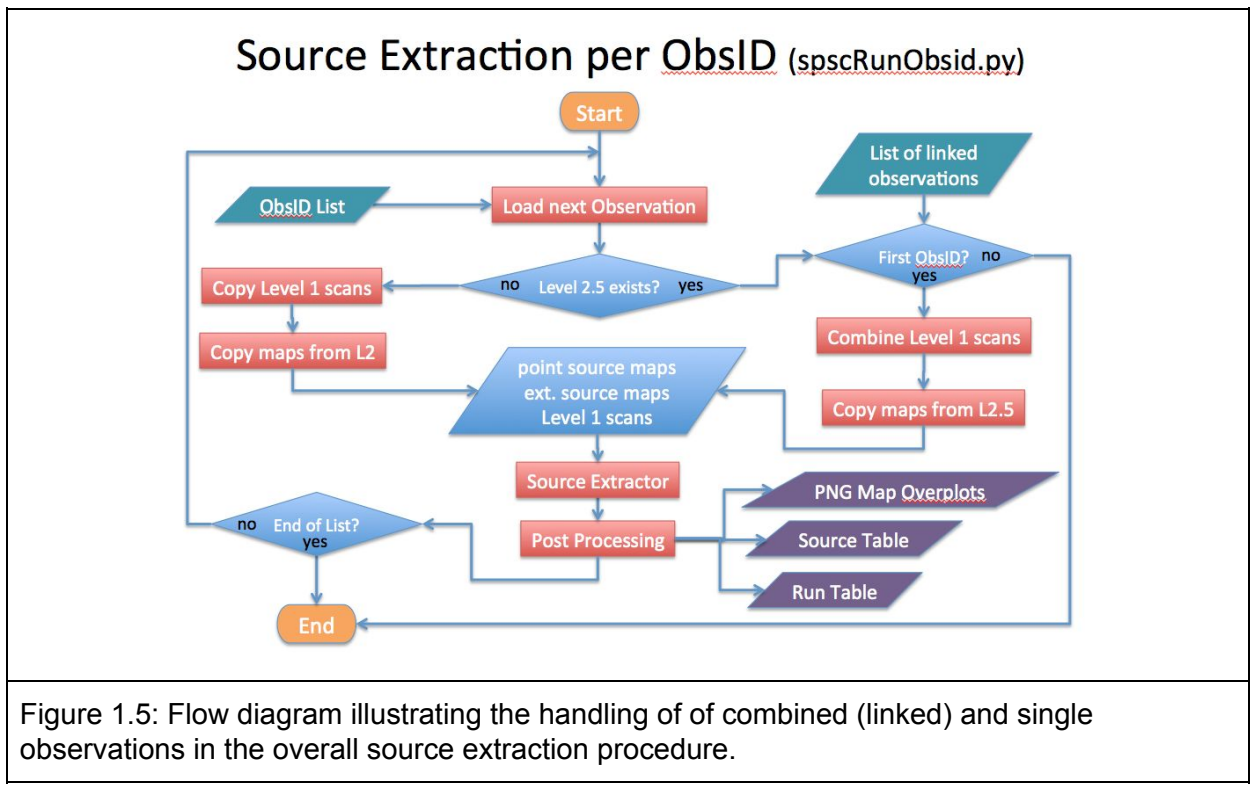

Figure 1.5: Flow diagram illustrating the handling of of combined (linked) and single observations in the overall source extraction procedure.

## Sussextractor

The procedure first applies a Bayesian source detection filter, followed by a Bayesian photometry stage (Savage & Oliver 2007). We operated the program with default parameters, except that we explicitly provided a Gaussian beam profile model with FWHM[1] 17.6", 23.6", 35.2" for 250μm, 350μm, and 500μm, respectively. The values are summarized in Table 1.1 below. By setting the flag "useSignalToNoise" and the respective parameters, only sources with a minimum SNR of three appeared in the output list. This list contains positions, fluxes, background levels, uncertainties, and a quality flag.

---

[1] We used the numbers from an early analysis. The numbers in the SPIRE Handbook are 17.9", 24.2", 35.4, and we verified that the impact on the photometry is negligible.

## Daophot

The Sussextractor positions are fed into the Daophot photometry tool (Landsman 1995), which is a port of the IDL Astrolib implementation into the HCSS. As an aperture photometry method, this procedure requires as input the extended source calibrated set of maps that are also provided in the SPIRE data products. With Daophot we use the same FWHM for the Gaussian beam profile model as in Sussextractor (see Table 1.1). The radius of the photometric aperture was chosen to reach to the first inner minimum of the beam profile, short of the first Airy ring with (22", 30", 42"), depending on filter. The background annuli start from there and extend to a factor of 1.5 of the inner radii, i.e., (33", 45", 63"), respectively. This less than optimal choice of the background annulus, covering the Airy rings, was forced by convergence problems of the HCSS implementation, experienced with larger radii. It is also important to mention that the "doApertureCorrection" flag was switched off. Instead, the corrections for flux lost outside of the photometric aperture and residual flux falling into the background annulus were done empirically by comparison with the Timeline Fitter results for SPIRE's prime calibrator, Neptune. The correction amounted to multiplication of the Daophot fluxes by factors of 1.114. 1.111, 1.099 for the three wavelengths, respectively.

| Parameter | PSW | PMW | PLW | Sussex-tractor | Daophot | Timeline Fitter | Remarks |
|---|---|---|---|---|---|---|---|
| fwhm | 17.6 | 23.9 | 35.2 | x | x | x | FWHM in arcsec |
| pixSize | 6 | 10 | 14 | x | x | N/A | Pixel size in arcsec |
| detThreshold | 3 | 3 | 3 | x | | | S/N detection threshold |
| rpeak | 22 | 30 | 42 | | x | x | Radius of central source aperture in arcsec |
| prfSize | 5 | 5 | 5 | x | | | Size of Sussextractor PRF in pixels |
| innerArcsec | 22 | 30 | 42 | | x | | Inner radius of background annulus in arcsec |
| outerArcsec | 33 | 45 | 63 | | x | | Outer radius of background annulus in arcsec |
| maxIter | 400 | 400 | 400 | | | x | Maximum number of iterations |
| rBackground[0] | 70 | 98 | 140 | | | x | Inner radius of background annulus in arcsec |
| rBackground[1] | 74 | 103 | 147 | | | x | Outer radius of background annulus in arcsec |

Table 1.1: Summary of the parameters that are used in the different stages of the source extraction procedure.

## Daophot Filtering

Sharpness compares the peak of a Gaussian with the peak of a 2-D delta function (Stetson 1987) and was originally designed to discriminate against high energy radiation events in CCD data. Roundness compares the peakedness of sources in the x- and y-direction of the map and is sensitive to asymmetries in those cardinal directions. Figure 1.6 shows examples of distributions measured in various SPIRE fields. The first is from an extragalactic field (ObsID 1342187711) that is dominated by point sources. The second diagram shows the distributions for a bright galactic field that contains a high proportion of bright extended sources. The latter peaks at about 0.35, while the point source dominated peak is at about 0.4. The third represents the sources from a selection of representative backgrounds that we used for source injection experiments. In this case the sharpness distribution is extended between 0.35 and 0.4, including both types of distributions. The roundness parameter shows a different width that is mainly governed by the average brightness of the sources. Figure 1.7 shows Daophot fluxes versus the shape parameters roundness and sharpness for a mix of point and extended sources from various representative environments.

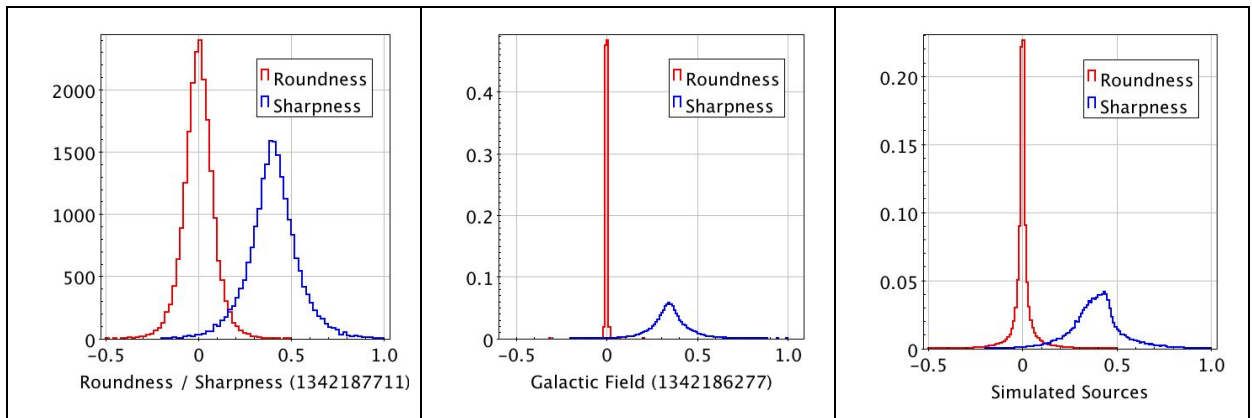

Figure 1.6: Distributions of roundness and sharpness parameters as delivered by Daophot. From left to right the diagrams correspond to 1) a range of different backgrounds, 2) an extragalactic field with predominantly weak point sources, 3) a Galactic field with a mix of bright extended and point sources.

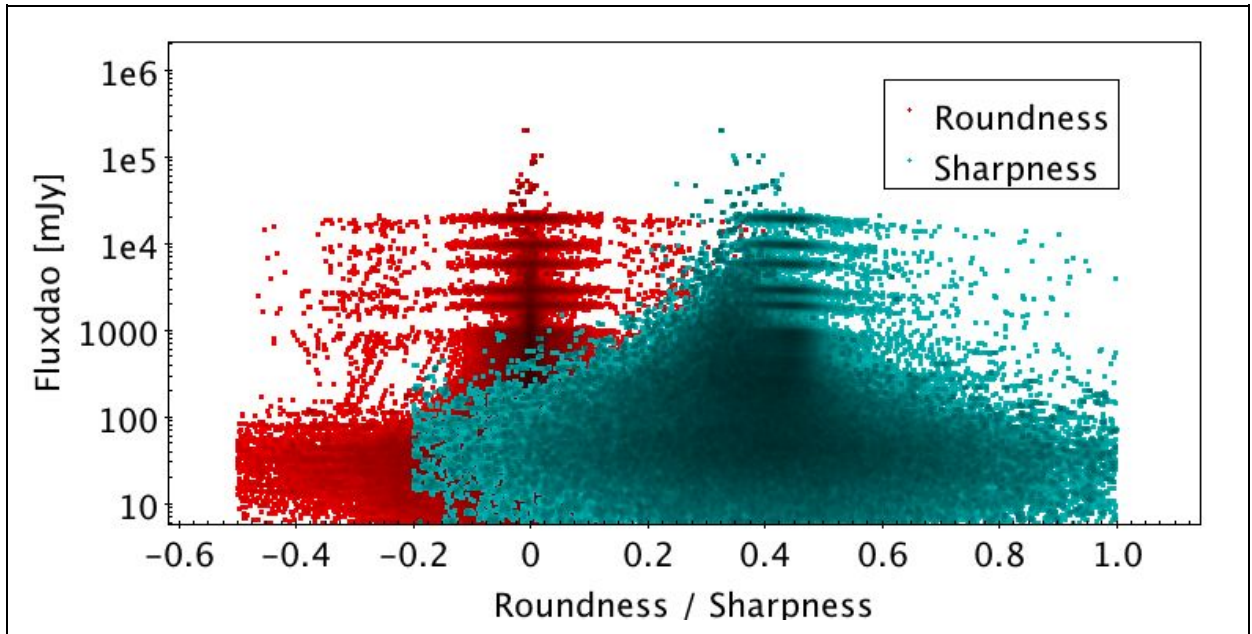

Figure 1.7: Diagram showing Daophot fluxes versus the shape parameters roundness and sharpness for a mix of point and extended sources from various representative environments.

The filter stage after Daophot is very conservative and eliminates only extreme outliers that are typically due to residual effects of high energy particle events and other instrumental artifacts. Based on empirical tests, we accept roundness values in the range from -0.5 to 0.5, while sharpness is accepted within values of -0.2 to 1.0. In addition, we eliminated all Daophot fluxes below 6 mJy, comparable to the one-sigma confusion limits for SPIRE (Nguyen et al. 2012), and required a Daophot quality parameter of at least 1.0.

## Timeline Fitter

The idea for this algorithm was somewhat inspired by the Scanpi procedure from IRAS, as well as by the desire to avoid the loss of positional information that goes with creating a pixelized map. SPIRE timeline data are sampled at rates of 10 (parallel mode) or 18.6 Hz (small or large map). Scan speeds are 30"/sec in small maps and nominal scan speeds in large scan maps. Large maps also were created via a fast mode of 60"/sec, while parallel mode maps scan with 20"/sec and 60"/sec. The possible combinations result in distances of 1.61", 2.0", 3.2", and 6.0" between samples in the sky, depending on observing mode and scan speed. Each sample is associated with a position on the sky, and the ensemble surrounding a point source can be fitted by a two-dimensional Gaussian beam profile. The Timeline Fitter does that, selecting the samples from within a central aperture around the source position and within an annulus further away from the center (see Figure 1.8 for an illustration).

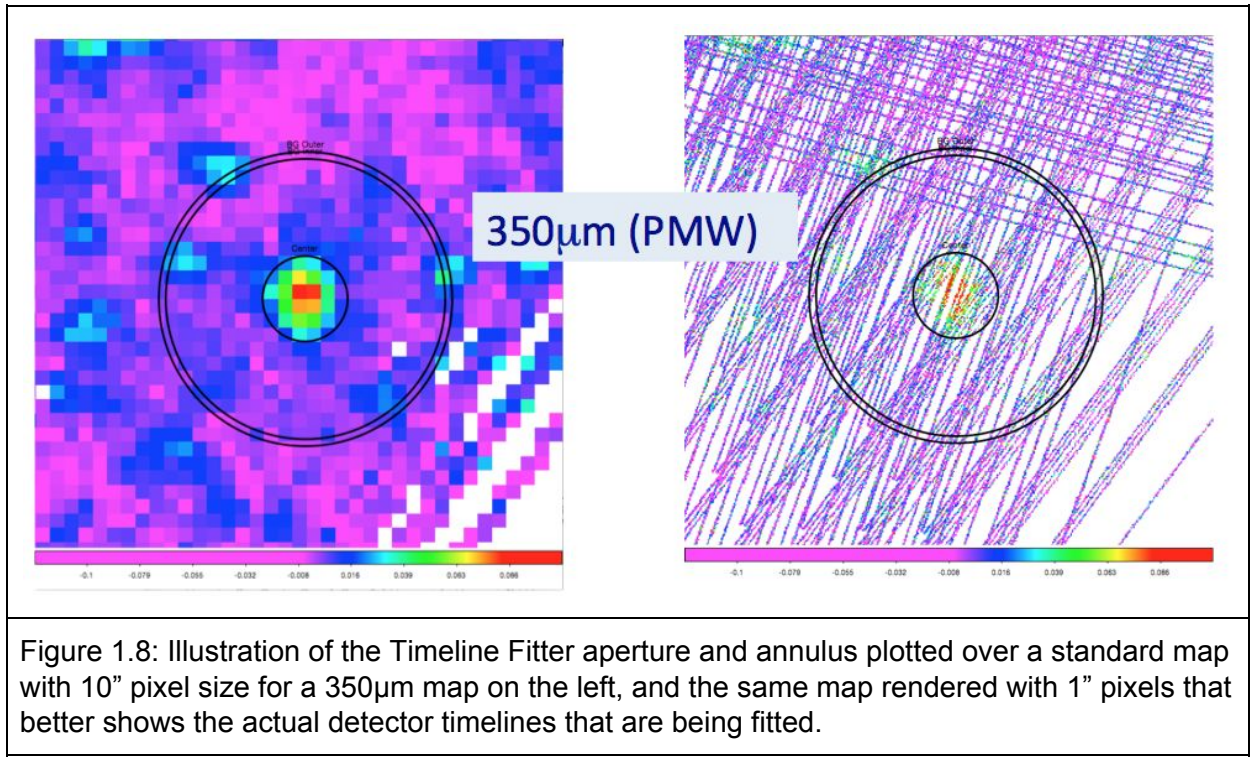

Figure 1.8: Illustration of the Timeline Fitter aperture and annulus plotted over a standard map with 10" pixel size for a 350µm map on the left, and the same map rendered with 1" pixels that better shows the actual detector timelines that are being fitted.

For this project, the Timeline Fitter is operated in two different modes, indicated by the acronyms TML and TM2. The first fits an idealized circular Gaussian beam profile model with 17.6", 23.9", 35.2" FWHM for the three filter bands, respectively. The second fits an elliptical Gaussian, leaving the two values for the FWHM and the rotation angle as free fitting parameters. It also allows for a tilted background plane. The background level is a free fit parameter in both modes. Both modes fit to the readouts within the central aperture and those inside the background annulus. The radii of the central apertures and the background annuli are listed in Table 1.1. Figure 1.8 illustrates how a standard map (left) with a point source for the 350µm filter looks in terms of detector timelines (right). This particular example shows a source close to the fringes of a map, where the orthogonal cross scans did not quite reach anymore, but where moving the detector to the next scan leg provided additional scans at a shallower angle.

Both fits start at the position found by Sussextractor. The position returned is refined during the fit. If the readouts do not provide enough constraints due to noise, the fit can go wrong and the position can drift away. This condition is checked later during the cleaning phase. Another failure mode of the fit is non-convergence. We found that convergence usually takes on the order of 10-20 iterations. All converged fits stayed below 377 iterations, so that the timeline fitters were operated with a maximum limit of 400 iterations.

Both Timeline Fitter runs return a new position, a flux, and values for the background, all with uncertainties. In addition they return the number of iterations performed, the number of readouts in the central aperture, and those in the background annulus. They further provide a flag whether the fit converged, a $\chi^2$ value, a normalized $\chi^2$, and a Bayesian evidence value.

All of these values were collected in a Postgres database table that eventually contained 9927348 records extracted from a total of 5743 maps (5202 Level 2 maps and 541 Level 2.5 combined maps). These were considered source candidates and had to be rigorously filtered in the source table cleaning phase.

## The Four Extractors in Comparison

Before continuing with the description of the next step, we want to highlight some of the differences between the source extractors, in particular their different reaction to slightly extended sources.

Sussextractor is only used for source detection and it is not sensitive to sources that are substantially wider than a point source. It uses a rather crude 5x5 pixel beam profile model for each filter, that operates on the map itself, not the detector timelines. This is illustrated in Figure 1.9, where that model is shown in blue, compared to a realistic beam profile in black. The flux estimates from Sussextractor generally have a larger scatter than those of the Timeline Fitter, but for fluxes smaller than 30 mJy, the uncertainties become similar, as the non-linear fit of the Timeline Fitter can be more easily thrown off-course by instrument and confusion noise.

Slightly extended sources are still detected by Sussextractor, but the flux returned is usually too small, compared to what would be obtained from integrating the extended beam profile (see Figure 1.10). The flux reported by Daophot is obtained from integrating within apertures, that are matched to the beam profile of a point source, as illustrated by the pink dashed lines in Figure 1.9. For extended sources the flux will be larger than that of Sussextractor, but still underestimated, due to the limited radius of the aperture and the background annulus. More flux outside the aperture is lost than if it were a point source, which is what we only correct for. On the other hand, source flux is spilling out into the background annulus, raising that level and diminishing the eventual flux difference further. Yet, a significantly elevated Daophot flux, compared to that of the Timeline Fitter or Sussextractor, is a good indicator of an extended source.

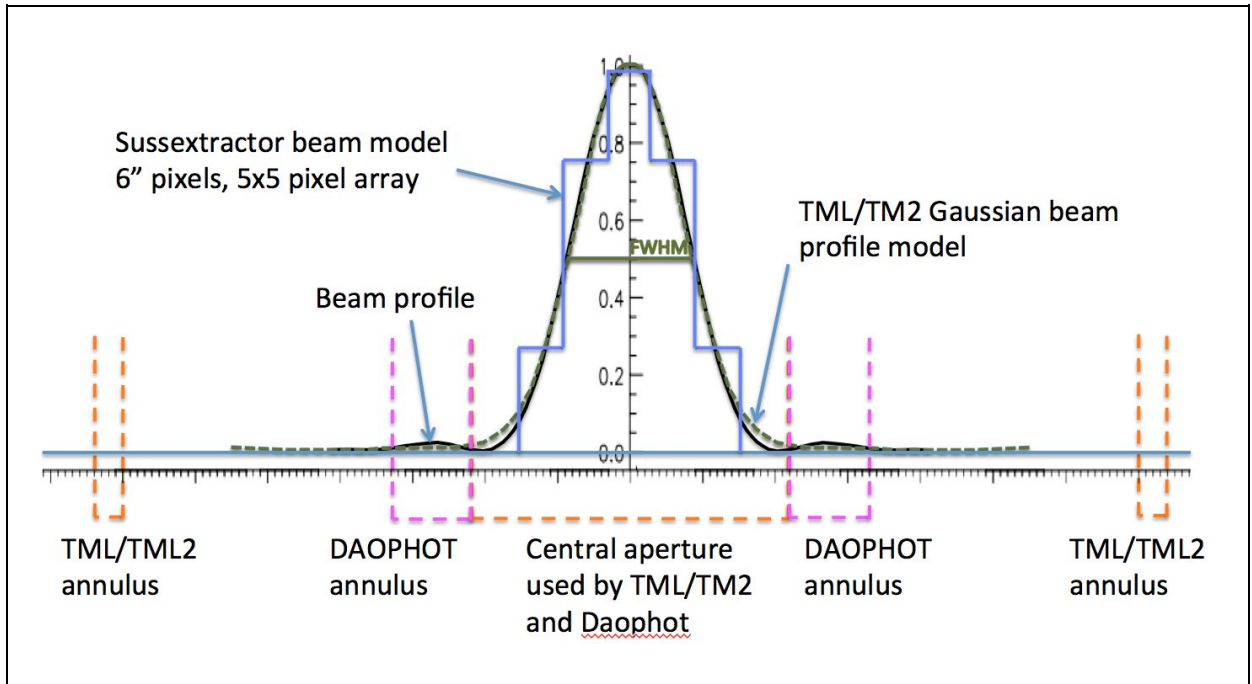

Figure 1.9: Illustration how the beam profile models, apertures, and background annuli used by the different source extractors match with a real instrument beam profile of a point source.

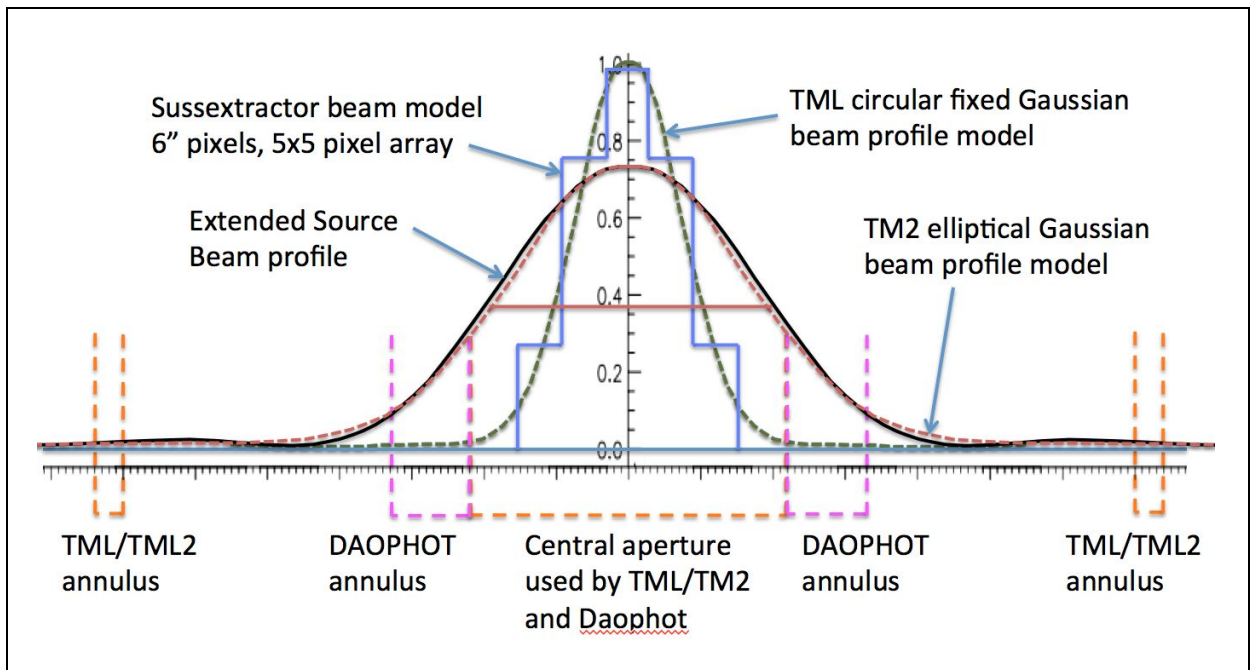

Figure 1.10: Illustration how the beam profile models, apertures, and background annuli used by the different source extractors match with an extended source that still resembles a Gaussian profile.

The Timeline Fitter, using a fixed circular Gaussian beam profile model, delivers the best flux estimates for point sources. It fits the Level 1 readouts within the central aperture, which is the same as that of Daophot, and the readouts inside the background annulus, which has a much larger radius than that of Daophot (see Figure 1.9, orange dashed lines). Similar to Sussextractor, the Timeline Fitter is underestimating the flux of an extended source, but tends to produce higher fluxes than Sussextractor in this case. At fluxes below 30 mJy we observe higher uncertainties due to background confusion and instrumental noise, and comparison with results from the other methods becomes more important.

Finally, the second Timeline Fitter run that uses an elliptical free FWHM Gaussian beam profile produces values very consistent with the Timeline Fitter for point sources, as expected. The scatter is similar or even slightly smaller than that of Sussextractor in comparison. For extended sources its values are generally larger than the other flux values, most pronounced if compared to those of Sussextractor. For this type of source, assuming it still fits the shape of a Gaussian profile, the TM2 run has the highest potential to produce a reliable flux estimate for an extended source. Unfortunately, many extended sources are actually groups of point sources that cannot be separated anymore by the source detection and extraction algorithms used here. These confused source groups can assume a variety of shapes and will require visual inspection and a more thorough analysis than our automated extraction can provide.

## Cleaning Source Tables

Before starting to consolidate source detections into object positions in the sky, the source table needed to be cleaned from spurious entries that occur due to bad timeline fits. The cleaning consists of resetting a master flag for the respective record if one of the following conditions is true: 1) One of the two Timeline Fitter runs did not converge; 2) the source positions returned by the Timeline Fitter is further away from the original Sussextractor position than half of the wavelength dependent FWHM (see Table 1.1); or, 3) the flux returned by any of the Timeline Fitters is zero or negative, or if the TML flux exceeds 10000 Jy. This step removed 1749188 source candidates.

In a second step an average FWHM $\overline{\delta}$ was calculated as:

$$\overline{\delta} = \frac{\sigma_1 + \sigma_2}{2} \; 3600 \; \sqrt{8 \; ln(2)}$$

for source candidates with a Timeline Fitter SNR > 5, where $\sigma_1$ and $\sigma_2$ are the Gaussian sigmas returned by the TM2 run in degrees. The result is expressed in units of arcseconds. Depending on flux interval and wavelength, all source candidates that have a smaller average FWHM than those given in Table 1.2 were eliminated by resetting their master flag.

| **Flux Range** | **250 μm** | **350 μm** | **500 μm** |
|---|---|---|---|
| S > 300 mJy | 14.2" | 18.7" | 26.9" |
| 300 mJy > S > 100 mJy | 12.4" | 16.4" | 23.7" |
| S < 100 mJy | 11.3" | 15.1" | 20.8" |
| Table 1.2: Lower FWHM limits in arcsec for a source entry not to be considered spurious and discarded. | | | |

The thresholds were determined from artificial point source injections into a map of the COSMOS field (obsid 1342195856), limiting the Timeline Fitter SNR to 5 and above. This filter removed another 35733 of the source candidates, leaving us with a total of 8142427 source detections that split into (3901124, 3170534, 1070769) for the three filter bands, respectively. We note that these limits are based on an older state of development of the point source discrimination than is used later for setting the flags that distinguish extended sources, point sources, and low FWHM sources. The method used is still acceptable as it only removes extreme unphysical cases.

## Map Position Corrections

The Herschel telescope pointing is generally good to within 2 arcsec (68% c.i.), although sometimes the pointing can be off by 5 or more arcsec. That is why it is useful to have an idea of the absolute astrometry and correct for possible offsets.

In order to derive the absolute astrometry for the SPIRE maps we make use of the WISE all sky catalog (Wright et al 2010, Cutri et al. 2012), using band W4 (22 μm) detections at more than 3 sigma. We build a stack with 11x11 pixel cutouts (=66"x66") from the SPIRE 250 μm map (the one with the best spatial resolution) centered on each WISE 22 μm source (see Figure 1.11). The stack signal is then fit with two circular Gaussian 2D models: one where we keep the FWHM fixed to the FWHM of a point source, and the second model in which it is left as a free parameter. Then we compare the two models in terms of the odds derived from the Bayesian evidence. This helps to identify maps where the stack signal is indeed a point-like source: i.e., those where the first model is preferred or the second one is preferred, but the best fit FWHM is equal, within the uncertainties, to the 250 μm beam FWHM. The derived offset of the stack signal from the stack centre is then the astrometry offset.

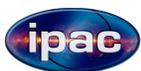

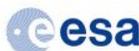
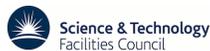

This method does not work in fields with extended emission, when the stack signal is dominated by sources more extended than the beam. In some cases one or a few strong resolved sources may dominate the signal and lead to unreliable results, which the Bayesian evidence approach is not capable to flag. Therefore a visual inspection is necessary in order to remove any dubious stack results.

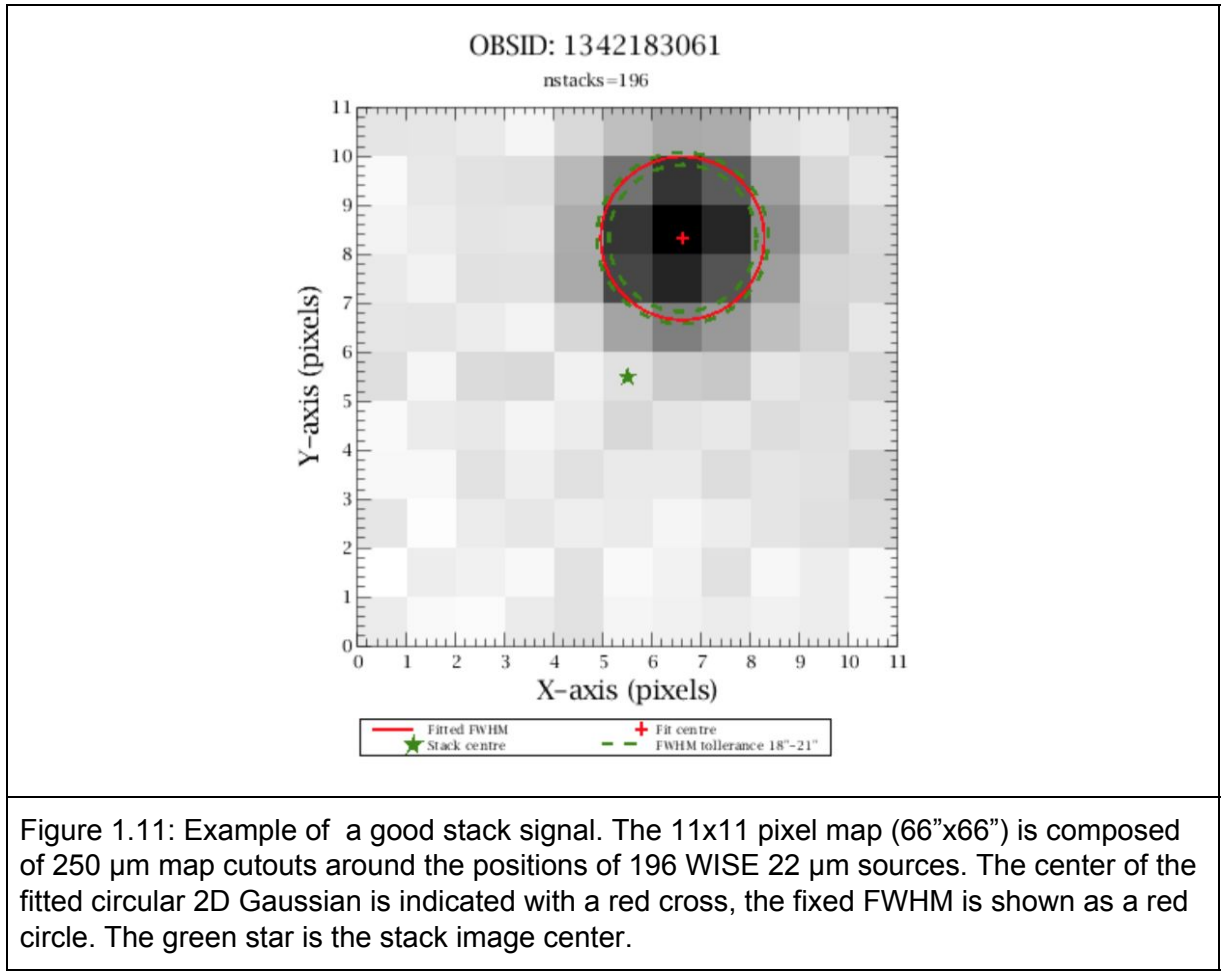

Figure 1.11: Example of a good stack signal. The 11x11 pixel map (66"x66") is composed of 250 μm map cutouts around the positions of 196 WISE 22 μm sources. The center of the fitted circular 2D Gaussian is indicated with a red cross, the fixed FWHM is shown as a red circle. The green star is the stack image center.

We applied the stacking method on all 6959 publicly available SPIRE 250 μm Level 2 maps. Note that we do not use merged maps (Level 2.5) for the stacking. The stack fit failed for 481 maps and it was "successful" for 6472 maps. Out of these results, by visually inspecting all maps with reported offsets greater than 5 arcsec, we identified 110 with good stack results and confident astrometry offsets.

For this version of the catalog we use this information to flag (ASTROM_FLAG) all objects that have at least one contributing detection that can be traced back to one of these 110 maps. In a potential future revised version the derived positional offsets can be used to correct the astrometry of Level 1 timelines, before the map reconstruction of combined maps, in order to avoid elongated or even duplicate sources.

## Object Consolidation

Of the total of 6878 usable SPIRE scan map observations, 1676 have been combined into 541 Level 2.5 maps to improve map quality and SNR. Although the majority of the detections is just detected once, there are still many maps that overlap each other. This is illustrated in Figure 1.12, which shows a histogram of the number of detections of the same object in different maps with that number going over 90 for some. Note that the numbers shown apply for the final list after all SNR and structure noise (STRN) cuts have been imposed (see below).

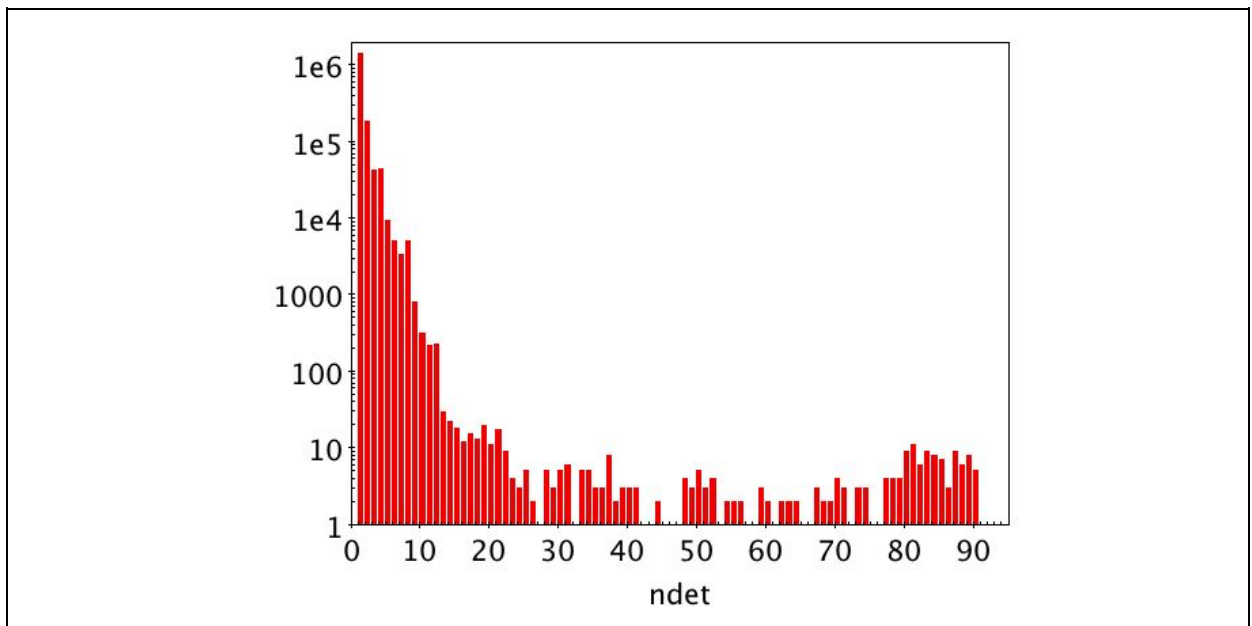

Figure 1.12: Histogram of the number of detections of an object (NDET) at the same position and filter band. Although the majority has only one detection, there is a sizeable number of multiple detections where maps overlap.

In order to consolidate the many overlapping source detections into discrete objects in the sky that are indistinguishable at the respective SPIRE filter dependent spatial resolutions, we used the fast spherical search and indexing plug-in Q3C V 1.4.23 for use with Postgres databases (Koposov & Bartunov 2006). The entire procedure was developed and implemented as a sequence of stored SQL procedures and functions. Its primary function is to identify groups of source detections that are located so close together that they cannot be distinguished within the beam FWHM of the respective SPIRE filter, and assign them a new identifier in a group table. We call such a record an object in our list.

The algorithm loops through every observation. We put database indices on the coordinates generated by the Timeline Fitter in the source table and clustered it for optimal search performance. For a given observation it first creates a temporary table that contains all valid sources within that observation that have not yet been assigned to a group ID. If that table is not empty, it goes on to match the positions of that table against all valid positions in the full source table. We used a search radius of half the FWHM of the respective filter band, increased by 6 arcseconds to allow for a 3-sigma pointing uncertainty of the spacecraft.

For each group of source detections within the same filter band and within the search radius, the procedure then creates a new unique group ID and sets the respective filter ID in the group table. It further enters the new group ID into the respective column of all detections in the source table that belong to the group, thus creating a foreign key that links several records in the source table to one record in the group table. The entries of the group table then represent the list of actual SPIRE objects in the sky, detected at the given wavelength. The group ID entry in the source table has the additional function of a flag, as it signals to the algorithm that a given detection has already been assigned to a group.

As long as the source positions and the spacecraft pointing are within their budgets, this procedure works quite well. However, if the pointing performance is poor or if the distance between two point sources is less than the search radius, the inherent asymmetry of our approach requires additional fixes. The issue is illustrated in Figure 1.13.

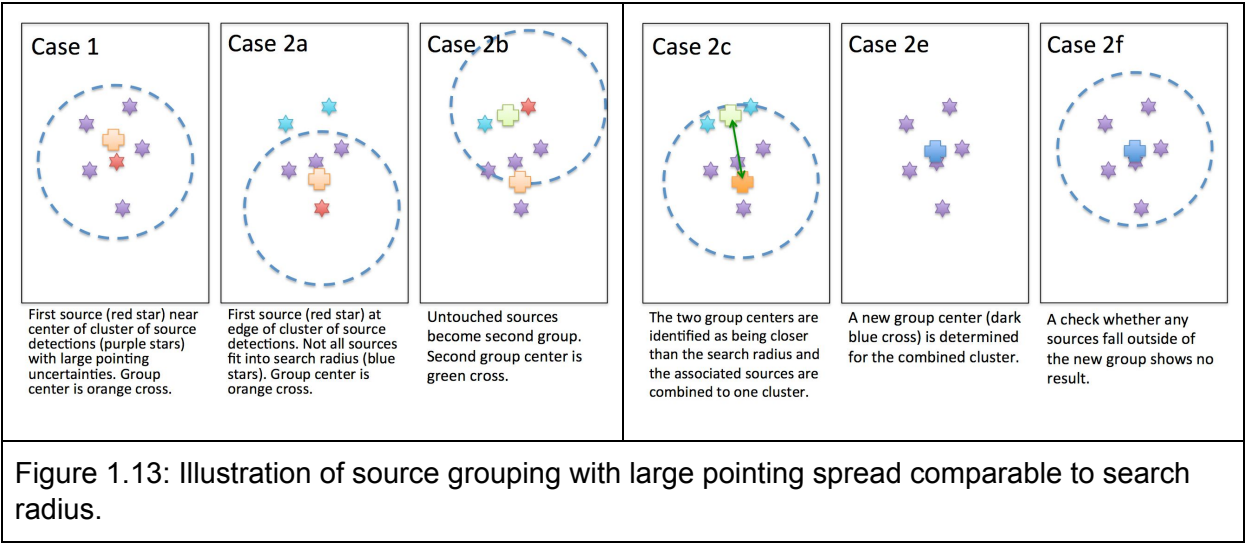

Figure 1.13: Illustration of source grouping with large pointing spread comparable to search radius.

Case 1 shows the situation where the algorithm starts with a source near the center of the cluster of source positions. All positions fit into the search radius, and the new group center represents the new position of the group (list object). In Case 2a the arbitrary starting point is a source near the edge of the cluster. Not all sources fall into the source radius, leading to the creation of a second group from the leftovers (2b). In a corrective step, group centers with distances below the search radius are identified (2c) and its group members are combined into one (2e). To make sure that the combination does not actually represent two sources, another check is performed in which sources outside of the search radius are removed from the group and made available for another group consolidation iteration (2f). We performed two iterations, ending with the group combination step for too close-by group centers (step "e").

Once the groups are determined, the final group positions and uncertainties are calculated, and the number of detections (ndet) parameter, as well as the position flag, are updated.

## Source Fluxes and Uncertainties

After grouping is complete, there are between one to several source detections per object. In order to obtain the best flux and uncertainty estimate for a given object, we decided to average the contributing fluxes, weighted by the uncertainties from their respective extractor runs. This is done for all four flux flavors, Sussextractor, Daophot, and the two Timeline Fitter runs, TML and TM2. Details are given in the descriptions of the individual columns further below. Each of these fluxes is accompanied by a propagated weighted uncertainty.

These uncertainties reflect well instrument noise and possible issues with the individual extraction on a relative basis, but they do not account for the total uncertainty due to the structure of the immediate background, i.e., the confusion noise. In order to obtain a better estimate of the total uncertainties, we undertook point source injection simulations and tied the variance we observed in the re-extracted fluxes with the Timeline Fitter to both the injected flux and the STRN around the source. We call the STRN and the flux dependent variation the Total Noise for a given detection and describe its detailed derivation later.

This Total Noise can be reduced to an estimate of the local confusion noise by quadratically subtracting the instrument noise for a given observation. We estimate the instrument noise from an empirical function that depends on the number of readouts in the central aperture of the Timeline Fitter. When combining several detections of the same object, all the local confusion noise estimates of the detections are averaged, weighted by the respective uncertainties, to yield the best estimate for the local confusion noise of the object.

The total flux uncertainty of an object (group of detections) is calculated by quadratically adding the estimated instrument noise, derived from the sum of all readouts of all detections within the central aperture of the Timeline Fitter, with the local confusion noise. This value is strictly only applicable to point sources and the TML fluxes and may need to be scaled up for application to extended flux estimates derived from the TM2 run.

This total uncertainty is the best uncertainty estimate for point sources that we have. In this list it is also used to define an SNR threshold that is required to be larger than 3 for a source to appear. In addition we require the ratio of TML-generated flux and uncertainty to be larger than 3 as well, to safeguard against glitches of the individual TML run. Both SNR thresholds improve reliability and provide for better flux estimates.

## Structure Noise

The STRN measures the intensity fluctuation around a given map pixel. It includes the spatial noise of the celestial environment, as well as the noise contribution from the instrument. For each pixel of the map the structure noise is calculated as the standard deviation of all flux differences between the pixel and all the surrounding pixels at a fixed distance, in a circular configuration. For more details of the theory we refer to Kiss et al. (2005).

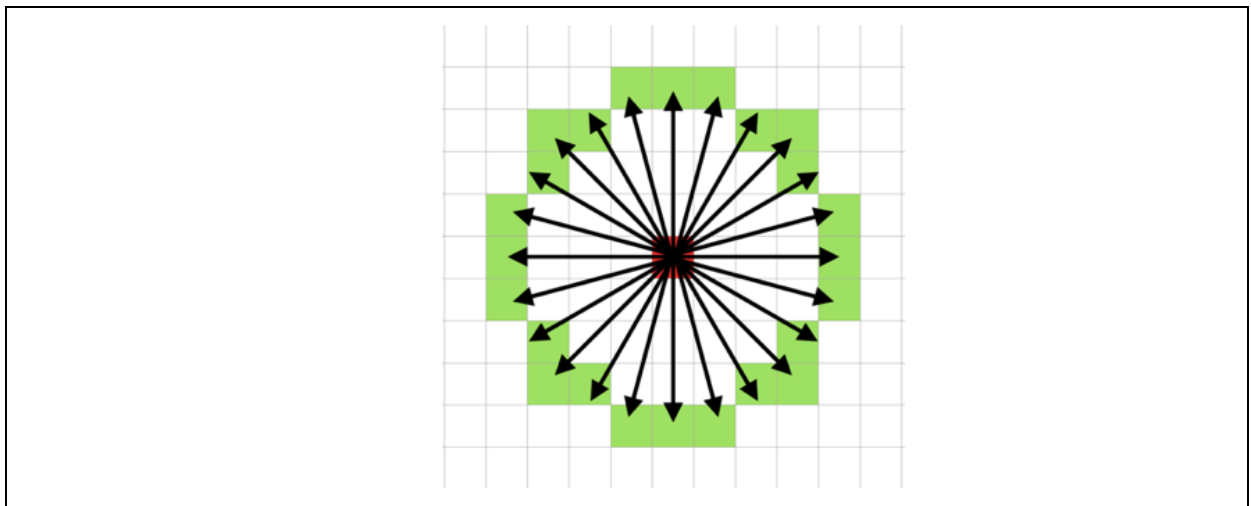

Figure 1.14: The circular configuration used for structure noise calculation. The target pixel is shown with red square. The neighbouring pixels at the predefined angular distance are shown with green squares.

We create the structure noise maps using an IDL code. The code itself reads the same point source flux calibrated FITS maps used for generating the initial list of detections with Sussextractor. For each pixel of the map (target pixel, shown as a red square in Figure 1.14) we are looking for neighbouring pixels in 24 directions (green squares in Figure 1.14) at a certain angular distance. If the distance is not big enough, then less than 24 pixels are found. In this case the unique pixels are selected. The next step is to calculate the absolute difference between the unique neighboring pixels and the target pixel. When the number of unique data points is larger than three, the standard deviation of these differences is stored as a pixel value in place of the target pixel. The pixel value is calculated according to :

$$\sigma_{strn} = \sqrt{\frac{1}{24} \sum_{i=1}^{24}(d_i - \mu)^2}$$

where $d_i$ are the differences of each of the fluxes of the 24 (or less) pixels and that of the central pixel, and $\mu$ is the average of all $d_i$. The resulting structure noise map is then stored as a standard FITS file with the header of the original map.

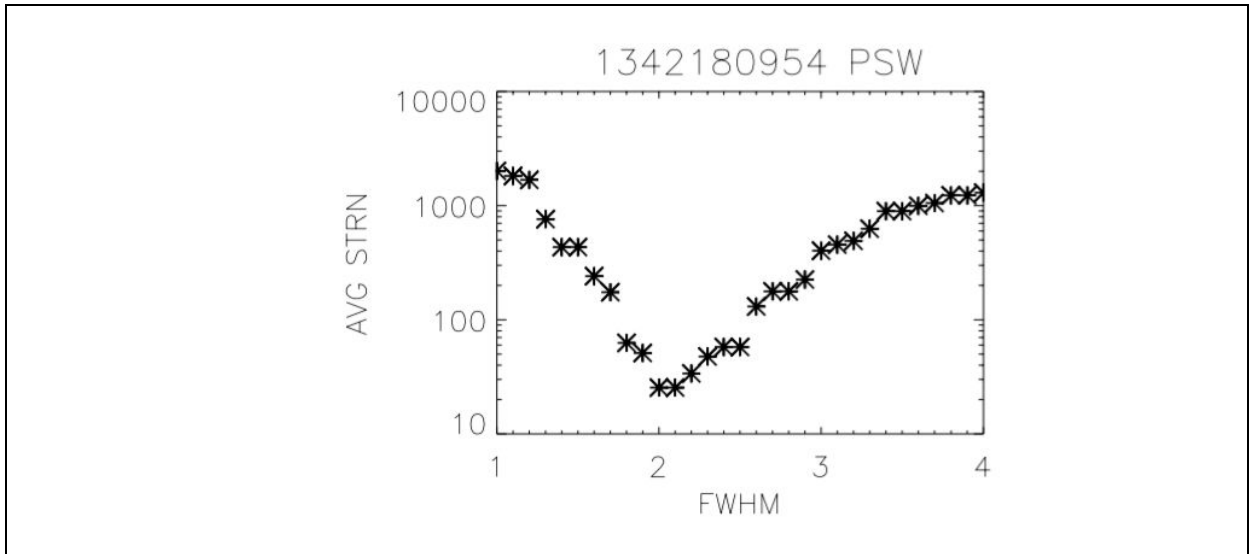

Figure 1.15.: The average structure noise value around the injected sources of 20 Jy into a 250μm map, as a function of the angular separation between the target pixel and the annulus, in units of the beam FWHM.

The distance between the target pixel and the neighbouring ones corresponds to a spatial frequency optimized for each filter band. Choosing too small a distance would include flux from the point-spread function (PSF) wings, causing an additional noise term that scales with the source flux. To minimise this effect, we created structure noise maps of simulations, in which we injected sources with 20 Jy flux and increased the angular distance between the target pixel and the neighboring pixels. On each structure noise map the structure noise value at the position of our artificial sources was calculated and checked. This test confirmed that the structure noise value depends on the spatial scale on which the structure noise maps are calculated. We found that the dependency has a minimum at 39" (2.21+/-0.37 FWHM), 47" (1.94+/-0.44 FWHM) and 64" (1.82+/-0.53 FWHM) in the 250μm, 350μm and 500μm bands, respectively. The structure noise value decouples from the source flux at these distances. We decided to use this angular separation to create our structure noise maps and to attach a structure noise value to each of our detections. Figure 1.15 shows an example of the 250µm structure noise value dependence on distance, expressed here in multiples of the FWHM of the beam.

The pixel value has the same units as the input map. To avoid so-called NaN-Donuts, single NaNs in the input map are interpolated before the calculation.

The structure noise value for a given source is calculated as follows: We place an aperture at the extracted position of each source, the diameter of which equals the corresponding beam FWHM, i.e., 17.6", 23.9" and 35.2" for the respective filters. The average structure noise is then calculated inside the aperture. Eventually the resulting STRN value is attached to each source in a separate database column. The derivation of the structure noise maps and the calculation of the structure noise value were performed in IDL, outside of HIPE.

## Structure noise based error

We have developed a method that gives a proper estimate of the error of the measured flux. As described in the section "Simulations", below, we injected sources into fields with various complexity. The same pipeline we used to detect our sources and collect photometry from real observations was used to detect our simulated sources and to measure their flux. Also, the structure noise values were collected in the same way. This procedure allowed us to compare the input flux to the measured flux as a function of the structure noise.

Since we know the theoretical flux $S_{in}$ for each injected source and we also have determined the complexity of their environment through the STRN $\sigma_{str}$, we could deduce the statistical uncertainty of our extracted fluxes as a function of these two parameters as follows: From our database of extracted artificial sources, for each injected flux level $S_{in}$ (in this example, 200 mJy), we selected the measured flux $S_{meas}$ and the STRN $\sigma_{strn}$. Binning the STRN into 20 mJy/beam intervals, we calculated for each bin the average measured flux $S$, its standard deviation $\sigma_S$, and the average $\sigma_{strn}$. We then calculated the Signal to Noise Ratio as $SNR = S_{in}/\sigma_S$ for each flux level and fitted the data with a power law of the form $SNR = A\,\sigma_{strn}^B$ to be able to calculate SNR values for any intermediate $\sigma_{strn}$. An example is shown in Figure 1.16 for the 200 mJy flux level. The data points are plotted as black crosses, while the fitted model is shown as the solid red curve. The uncertainty of the data points used in the fit was calculated as 1/N, where N is the number of data points in each bin.

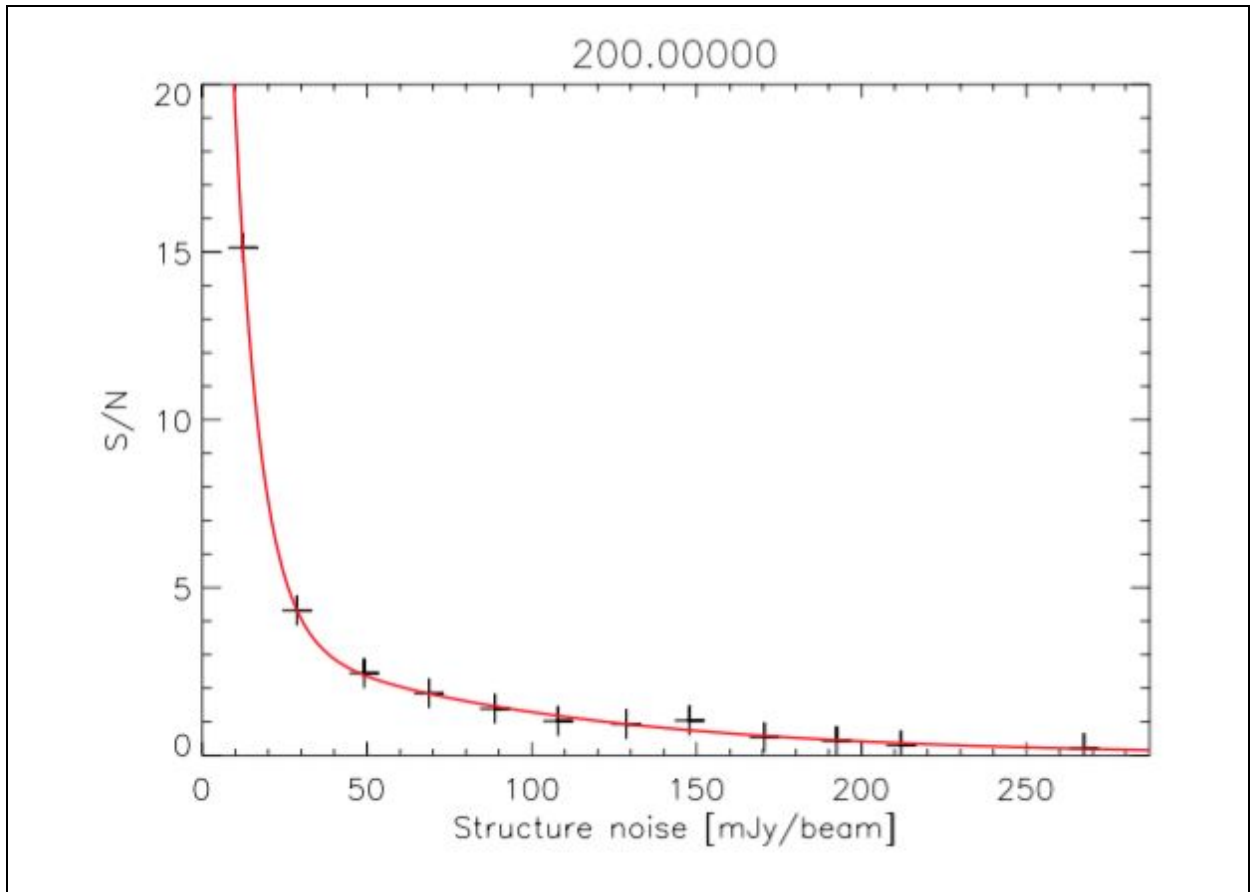

Figure 1.16: The SNR values as a function of the structure noise for sources with 200 mJy injected flux (black crosses). The SNR value is large at small structure noise values and drops rapidly for higher $\sigma_{strn}$. The fitted power function is presented with the solid red line.

The same procedure was repeated for all the flux levels. For each flux level the SNR values were stored for $\sigma_{strn}$ values between 0 and 2000 mJy/beam in intervals of 1 mJy/beam. In order to achieve a better resolution along the $S_{in}$ dimension we fitted the data in this direction as well. Figure 1.17 shows the logarithm of the SNR values as a function of the logarithm of the input flux values at a $\sigma_{strn}$ level of 200 mJy/beam. As the figure shows the SNR values increase rapidly up to ~1000 mJy, but a plateau-like feature develops towards the higher flux values, where the increase is much less significant. To handle this behaviour, we fitted a function of the form

$$SNR = C + \frac{D}{E + e^{\frac{F-S_{in}}{G}}}$$

to each STRN level, where C to G are fit parameters.

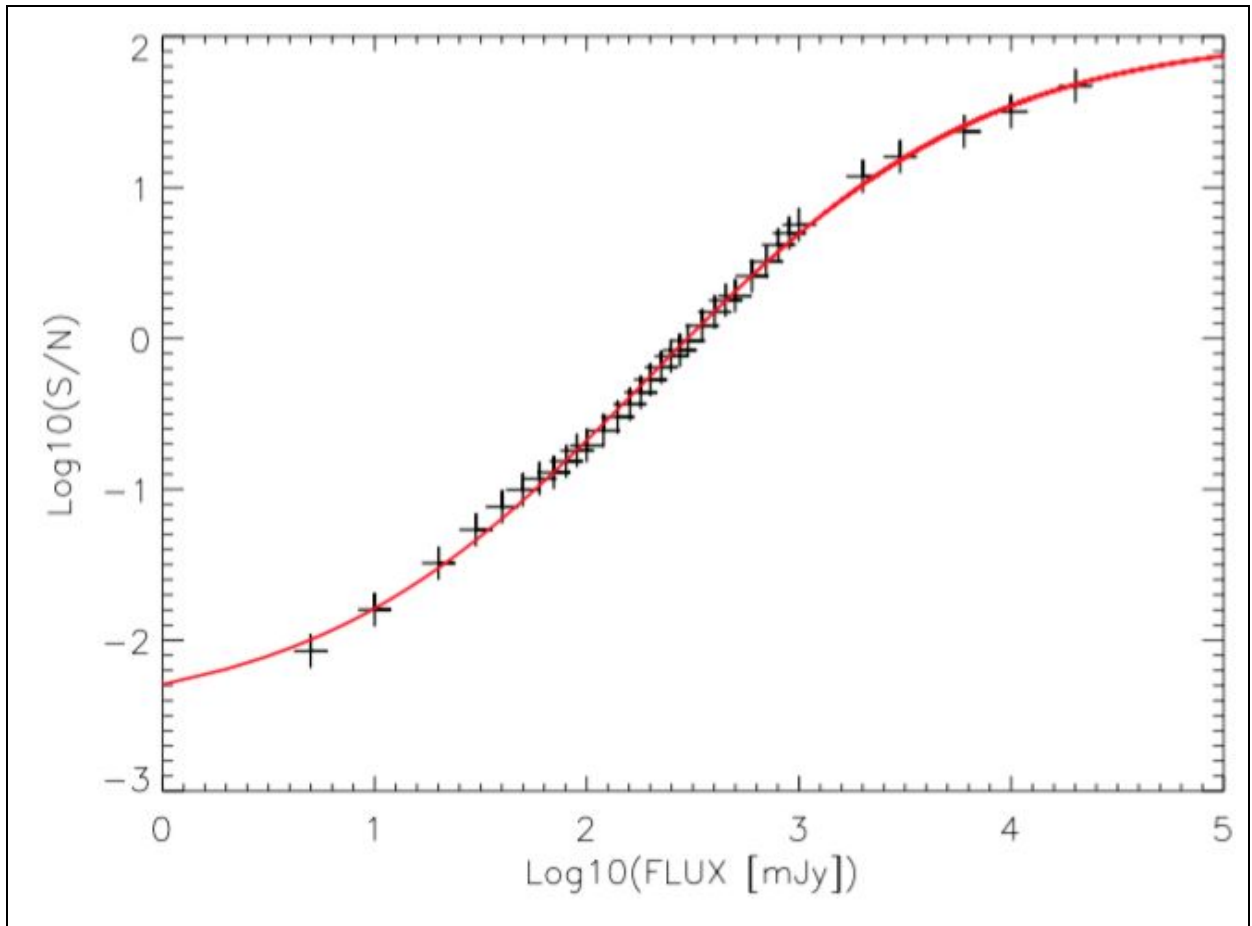

Figure 1.17: The SNR values as a function of input flux (black crosses) at a fixed structure noise level of 200 mJy/beam. The fitted exponential function is shown as red solid line.

After completing the fit procedure in both $\sigma_{strn}$ and $S_{in}$ directions, we ended up with SNR surfaces that covered a flux range between 0 and 100 Jy and covered the $\sigma_{strn}$ range between 0 and 2000 mJy/beam (see Figure 1.18). This grid has intervals of 1 mJy in the STRN direction and 1 mJy in the flux direction. The resulting arrays (SNR surfaces) have dimensions of 100,000 x 2,000 data points. The surfaces for all three bands were then used to calculate the SNR value for all of our source detections. The corresponding SNR value was calculated by the IDL "interpolate" function. Such a fine grid allowed us to interpolate with an acceptable accuracy. Note that all total flux uncertainties of sources fluxes above 100 Jy and from STRN values above 2 Jy are extrapolations from our actual simulations and are to be considered carefully.

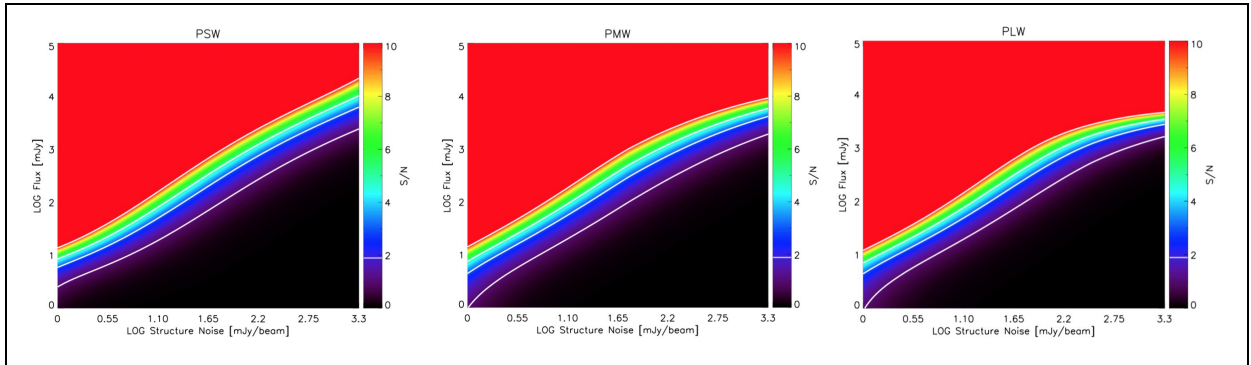

Figure 1.18: Log-log contour plot of the SNR values as a function of structure noise and source flux for the PSW, PMW and PLW arrays (left to right). SNR levels of 1, 3, 5 and 10 are overplotted with white contour lines.

## Instrument Noise

The total flux uncertainty is composed of two components, the instrument noise and the astronomically interesting background confusion noise, which can be extracted by removing the former. There have been no hints of strong variations of instrumental noise, and high energy events in the signal timelines are eliminated by the two-pass de-glitching scheme in the data reduction pipeline. In addition, in most cases the signal uncertainty is dominated by background confusion, so it is sufficient to estimate the instrument noise empirically, based on the instrument configuration. The important variables are: Mapping mode, scan speed, readout frequency, number of repetitions, detector array, and detector bias. The first four all affect one fundamental parameter, the number of readouts per sky area, which can be determined directly from the data. This leaves only the detector array (one of the three filter bands) and the detector bias (nominal and bright mode) as additional free parameters.

We derived the instrument noise using twelve "large map" (single cross-scan) observations of the COSMOS field, covering the same sky area. This dataset allowed us to combine multiple maps in order to simulate an increasing number of repetitions, analogous to an increasing integration time. 100 sources were injected into the timeline data of the 12 different observations at the same sky positions with 300 mJy source flux. We then reconstructed 12 maps from those data, successively including the data of one more observation into each map, such that the Nth map has N observations combined. (In the first one, only the first observation was used; in the last one, all 12 observations were used.) From each map we created a source table with the same pipeline used for the catalog generation. We then identified the injected sources and calculated their average flux and the flux uncertainty by calculating the standard deviation of the measured flux. From such an exercise the instrument noise is expected to converge towards zero as it is decreasing by $\frac{1}{\sqrt{N}}$, where N is the number of observations combined (equivalent to the integration time). The convergence continues towards the confusion floor, for which only the sky fluctuation is present. For each source we determined the number of readouts in the central aperture of the Timeline Fitter $n_{readout}$ that describes the depth of the map from which the photometry was derived. Figure 1.19 shows the flux uncertainty σ as a function of average readout number for the detected sources in each map. The fitted power laws gave us the functions that we used to calculate the instrument noise portion of the flux error for each source as a function of the number of readouts $\sigma_{inst} = a\, n_{readout}^{b}$. The coefficients for the three arrays are listed in Table 1.3.

| **Filter band** | a | b |
|---|---|---|
| **250 μm** | 423.6 | -0.7909 |
| **350 μm** | 148.9 | -0.6020 |
| **500 μm** | 123.7 | -0.5305 |

Table 1.3: Parameters used to estimate the instrumental noise based on the number of readouts found in the central TML aperture.

The confusion floor we derived from the exercise is 4.8, 4.4 and 4.8 mJy for the 250μm, 350μm, and 500μm arrays, respectively. In contrast, the values calculated by Nguyen et al. (2010) appear to be slightly higher (5.8, 6.3 and 6.8 mJy/beam).

The maps used for the simulations strictly only cover the "nominal" bias mode. For the 101 valid observations that were performed in high bias (bright) mode, a separate assessment would have to be done. Given that the instrument noise portion is only a small component, especially for brighter sources, and that only 715 of the entire list of objects are actually affected, we only flagged objects for which bright mode detections are among the contributing ones and for which the total error and the confusion noise value may be affected by the underestimated instrument error.

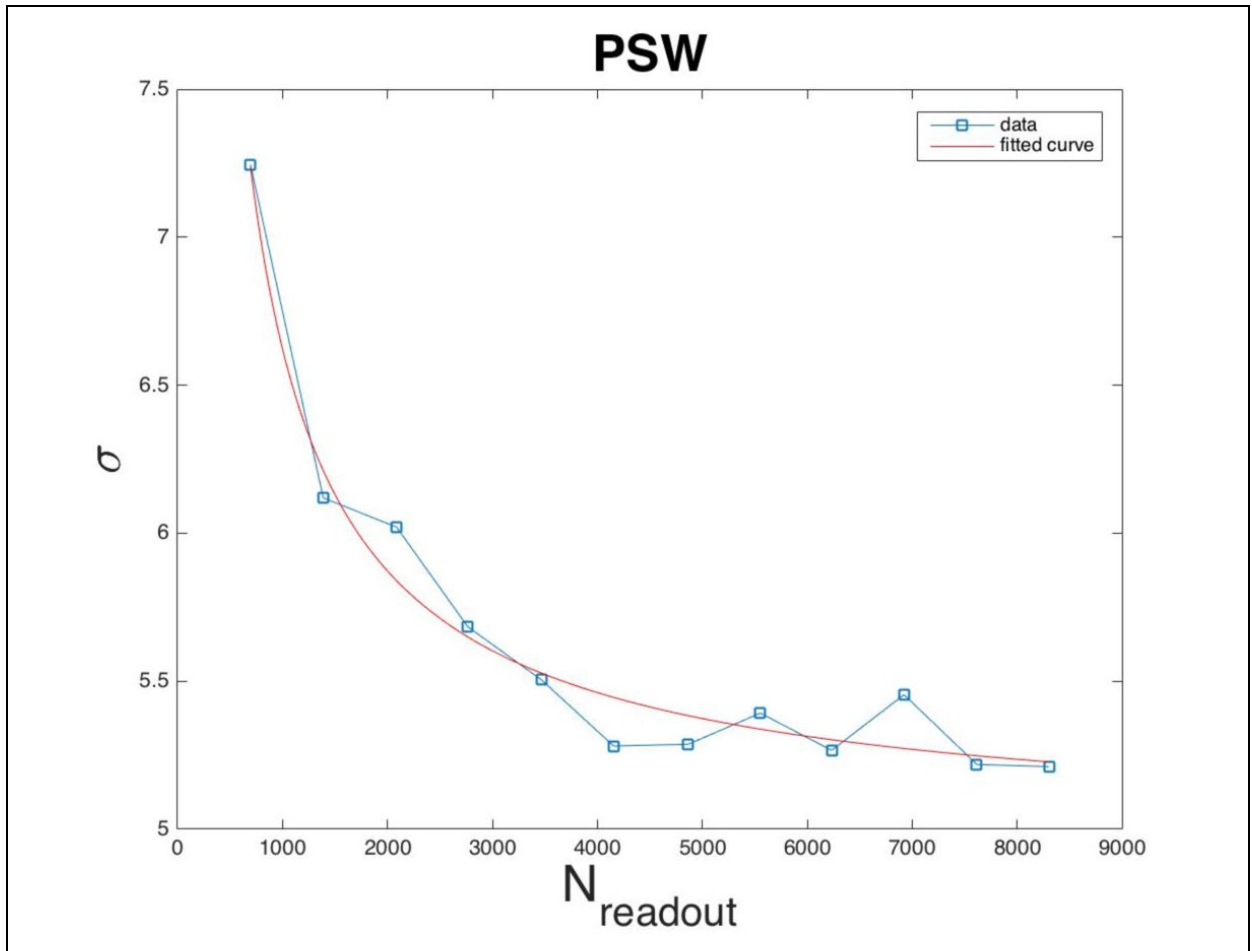

Figure 1.19: The flux uncertainty (σ - standard deviation of the measured flux) values as a function of the average readout number (blue squares). The uncertainty is decreasing according to a power law. The fitted function is shown as a solid red line.

## Structure Noise Threshold

From the start of the project it was clear that point source extraction and photometry would be difficult in the regions with strongly structured background, most prominently in the Galactic Plane. We adopted the concept of structure noise to identify such regions and as a handle on the uncertainties. During the validation of our list it became clear that most of the sources we found in strongly confused regions were extended, that the detection and cleaning stages missed many sources, and that the extended source fluxes were often different from existing extractions from other groups, in particular the Hi-GAL survey.

Taking into account that this project was started with point and point-like sources in mind, and that a good understanding of the properties of the source extraction in highly confused regions would certainly require much more work and potentially a different detection and extraction method, we decided to exclude regions on the sky from our catalog that rise above a certain confusion threshold. By excluding contiguous regions in a way as if they were not observed, rather than excluding individual sources based on their local STRN value, we avoid potential additional statistical biases that are poorly known.

We define the excluded regions using the tiles defined by the Q3C indexing scheme (Koposov & Bartunov 2006) at the 22 bit level. These tiles do not guarantee an equal area of the tiles, as does HEALPix (Górski et al. 2005), but this was not important for our purposes, and we had already used Q3C for database indexing. The tiles at the 22 bit level measure about 16.7 arcmin from corner to corner, and their locations are well defined. Each tile was assigned a STRN value determined as the median STRN at all object positions that fall within the boundaries of the tile. All 6.7M object positions before the SNR threshold application were used for maximum coverage. The histogram is shown in Figure 1.20.

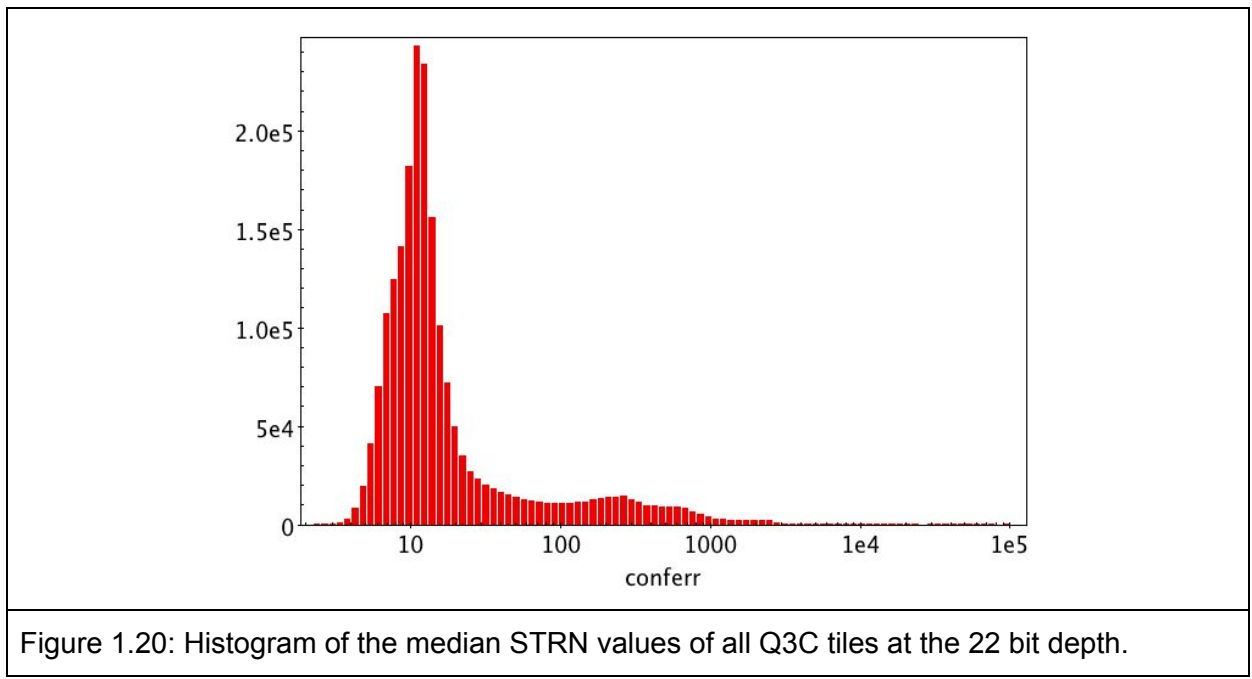

Figure 1.20: Histogram of the median STRN values of all Q3C tiles at the 22 bit depth.

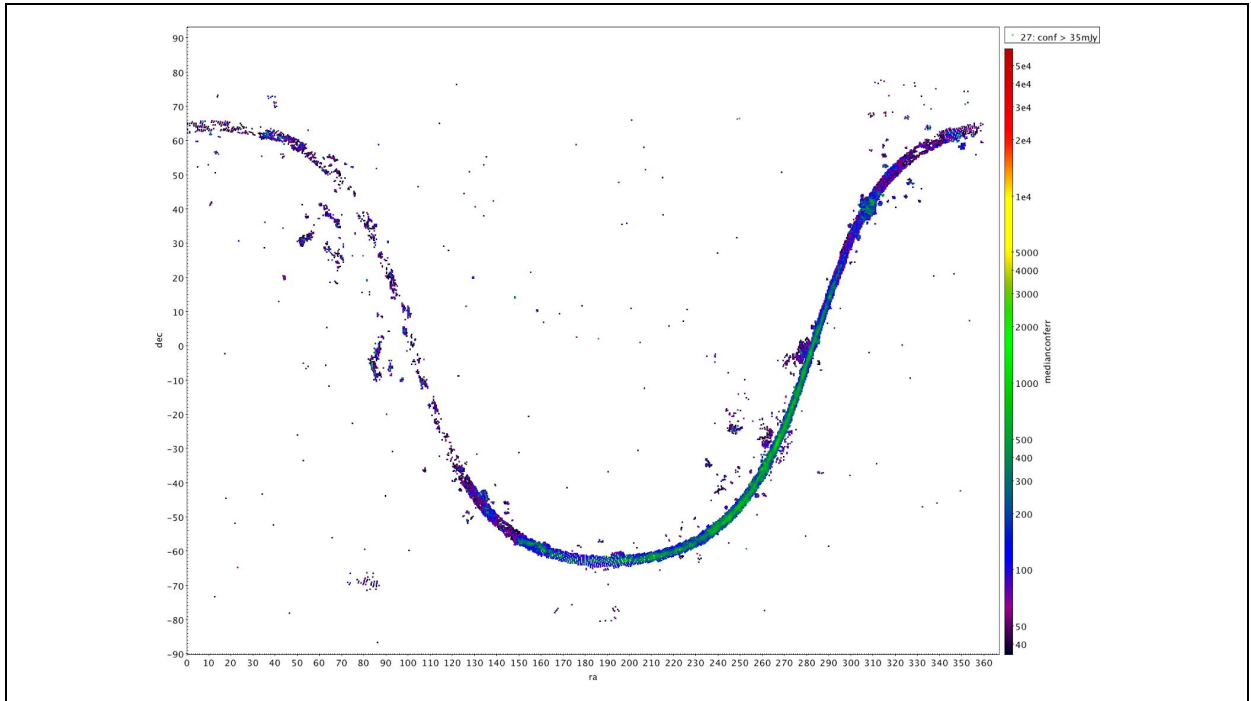

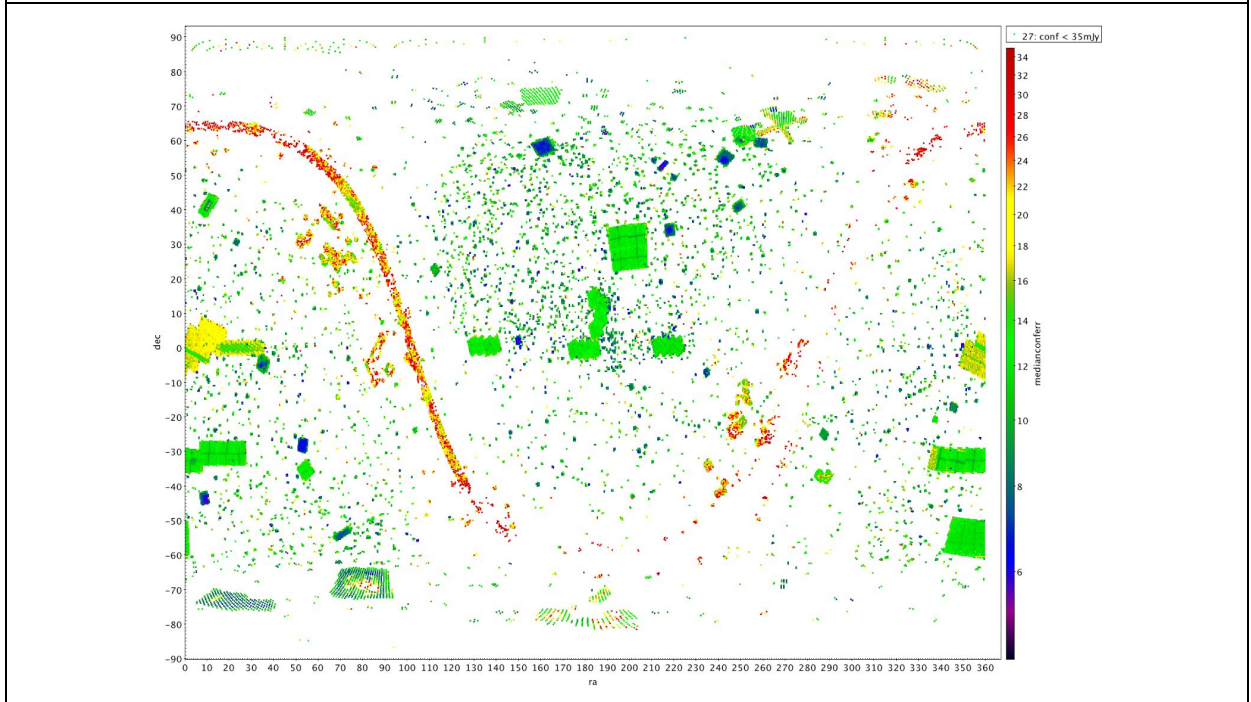

Figure 1.21: Top: Distribution and median STRN of eliminated tiles. Bottom: The distribution of the remaining Q3C tiles (22 bit depth) across the sky where positions of this catalog are found.

To eliminate the tail of high STRN regions, we imposed a threshold of 35 mJy on the median tile STRN. This threshold excludes a major part of the Galactic Plane and some additional regions as shown in Figure 1.21, top panel, as expected. The bottom panel shows the distribution of the remainder of the tiles across the sky. The median STRN is color coded. The cut removes about 15.6% of the consolidated objects, leaving them for a more thorough analysis at a future time.

## Flags and Qualifiers

The quality of derived photometry and the reliability of extracted sources are affected by several factors. Different quality flags have been derived to denote when effects degrading the photometric or astrometric quality are present.

### The Position Flag

The final celestial position we list for an object is the average of the positions of all contributing detections, executed in Cartesian space and then back-projected onto the sphere. The positional uncertainties are determined as the larger of the standard deviations of all positions, or the quadratic mean of all position uncertainties provided by TML, both divided by the square root of the number of contributing detections. In addition we determine the maximum distance between all contributing positions, called range. If either range or positional uncertainty in either the RA or Dec direction are greater than the search radius used for object consolidation, the position flag is set, to indicate an unusually high positional uncertainty. This condition exists 7242 times in this list.

### The Astrometry Flag

As we have described in the Map Position Corrections section above, we used the WISE all sky catalog to derive the absolute astrometry of SPIRE maps. All objects that have at least one contributing detection that can be traced back to one of 110 maps with offsets greater than 5 arcsec are flagged. 140932 objects possess this condition in the catalog (see section Map Position Corrections for details).

### The Duplication Flag

Normally, the Sussextractor source detection does not find sources closer than FWHM/2, however, it is possible in rare cases. Also, if two sources are located relatively close together, the TML position refinement may move nearby source positions even closer. In such a case it is also possible that the Timeline Fitter, starting with the fainter source, finds the other one better fit and jumps, adopting it instead of the initial weaker source with which it started. In all those cases two source detections can end up in the same group contributing to the same object. We flag this condition that exists for 360 objects.

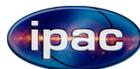 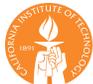 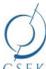 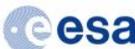 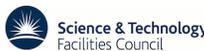 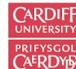

## The Instrument Error Flag

The instrument noise is a function of the number of readouts in the central aperture of the Timeline Fitter. For 6137 detections the Timeline Fitter did not produce a count, so we reverted to using an average estimate based on the instrument setup, multiplied by the repetition factor of the observation. The numbers are listed in Table 1.4. To warn of this less accurate method to derive the instrument noise, we set the Instrument Error Flag. This flag was set for 2565 records.

As described above, objects that have bright mode detections contributing to their parameters were marked with this flag as well, again signalling a less reliable instrument noise estimate, that in these cases is an underestimate. This additional condition raises total number of flags to 3205.

| Instrument Mode | 250 µm | 350 µm | 500 µm |
|---|---|---|---|
| Slow Parallel Mode | 631 | 742 | 686 |
| Fast Parallel Mode | 212 | 250 | 232 |
| Nominal Large Scan Map | 690 | 778 | 744 |
| Fast Large Scan Map | 350 | 401 | 374 |
| Small Scan Map | 619 | 688 | 675 |

Table 1.4: Median number of readouts of an observation operated in a specific instrument mode with repetition rate of one. These numbers multiplied with the number of repetitions was used to estimate the instrument noise if the number of readouts in the TML aperture had not been reported.

## Point Source / Extended Source / Low FWHM Flags

Most SPIRE maps, in particular those of the extragalactic sky, at first sight appear littered with lots of unresolved sources down to the point where the beam profiles overlap and distinction becomes difficult. More detailed analysis reveals that a large fraction is actually slightly extended, compared to a standard beam profile. Reasons for this range from seeing several distant point sources in the same direction merging into one, to seeing the actual extent of the cold dust of a galaxy, or a YSO or other dust concentration, that is close enough to appear spatially resolved.

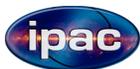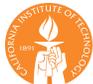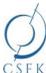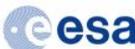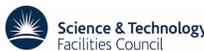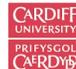

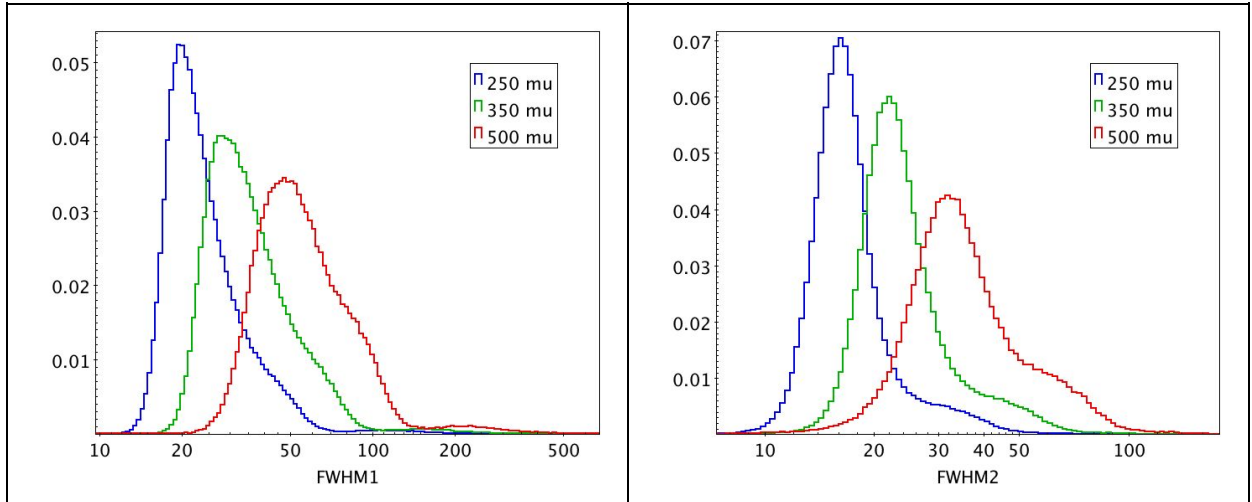

Figure 1.22: Overall distributions of FWHM1 and FWHM2 in arcsec for the three filter bands. Note the difference in the scale of the x-axes. Note that these distributions still contain sky areas with high structure noise that were excluded for the final catalog.

We added the second Timeline Fitter run (TM2) to be able to distinguish slightly extended sources, with the additional benefit of getting good flux estimates, provided that the source profile still resembles an elliptical Gaussian. Apart from the peak, the fit provides the major and minor axes, as well as the rotation angle. The distributions of the derived FWHM in both axes are shown in Figure 1.22. The peaks are consistent with the expected values from the known beam profile dimensions, but there are significant extensions to the right suggesting a sizeable fraction of extended sources.

To make a distinction, we show in Figure 1.23 the FWHM versus the TM2 fluxes for 3648, 4206, 2855 artificial point sources for the three filters, respectively, that were extracted from a low background and low STRN map (obsid = 1342195856). The distributions converge at high fluxes towards the injected FWHM values of 17.6", 23.9", 35.2", respectively, but the distribution widens towards low fluxes, as the fit results are impacted by diminishing SNR.

The distributions were fitted by upper and lower power laws to establish a region in the diagram that is dominated by point sources.

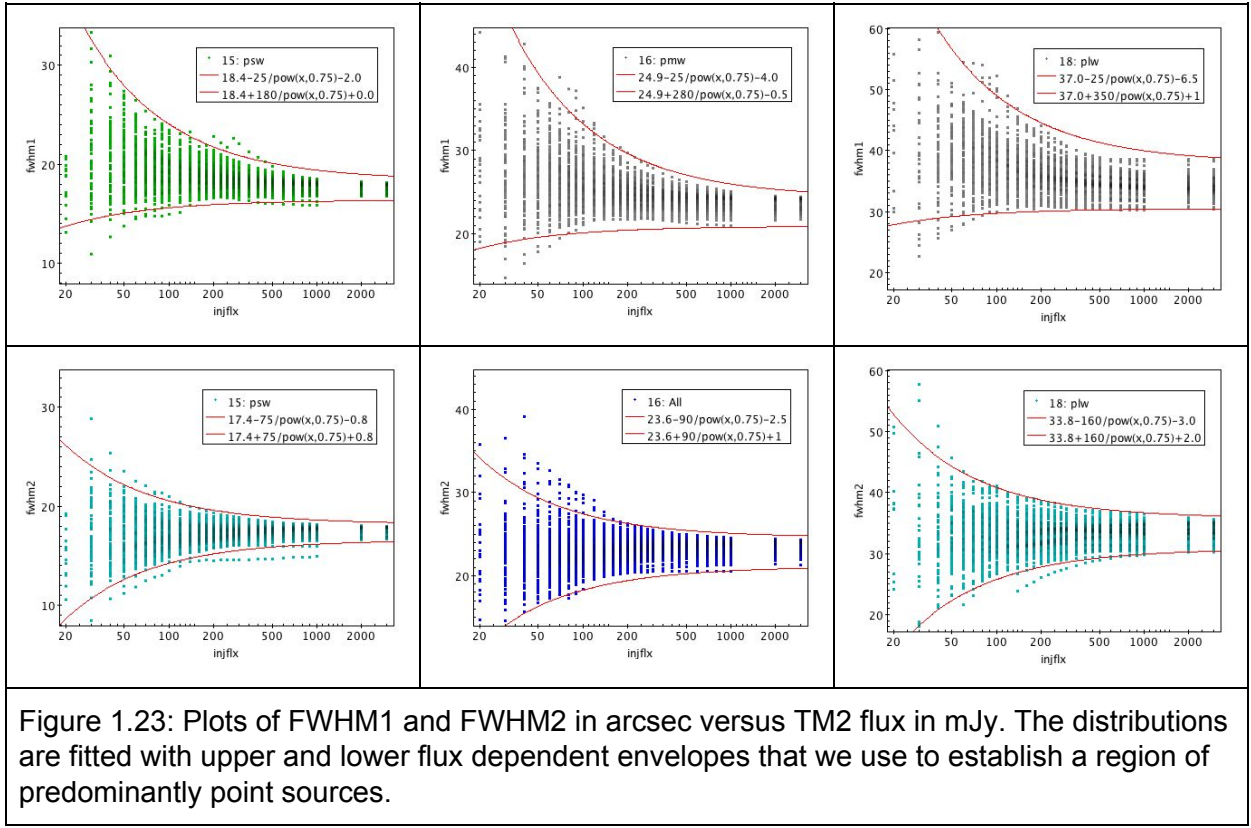

Figure 1.23: Plots of FWHM1 and FWHM2 in arcsec versus TM2 flux in mJy. The distributions are fitted with upper and lower flux dependent envelopes that we use to establish a region of predominantly point sources.

To account for the difference between the injected circular beam profiles and the actual slightly elliptical beam profiles, we corrected the respective functions by adding or subtracting the differences. The actual values for the major and minor axes of the beam profiles are given in Table 1.5 as $\delta_{0maj}$ and $\delta_{0min}$. We then set a point source flag for the major and minor FWHM, $\delta_{maj}$ and $\delta_{min}$, of a given object if the following conditions are satisfied:

$$\delta_{maj} \begin{cases} \leq \delta_{0maj} + \dfrac{k_{maj+}}{S_{tm2}^{3/4}} + l_{maj+} \\ \geq \delta_{0maj} - \dfrac{k_{maj-}}{S_{tm2}^{3/4}} - l_{maj-} \end{cases}$$

$$\delta_{min} \begin{cases} \leq \delta_{0min} + \dfrac{k_{min}}{S_{tm2}^{3/4}} + l_{min+} \\ \geq \delta_{0min} - \dfrac{k_{min}}{S_{tm2}^{3/4}} - l_{min-} \end{cases}$$

where $S_{tm2}$ is the flux derived from the TM2 run, and the parameters $\delta_{0maj}$, $\delta_{0min}$, $k_{maj+}$, $k_{maj-}$, $k_{min}$, $l_{maj+}$, $l_{maj-}$, $l_{min+}$, $l_{min-}$ are given in Table 1.5.

| Parameter | 250 μm | 350 μm | 500 μm |
|---|---|---|---|
| $\delta_{0maj}\,['']$ | 18.4 | 24.9 | 37.0 |
| $\delta_{0min}\,['']$ | 17.4 | 23.6 | 33.8 |
| $k_{maj+}\,['' \, mJy^{3/4}]$ | 180 | 280 | 350 |
| $k_{maj-}\,['' \, mJy^{3/4}]$ | 25 | 25 | 25 |
| $k_{min}\,['' \, mJy^{3/4}]$ | 75 | 90 | 160 |
| $l_{maj+}\,['']$ | 0.8 | 0.5 | 2.8 |
| $l_{maj-}\,['']$ | - 1.2 | - 3.0 | - 4.7 |
| $l_{min+}\,['']$ | 0.6 | 0.7 | 0.6 |
| $l_{min-}\,['']$ | - 1.0 | - 2.8 | - 4.4 |

Table 1.5: Parameters defining the flux dependent point source thresholds for the FWHM of minor and major axes of the TM2 fitted elliptical Gaussians.

The same functions are used to set the extended source flag and the low FWHM flag. The extended source flag is set if either $\delta_{maj}$ or $\delta_{min}$ satisfy one of the following conditions:

$$\delta_{maj} > \delta_{0maj} + \frac{k_{maj+}}{S_{tm2}^{3/4}} + l_{maj+}$$

$$\delta_{min} > \delta_{0min} + \frac{k_{min}}{S_{tm2}^{3/4}} + l_{min+}$$

For cases where $\delta_{maj}$ or $\delta_{min}$ are too small, the low FWHM flag is set as a warning that the FWHM values are less trustworthy. The condition for that is one of the following:

$$\delta_{maj} < \delta_{0maj} - \frac{k_{maj-}}{S_{tm2}^{3/4}} + l_{maj-} \quad \text{or}$$

$$\delta_{min} < \delta_{0min} - \frac{k_{min}}{S_{tm2}^{3/4}} + l_{min^-}.$$

The solution of defining simple regions in the FWHM / flux diagram may appear simplistic, but more sophisticated approaches, that are based on machine learning, were considered but found to be outside the means and scope of our effort. We also found that the stability of the Timeline Fitter at low flux levels is decreasing, delivering sometimes unrealistic or unphysical FWHM values, although the source does appear as a good, albeit faint point source upon visual inspection. Thus these three flags, as well as all the shape parameters, are best treated as reasonable indicators that the FWHM can be off occasionally.

The point source flag is set 1075298 times, the extended source flag is set 576944 times, and the low FWHM flag is set 50108 times.

## The Edge Flag

The position of a source with respect to the edges of the image affects the quality of the photometry, because the extraction procedure needs a large enough area to properly compute the flux density and the background to be subtracted.

The edge flag for each source is obtained using a Jython procedure that defines a subimage from the corresponding map (level 2 or level 2.5), using the background external aperture radius used by the Timeline Fitter (74", 103", 147") for 250μm, 350μm, and 500μm respectively, centered on the source. We count the total number of pixels and the number of good values (not NaNs) in this subimage, obtaining the ratio between the two quantities. The conditions to set the edge flag are:
1. The external radius is off the map
2. The ratio is less than 0.9

The flag is set first for all source detections. In the final table of objects, the flag is set if any of the contributing sources has the flag set. This condition is fulfilled for 82691 objects.

## The Large Galaxy Flag

If a detected source is within the ellipse of a large galaxy, there is a probability that it is not a field object, but a part of a galaxy. We have derived a 'large galaxy' table from a NED search of the 2MASS Large Galaxy Atlas (Jarrett et al 2003) for galaxies with major axis bigger than 1 arcmin. This table contains RA, Dec coordinates and ellipse parameters for each galaxy. To set the flag, the detected source must be within the area defined by the galaxy ellipse. In this list 24440 sources are flagged.

## The Solar System Object Flag

By-passing solar system objects may cause false detections and affect the photometry of the detected sources. To characterise this, we have carried out a detailed check of SSO contamination in all SPIRE Point Source Catalog fields. This test consisted of two main points: (1) determination of the position of all known solar system bodies in a specific field at the time of the observations and (2) estimation of the object's thermal flux for the SPIRE photometric bands. As the thermal emission of even the farthest solar system bodies peaks shortward of the SPIRE central wavelengths, we used the 250 μm flux as a sole selection criterion.

As a first step, we used the latest available MPCorb database as input (version of April, 2016) which contained orbital elements for 713 289 objects. For a pre-selection of objects we calculated a "worst-case" thermal flux for all targets in this database. For all objects the "worst-case" geometry is defined as when the target is the closest to the observer at maximum possible solar illumination (e.g. a main belt asteroid is at its perihelion at zero phase angle with respect to the observer). For each object we used the Near-Earth Asteroid Thermal Model (Harris et al., 1989) to calculate the wavelength dependent flux densities of the thermal emission for the observing geometry above. We fixed the beaming parameter to 0.756 (a canonical average value for main belt asteroids) and used a fixed geometric albedo of 0.05 and the $H_V$ absolute magnitude given from the MPCORB input file. The observing geometry, beaming parameter and geometric albedo setting result in a definite overestimation of the thermal flux for the vast majority of the objects. We selected those objects for further study for which the 250 μm thermal flux exceeded 6 mJy, the approximate extragalactic confusion limit at this wavelength. This reduced the number of objects to 97663.

To check whether an SSO crossed an actual SPIRE map, we used the SPICE Toolkit (https://naif.jpl.nasa.gov/naif/toolkit.html) in IDL. We generated SPK kernel (ephemeris) files for the selected objects using the *smb_spk* script, written by Jon D. Giorgini (ftp://ssd.jpl.nasa.gov/pub/ssd/smb_spk) for the operation time of the Herschel Space Observatory, and the SPK kernel file for the Herschel Space Observatory itself.

In the next step we calculated for each scan map the position of the selected SSO-s for three epochs: the beginning of scan, the midtime and at the end, using the *CSPICE_SPKEZR* task with Converged Newtonian light time correction (see documentation here: https://naif.jpl.nasa.gov/pub/naif/toolkit_docs/IDL/icy/cspice_spkezr.html). Then we also calculated the position of the Herschel spacecraft for these epochs and compared the actual FOV (derived from the images) and the RA,DEC position of the asteroids, as seen from Herschel. For each map the FOV was approximated by the four corners of a rectangle enclosing the map, in equatorial coordinates (this resulted in a slight overestimation the area in all cases). If the SSO was found to be in the FOV we included this record in our database, if not, we rejected it.

In the next step we checked for every map whether an SSO was present at least two times. If yes, we checked whether it was on the map itself, or just in the surrounding area. In case the target was on the image, the start and end positions are provided as the final start/end equatorial coordinates (R.A.,DEC). If the object was just present once, or if it was out of the image area at both epochs, we calculated a trajectory for the object with a simple linear fit. If there was no intersection with the actual image area, we discarded the SSO. In the case of an intersection, we used the "ingress" and "egress" points of the image, and provide them as start/end coordinates. If an SSO was present only once in the larger area, we performed a new position query with an adequate time shift (e.g. half hour later/earlier) to determine a second position, and we proceeded with the ingress/egress point determination as above.

In some cases a map is consisted of scans which were taken at very different epochs. In these cases we took the unique scans (OBSIDs) and performed the search for these OBSIDs individually.

We cross-checked our result with HORIZONS/ISPY to test the accuracy of our calculated positions and we found that the difference between our calculations and the e-mail query was negligible at the scales of the SPIRE spatial resolution. The typical errors are 0.41±0.42 arcsec in R.A. and 0.13±0.18 arcsec in DEC, but showing very skewed distributions in both coordinates. The maximum differences we obtained were ~1.5 arcsec, significantly smaller than the pixel scale of the SPIRE photometer channels.

Knowing the the actual position (R.A., DEC and longitude/latitude in helioecliptic coordinate system) and time (julian date) of an object, we could perform a more accurate NEATM calculation to obtain the flux density at 250 μm. We used the absolute magnitude, size and albedo from the Horizons, MPCORB or NEOWISE database, when these parameters were available, otherwise $H_V$ was estimated from the measured, apparent brightness and the observing geometry (heliocentric and observer distance and phase angle) at the time of the observations. We assumed population-average albedos and a standard beaming parameter of η = 0.756.

We consider that the flux of a source can be contaminated by a SSO when its position in the correspondent image is within a rectangle defined by initial and final coordinates of a solar object (SSO trail) plus the correspondent array FWHM. To set the flag, an additional condition has been imposed on the SSO flux. Only if the SSO flux is bigger than the confusion limit (5.8, 6.3 and 6.8 mJy for 250, 350 and 500μm, respectively), the object is flagged as True.

In the catalog, if any of the contributing sources for an object has the flag set, the object has flagged as True.

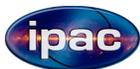

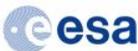
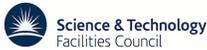

# 3 Column Descriptions

In the following we describe the columns of the catalog, arranged in logical groups. These groups are: 1) **Identification** for object naming and specification of wavelength and position in the sky, 2) **Detection** providing information on the number of independent detections 3) **Photometry** giving fluxes and uncertainties obtained with different methods 4) **Shape** providing parameters of fitted elliptical Gaussians and 5) **Other Flags** providing additional information.

## Identification Columns

### SPSCID

The object identifier consists of two parts, acronym and sequence, and is formatted as "HSPSCwwwA_Jhhmm.mm+ddmm.m" with the exception of the planets Uranus and Neptune, which are following the format "HSPSCwwwA_ppp_000". "HSPSC" stands for Herschel SPIRE Point Source Catalog, "www" is one of [250,350,500], identifying the wavelength in microns, and the letter A indicates the first public version of the catalog. The sequence has the format "Jhhmm.mm+ddmm.m", which is derived from the object coordinates with J indicating a J2000 reference system, "hhmm.mm" indicating the Right Ascension in hours and minutes to an accuracy of two digits after the decimal, and "+ddmm.m" indicating Declination in signed degrees and minutes to an accuracy of one digit after the decimal. In the case of the planets the sequence is formatted as "ppp_000", where "ppp" stands for one of the strings "URA" or "NEP" indicating Uranus or Neptune respectively, followed by a running number "000" indicating the observation. The spatial resolution of the coordinate based serial number is well matched to the size of SPIRE's beam profiles and will keep changes in object names to a minimum if another version of the catalog is generated.

### DET

This is one of the strings "PSW", "PMW", "PLW" identifying the SPIRE detector array names for the filter bands centered at 250, 350, 500 μm respectively. This column is for convenience for those more familiar with the instrument specific naming. Otherwise the wavelength is indicated in the object identifier.

## RA / DEC

These two columns are the Right Ascension and Declination coordinates in a J2000.0 reference frame, calculated as the average of the Right Ascensions of all the contributing source detections in different maps. The positions used result from the refined Timeline Fitter positions and their uncertainties. The average is calculated in Cartesian space and transformed back to spherical coordinates to avoid problems near the celestial poles.

## RA_ERR / DEC_ERR

These two columns show the positional uncertainties in arcsec of each catalog object. They are derived as the larger of either quadratic mean of all the individual TML uncertainties of all contributing source detections or the standard deviation of all source positions, divided by the square root of the number of maps the source is detected in.

## POS_FLAG

This flag indicates a potential problem with the position of this source if set to True. It is set if the uncertainties associated with the position are larger than the search radius used by the consolidation algorithm. It is also set if the maximum distance between the positions of all contributing detections is larger than the search radius used by the consolidation algorithm. The search radius is chosen as half of the FWHM of the beam profile for the respective source (See Table 1.1) increased by 6" as allowance for the 3 sigma pointing uncertainty of Herschel.

## ASTROM_FLAG

Astrometry flag indicating that one or more of the contributing source detections came from a map with a positional offset greater than 5" derived from stacking at WISE 24 µm catalog positions. A total of 140573 records was flagged in this way.

# Detection Columns

## NMAP

The number of times a source should have been detected at the list position and in the same band pass, based on the coverage maps. This value is derived by testing a 3x3 pixel square centered on the source in each coverage map. If the coverage anywhere within this square is greater than zero, the NMAP value for this object is incremented by one.

### NDET

This column contains the actual number of detections in separate maps of this object at this position and filter band. Combinations of maps into a Level 2.5 map are counted as one. There are rare cases where two close, but separate detections in the same map become part of the same object because the grouping algorithm was not able to distinguish them. In this case the source duplication flag is set.

Ideally the NDET value is equal to the NMAP value, indicating that in all maps the same source was found. The ratio NDET/NMAP decreases towards fluxes that come closer to the local confusion noise and instrument noise. In general the confusion noise dominates. Other reasons for this ratio to be smaller than one are source that are located close to the map edge where the coverage is not anymore sufficient for the Timeline Fitter to work well. Another reason can be the extendedness of the source when it is close to the acceptable threshold.

### DUPL_FLAG

This flag is False by default and set to True if more than one source from the same map were identified as part of this object by the object consolidation algorithm. This is a rare occurrence and mainly due to limitations of the algorithm used. The flag was set 1624 times (960,497,167 for the filter bands respectively).

## Photometry Columns

### FLUX

This column shows the average of the TML derived fluxes of all contributing sources, weighted with the respective TML generated uncertainties. This procedure of fitting a fixed width circular Gaussian beam profile model has proven to be superior to a number of other common methods used with SPIRE data in terms of reproducibility and photometric accuracy, down to fluxes of 30 mJy (Pearson et al. 2014). The units are in mJy, calculated as

$$\frac{\sum S_i . \sigma_i^{-2}}{\sum \sigma_i^{-2}}$$

where $S_i$ are the contributing fluxes and $\sigma_i$ are the respective TML uncertainties.

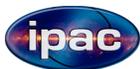

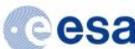
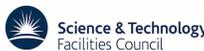

## FLUXTML_ERR

This column contains the propagated weighted error produced by the TML for all contributing source detections. They are expressed in mJy and calculated as

$$\sqrt{\frac{1}{\sum_i \sigma_i^{-2}}}$$

where $\sigma_i$ are the respective TML uncertainties.

## CONF_ERR

This column indicates the local confusion uncertainty in mJy. It is the average of the individual values of all contributing sources, weighted with the respective TML uncertainties. The individual confusion uncertainties are calculated as

$$\sigma_{conf} = \sqrt{\sigma_{total}^2 - \sigma_{inst}^2}.$$

The total uncertainty of the source $\sigma_{total}$ is derived from the local structure noise of the map around the source and the timeline fitter flux from a function that was established by artificial source injection and retrieval experiments in real data (see details above). The instrument noise $\sigma_{inst}$ of the observation is an estimate, based on the number of readouts $n_{readout}$ found in the central TML aperture according to

$$\sigma_{inst} = a \; n_{readout}^b$$

where $a$ and $b$ are parameters that depend on the filter band. The parameters were derived empirically from observations as explained above and are listed in Table 1.4. Note that we didn't treat the case of high bias (bright) mode separately. For these few observations the instrument noise will be underestimated, however in practise this is rarely a factor when bright sources dominate.

## FLUX_ERR

This column contains an estimate of the total flux uncertainty for the object in mJy. It is calculated as

$$\sigma_{total} = \sqrt{\overline{\sigma}_{conf}^2 + (a \; (\sum n_{rd})^b)^2}$$

where $\overline{\sigma}_{conf}$ is the averaged confusion noise derived from all contributing source detections and $\sum n_{rd}$ is the sum of readouts in the TML apertures of all contributing detections. This way the confusion noise remains the ultimate lower limit that does not decrease with the number of contributing sources, unlike the instrument noise component.

Although the formal uncertainties of the different photometry algorithms often indicate lower values, the uncertainty arising from overlapping sources in the background that can not anymore be easily distinguished, constitutes a fundamental limit to the uncertainty applicable to the measured source fluxes. This column can be considered a conservative estimate of the overall statistical uncertainty. It does not take into account the 4% systematic uncertainty due to the limits on the prime calibrator Neptune (Bendo et al. 2013).

### SNR

This value has been added for convenience and expresses the signal to noise ratio associated with a given object as the ratio between the Timeline Fitter flux (FLUX) and the total flux error (FLUX_ERR).

### INSTERR_FLAG

This flag indicates that the TML returned no count for the number of readouts in the central aperture. To still be able to derive an instrument noise component, a typical number was used, based on the instrument configuration (see Table 1.3) and multiplied with the number of repeats of the observation.

### FLUXSUS

This column shows the average of the Sussextractor derived fluxes of all contributing sources, weighted with the respective Sussextractor generated uncertainties. The units are in mJy, calculated as

$$\frac{\sum S_i . \sigma_i^{-2}}{\sum \sigma_i^{-2}}$$

where $S_i$ are the contributing fluxes and $\sigma_i$ are the respective Sussextractor uncertainties.

This procedure was used to generate a first list of source positions for an observation that was then passed on to the subsequent extractors. The flux repeatability of this algorithm is not as good as Timeline Fitter and applies only to point sources, however, it does a better job for faint sources below 30 mJy (Pearson et al. 2014).

### FLUXSUS_ERR

This column contains the propagated weighted error produced by Sussextractor for all contributing source detections. They are expressed in mJy and calculated as

$$\sqrt{\frac{1}{\sum_i \sigma_i^{-2}}}$$

where $\sigma_i$ are the respective Sussextractor uncertainties.

### FLUXDAO

This column shows the average of the Daophot derived fluxes of all contributing sources, weighted with the respective Daophot generated uncertainties. The units are in mJy, calculated as

$$\frac{\sum S_i . \sigma_i^{-2}}{\sum \sigma_i^{-2}}$$

where $S_i$ are the contributing fluxes and $\sigma_i$ are the respective Daophot uncertainties.

This procedure was used to generate a first list of source positions for an observation that was then passed on to the subsequent extractors. The flux repeatability of this algorithm is not as good as Timeline Fitter and applies only to point sources, however, it does a better job for faint sources below 30 mJy (Pearson et al. 2014).

### FLUXDAO_ERR

This column contains the propagated weighted error produced by Daophot for all contributing source detections. They are expressed in mJy and calculated as

$$\sqrt{\frac{1}{\sum_i \sigma_i^{-2}}}$$

where $\sigma_i$ are the respective Daophot uncertainties.

### FLUXTM2

This column shows the average of the TM2 derived fluxes of all contributing sources, weighted with the respective TM2 generated uncertainties. This procedure fits an elliptical Gaussian beam profile model where the major and minor axes have been left as free fit parameters. The configuration also allows for a tilted background plane.

The units are in mJy, and the weighted average is calculated as

$$\frac{\sum S_i . \sigma_i^{-2}}{\sum \sigma_i^{-2}}$$

where $S_i$ are the contributing fluxes and $\sigma_i$ are the respective TM2 uncertainties.

Note that we calculate the individual fluxes based on the integral under the elliptical Gaussian as

$$S = \frac{S_{peak}\ \delta_{maj}\ \delta_{min}}{\delta_{nom}^2}$$

where $S_{peak}$ is the flux estimate returned by the Timeline Fitter, which is the difference between background and the peak of the Gaussian in units of mJy, $\delta_{maj}$ and $\delta_{min}$ being the FWHM of the major and minor axis, and $\delta_{nom}$ being the nominal FWHM of a point source. This results in a more realistic flux estimate for extended sources, provided their shape is still reasonably well described by an elliptical Gaussian.

### FLUXTM2_ERR

This column contains the propagated weighted error produced by TML2 for all contributing source detections. They are expressed in mJy and calculated as

$$\sqrt{\frac{1}{\sum_i \sigma_i^{-2}}}$$

where $\sigma_i$ are the respective TML2 uncertainties that are individually calculated as

$$\sigma = \frac{S_{peak}\ \delta_{maj}\ \delta_{min}}{\delta_{nom}} \sqrt{\left(\frac{\sigma_{S_{peak}}}{S_{peak}}\right)^2 + \left(\frac{\sigma_{\delta_{maj}}}{\delta_{maj}}\right)^2 + \left(\frac{\sigma_{\delta_{min}}}{\delta_{min}}\right)^2}$$

Where $\sigma_{S_{peak}}$ is the uncertainty of the peak flux, and $\sigma_{\delta_{maj}}$ and $\sigma_{\delta_{min}}$ are the uncertainties of the FWHM of the major and minor axes respectively.

## Shape Columns

### FWHM1

This column contains the average FWHM of the major axes of the elliptical Gaussian beam profile models fitted during the TM2 run to all the contributing detections, weighted by their respective uncertainties. The values are given in arcsec and are referenced also as $\delta_{maj}$.

### FWHM2

This column contains the average FWHM of the minor axes of the elliptical Gaussian beam profile models fitted during the TM2 run to all the contributing detections, weighted by their respective uncertainties. The values are given in arcsec and are referenced also as $\delta_{min}$.

### FWHM1_ERR / FWHM2_ERR

These columns contain the respective uncertainties of the weighted averages of the major and minor axes of the fitted elliptical Gaussian beam profile in arcsec.

### ROT

This column contains a robust average rotation angle of the elliptical Gaussian beam profile models fitted during the TM2 run to all the contributing detections. Robust meaning that all angles with uncertainties that are larger than a factor 1.5 of the median of all uncertainties of the contributing detections are removed before averaging. The rotation angle on the sky is measured from West counter-clockwise in degrees. Possible values start from 0 and are smaller than 180 deg. The average of the angles is calculated in Cartesian space to avoid problems when crossing the 180 deg threshold. Since a position angle of 180 deg is equal to 0 deg, we multiplied the position angles by 2 before averaging and divided by 2 after the back-transformation. A constant of 180 deg was added when the angle was smaller than 0 to stay within the allowed range.

### ROT_ERR

This column contains the propagated error of the rotation angle for all contributing source detections. They are expressed in degrees and calculated as

$$\sqrt{\frac{1}{\sum_i \sigma_i^{-2}}}$$

where $\sigma_i$ are the respective uncertainties.

### PNTSRC_FLAG

The flag is set for all objects, where both the major and minor FWHM parameters are still compatible with values that are observed from point sources at a given flux level. As such it is an indicator but not conclusive, in particular not for small fluxes. This flag is set if the following conditions are met:

$$\delta_{maj} \begin{cases} \leq \delta_{0maj} + \dfrac{k_{maj+}}{S_{tm2}^{3/4}} + l_{maj+} \\ \geq \delta_{0maj} - \dfrac{k_{maj-}}{S_{tm2}^{3/4}} - l_{maj-} \end{cases}$$

$$\delta_{min} \begin{cases} \leq \delta_{0min} + \frac{k_{min}}{S_{tm2}^{3/4}} + l_{min+} \\ \geq \delta_{0min} - \frac{k_{min}}{S_{tm2}^{3/4}} - l_{min-} \end{cases}$$

where $\delta_{maj}$ and $\delta_{min}$ are the major and minor FWHM parameters, $S_{tm2}$ is the flux derived from the TM2 run. The parameters $\delta_{0maj}$, $\delta_{0min}$, $k_{maj+}$, $k_{maj-}$, $k_{min}$, $l_{maj+}$, $l_{maj-}$, $l_{min+}$, $l_{min-}$ and details about their derivation were given already above (see Table 1.5).

### EXTSRC_FLAG

This flag is set if any of the major or minor FWHM, $\delta_{maj}$ or $\delta_{min}$ is larger than the upper flux dependent threshold for point sources, i.e. if any of the two following conditions is met

$$\delta_{maj} > \delta_{0maj} + \frac{k_{maj+}}{S_{tm2}^{3/4}} + l_{maj+}$$

$$\delta_{min} > \delta_{0min} + \frac{k_{min}}{S_{tm2}^{3/4}} + l_{min+} \quad .$$

These conditions are more definite than the condition for a point source. Objects with this flag set to True are very likely to be extended or being a combination of more than one point source.

### LOWFWHM_FLAG

This flag is set if any of the major or minor FWHM, $\delta_{maj}$ or $\delta_{min}$ is smaller than the lower flux dependent threshold for point sources, i.e. if any of the two following conditions is met

$$\delta_{maj} < \delta_{0maj} - \frac{k_{maj-}}{S_{tm2}^{3/4}} + l_{maj-}$$

$$\delta_{min} < \delta_{0min} - \frac{k_{min}}{S_{tm2}^{3/4}} + l_{min-}$$

These conditions either indicate a statistical outlier that is still a point source, but they can also indicate a spurious source due to a map artifact and warrant closer inspection.

# Additional Flags

## LARGEGAL_FLAG

This flag is set when the catalog position is within one of the ellipses given in the 2MASS Large Galaxy Atlas (LGA) (Jarrett et al 2003) for galaxies with major axis > 1arcmin. The values are described as 'K_s (LGA/2MASS "total")' and represent the standard 2MASS aperture. If the flag is set there is a high probability that the source is part of the respective large galaxy rather than a background object.

## MAPEDGE_FLAG

The map edge flag is set when at least one of the contributing sources shows 10 or more per cent of map pixels without a flux value (NaN) within the outer radius of the background annulus used by the Timeline Fitter (74", 103", 147") for 250µm, 350µm, and 500µm respectively. This condition flags catalog objects where its detection or the flux estimate may have been affected by at least one contributor being close to a map edge. 24313 objects are the edge flag set.

## SSOCONT_FLAG

This flag is set as a warning if any of the known contributors to the catalog object is within a rectangle defined by the tracklet of a known asteroid at the time of the observation of the respective map, and the beam FWHM at the wavelength of the map. The estimated flux using a simple asteroid standard model needs to be at least at the one sigma confusion limit of Nguyen et al. (2010) of 5.8, 6.3, 6.8 mJy for 250µm, 350µm, and 500µm respectively. 810 objects have this flag set.

# Q3C Tile Identifier

## TILE

As the exclusion criterion for regions with highly structured background emission is determined on the basis of the median STRN within a Q3C tile at the 22 bit level, we add the respective Q3C tile identifier to each object. It is a 19 digit number that is best interpreted using the Postgres implementation of Q3C.

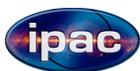 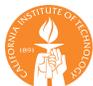 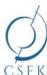 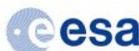 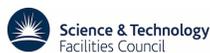 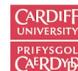

# 4 Cross Reference Matrix

One additional file besides the main catalog table serves as a cross reference matrix, connecting each catalog object (SPSCID) to the observations that contributed to it, identified by their 10 digit OBSID. In other words, if sorted by observation (OBSID), this table lists all catalog sources that contain contributions from source detections in a given observation.

Since more than one observation can contribute to one object, and typically more than one object are part of an observation, the identifiers in neither of the two columns are unique.

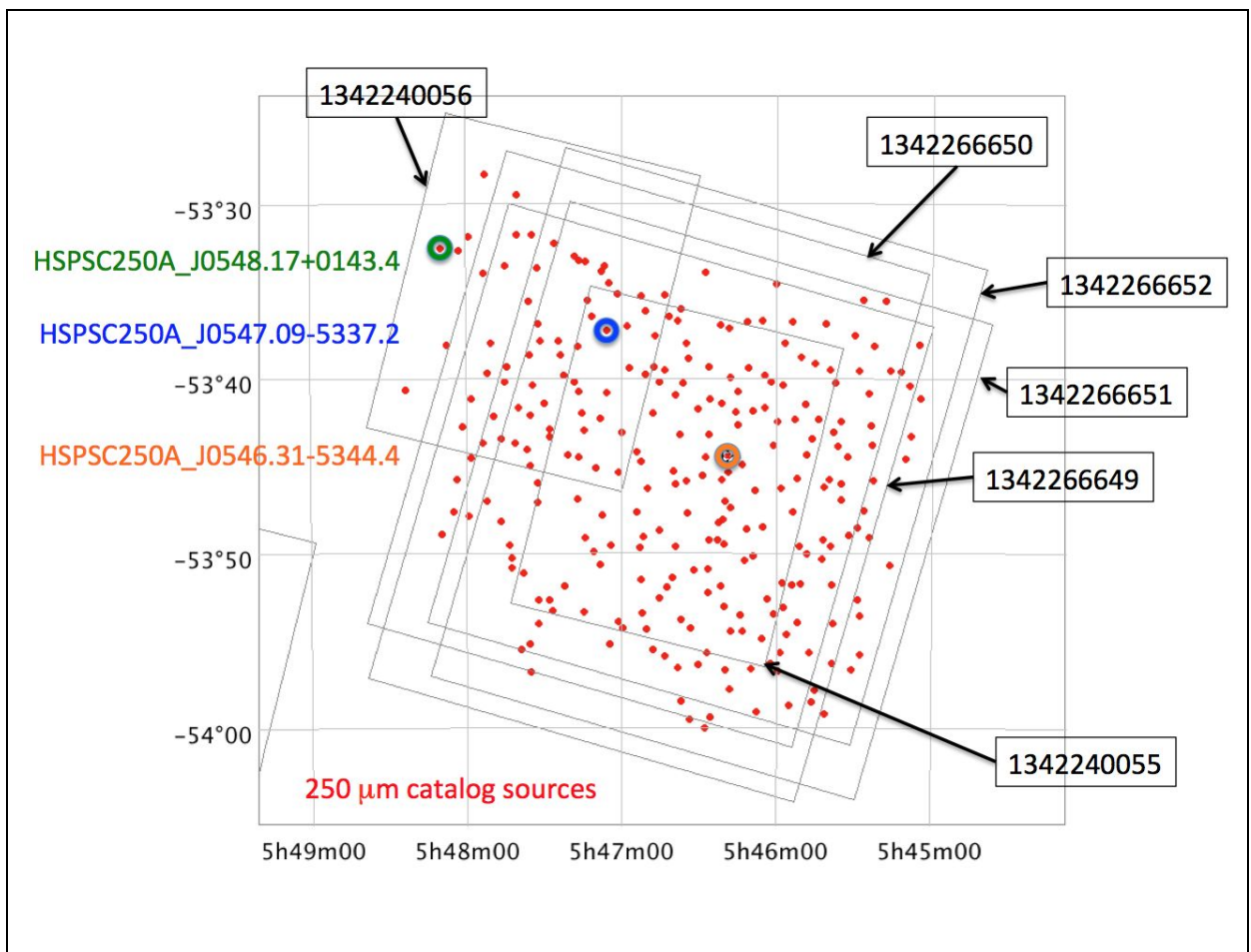

Figure 4.1: Illustration of how three exemplary catalog objects from the 250µm table are related to the SPIRE maps and their observation IDs used in the Herschel Science Archive. See text for details.

The correlations can sometimes be confusing, so we will illustrate the situation for three examples drawn from the 250µm table.

The first catalog object HSPSC250A_J0546.31-5344.4, shown encircled in orange in Figure 4.1, is located close to the center of a cluster of sources that come from a total of 5 SPIRE observations. The Cross-ID table lists the following:

```
1342266649   HSPSC250A_J0546.31-5344.4
1342266650   HSPSC250A_J0546.31-5344.4
1342266651   HSPSC250A_J0546.31-5344.4
1342266652   HSPSC250A_J0546.31-5344.4
1342240055   HSPSC250A_J0546.31-5344.4
```

This is consistent with the known extents of the maps, shown by grey rectangles in Figure 4.1. Note that this source was extracted from only 2 maps, as 4 observations are combined into one map, i.e. the nmap and ndet values are 2 in the catalog.

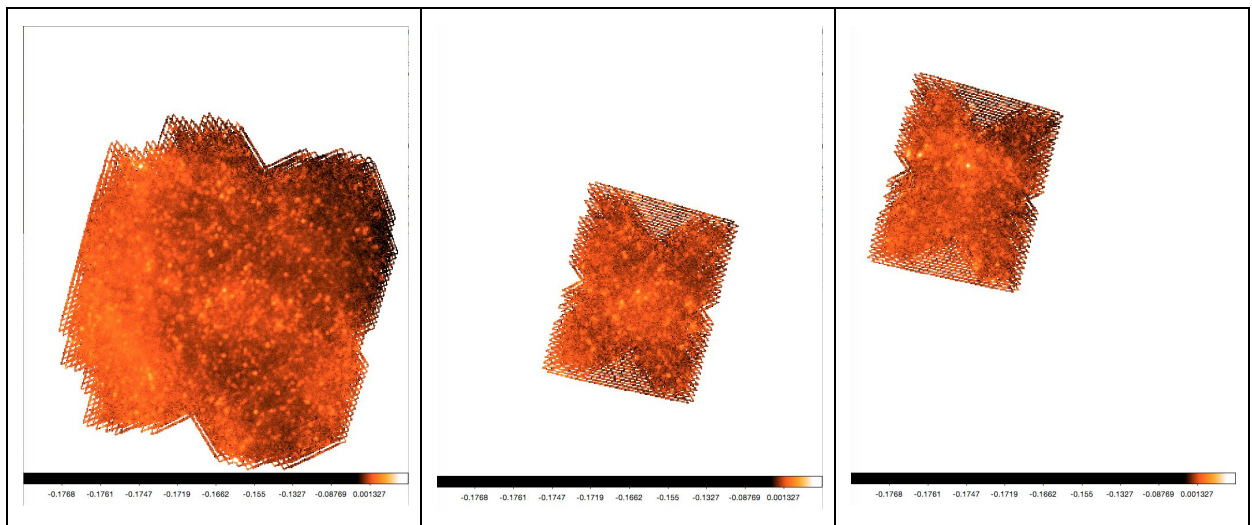

Figure 4.2: The three maps that contributed to the source cluster shown in Figure 4.1. To the left is the combined map consisting of observations 1342266649, 1342266650, 1342266651, and 1342266652. The center shows Observation 1342240055 and the right image shows Observation 1342240056.

The second example is HSPSC250A_J0547.09-5337.2 that from the information shown in Figure 4.1 would be expected to have detections in all 3 maps, i.e. 6 observations as one of the maps is a combination of 4. Also the nmap number, the expected number of detections in the catalog, is 3. However, there is only one detection recorded in the NDET value, and we find the 4 combined observations only listed in the Cross-ID table as follows:

>   1342266649   HSPSC250A_J0547.09-5337.2
>   1342266650   HSPSC250A_J0547.09-5337.2
>   1342266651   HSPSC250A_J0547.09-5337.2
>   1342266652   HSPSC250A_J0547.09-5337.2

Inspecting the three maps shown in Figure 4.2 reveals that this source was not well covered by the scans in both point source mode maps and failed to be detected there. Only the combined map to the left in Figure 4.2 contributed to this catalog object.

The last example HSPSC250A_J0548.16-5332.5 shows an object that has nmap and ndet values of 1 and is traced to only one of the three maps.

>   1342240056   HSPSC250A_J0548.16-5332.5

Thus care should be taken when comparing expected map coverages with actual results. There are many reasons why a catalog object may not have been detected in an overlapping map. Although the nmap values are drawn from the coverage maps, there is no guarantee for a detection if the position is in a region with a low sampling density. The Cross-ID table only lists observations that eventually ended up contributing to a given object.

Note that the cross-identification table is not a tool to find all catalog sources in the sky area covered by a certain observation, as there could for instance be another overlapping deeper observation, that contains more detections than the one in question, generating additional catalog objects that are not seen in the first map, and thus not recorded as related to it in the cross-identification table.

# 5 Validation

Although a number of catalogs were already produced by various science projects based on Herschel-SPIRE data, there is still room for an overall SPIRE Point Source Catalog. Apart from extracting all sources from those observations that originally just focused on one single object, this catalog also helps in making results more homogeneous and comparable. Given that the source extraction strategy followed here is based solely on a single wavelength without priors, it is also clear that it will not reach the depth of work based on already existing catalogs from other instruments that already can make assumptions about the scientific nature of the sources and even allow disentangling SED components of multiple sources that merge at longer wavelengths.

With that said, the goal for the validation of the SPSC is to verify that the vast majority of the sources in the catalog are in the positions as specified, have a high degree of reliability, have fluxes within the uncertainties, and are accompanied by flags that are useful and consistent.

There are two approaches, an inner validation and an outer validation. The first is looking at the products without comparison to additional external data. This includes analysis of the general properties of the various data columns, checking whether they make sense, and whether they are within expectations. It also includes exercising the source extraction method by verification through injection and retrieval of artificial sources in real SPIRE scan maps. To some extent this work was already part of the catalog construction as it yields uncertainties, completeness, and derives limits for point source discrimination.

The outer validation is based on comparisons with other catalogs. Given that Herschel SPIRE has quite a unique wavelength coverage, most of our outer validation efforts focus on comparisons to catalogs derived from the same SPIRE data, but by other groups using different techniques. These comparisons, especially those with surveys of the Galactic regions, showcased the limitations that our source extraction methods have on strongly structured backgrounds, eventually resulting in the decision to implement a structure noise threshold.

## General Catalog Properties

The three main tables, one for each of the three SPIRE filter bands, contain 950688, 524734, and 218296 records for the 250µm, 350µm, and 500µm bands respectively.

## Positions

We checked that the J2000 coordinates indeed stay within $0 \leq \mathrm{RA} \leq 360$ and $-90 \leq \mathrm{Dec} \leq 90$. That appears trivial, but becomes an issue when positions of several detections are averaged into one catalog object position. The positions of the objects are somewhat inhomogeneously distributed across the sky depending on the regions observed by the respective science programs and depending on the exclusion threshold we applied for higher background confusion levels (see Figure 3.1).

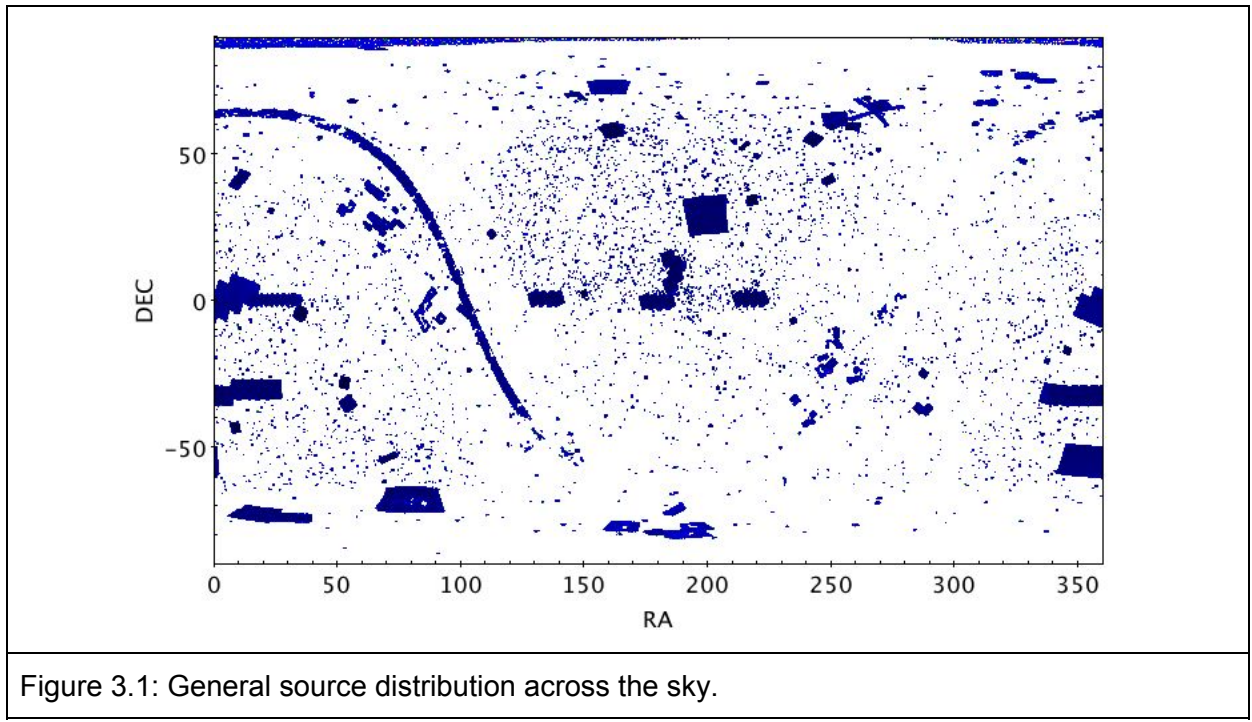

Figure 3.1: General source distribution across the sky.

Checking the extremes of the positional errors the Dec maximum is 7.3", 9.1", 11.7" for the three wavelengths, while the RA maximum is 1799.6s, 2271s, 2457.5s. We verified that the larger maximum uncertainties for RA are correlated with their proximity to the celestial poles. The uncertainties in RA and Dec are plotted against Dec in Figure 3.2.

A total of 0.37%, 0.41%, 0.69% records have the position flag set, while 8.5%, 8.1%, 7.9% of the objects are affected by maps with mispointings larger than 5" (astrometry flag).

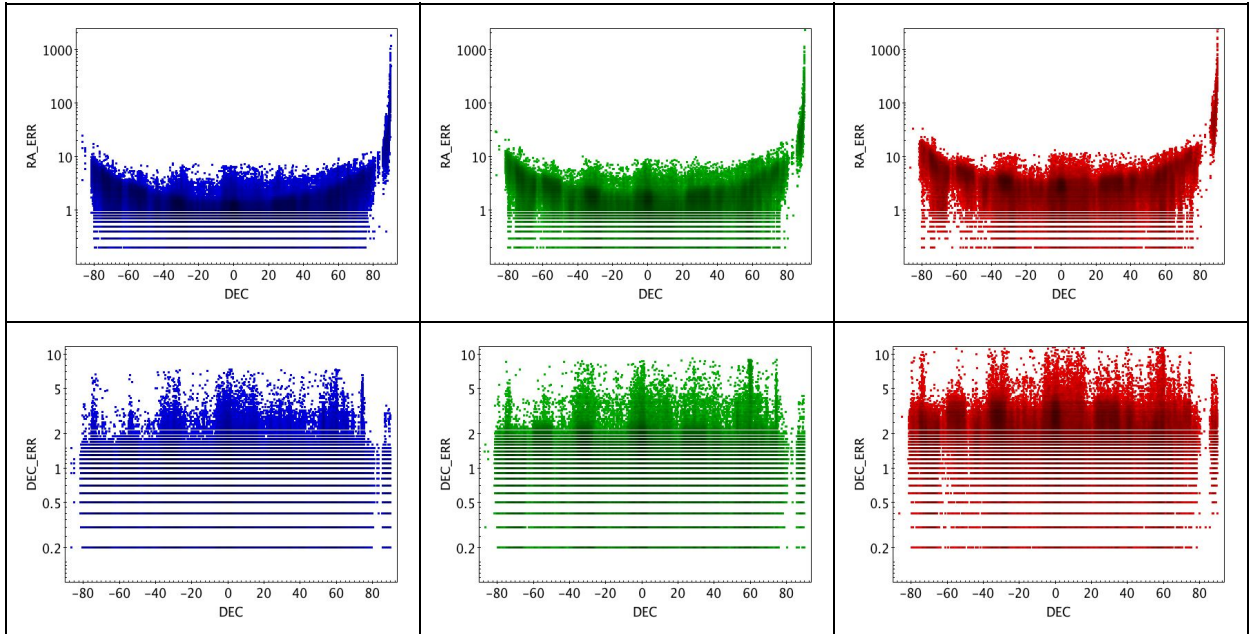

Figure 3.2: The top row shows RA uncertainties plotted against Dec, showing the expected increase towards the celestial poles. The bottom row shows the uncertainties in Dec. Wavelengths are 250μm, 350μm, and 500μm from left to right.

## Detections

The maximum number of detections per object is 96, 94, 89 depending on wavelength, but the vast majority of objects have just one or two detections. The histograms are shown in Figure 3.3.

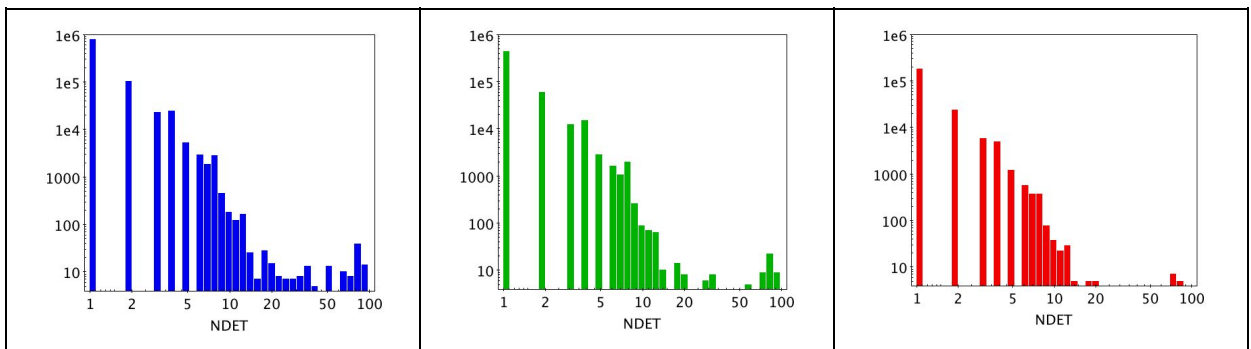

Figure 3.3: Histograms of the number of detections that contributed to a given object with wavelenth increasing from left to right.

To obtain an indication of reliability, we determined for each object position, based on the coverage maps, the maximum number of times an object at that position could have been detected in a map. In the catalog table this number is recorded in the column "nmap". The large majority of objects (89%, 91%, 84% for 250 respectively) has a ratio of ndet/nmap = 1. Plotting this ratio against the Galactic latitude not against Equatorial Declination does not reveal any unexpected peculiarities (see Figure 3.4).

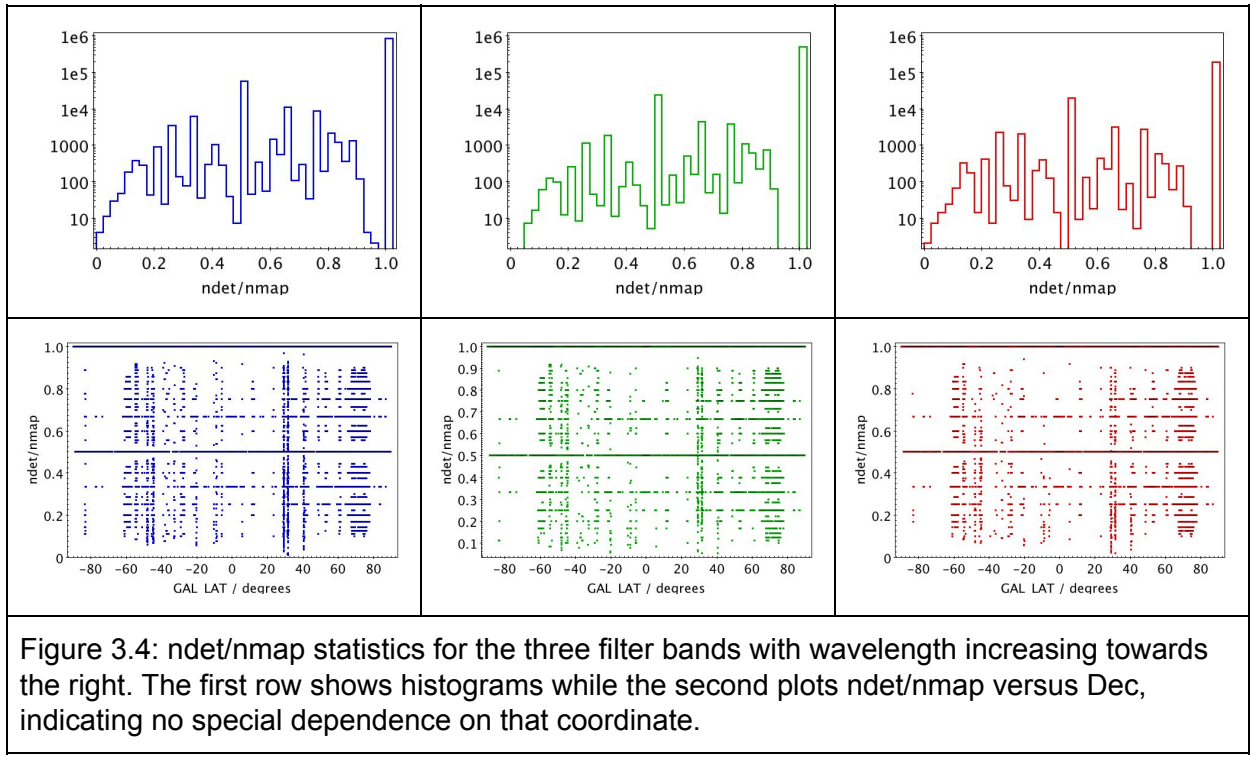

Figure 3.4: ndet/nmap statistics for the three filter bands with wavelength increasing towards the right. The first row shows histograms while the second plots ndet/nmap versus Dec, indicating no special dependence on that coordinate.

## Fluxes

The distributions of the fluxes derived by all 4 methods are shown in the upper row of Figure 3.5. They show that the TM2 fluxes, although they are the more realistic fluxes for extended sources, need to be treated with caution. The fitting process in this mode is less stable and produces a small population of outliers. In particular some extremely low fluxes below 10 mJy are reported that are not supported by any of the other methods and are typically a result of a bad fit under low SNR conditions.

The majority of the fluxes is between 10 and 100 mJy for all three filter bands. The minimum and maximum Timeline Fitter fluxes appearing in the catalog for objects classified as point sources, are given in the first row of Table 3.1 for the three filters. The flux ranges generated by Sussextractor and Daophot are given as well and show good agreement, especially at the high flux end. For sources classified as extended, the Daophot and TM2 fluxes are the most relevant.

Their ranges are given in the same table. Note that the extended source fluxes are much more uncertain which is also indicated by the discrepancies in the ranges of the two methods.

| Flux Ranges in Jansky | 250µm Min | 250µm Max | 350µm Min | 350µm Max | 500µm Min | 500µm Max |
|---|---|---|---|---|---|---|
| Point Sources (TML) | 0.0084 | 395 | 0.0105 | 254 | 0.0112 | 150 |
| Point Sources (Sussextractor) | 0.0061 | 400 | 0.008 | 244 | 0.0087 | 148 |
| Point Sources (Daophot) | 0.0067 | 397 | 0.0067 | 252 | 0.0066 | 150 |
| Extended Sources (Daophot) | 0.0067 | 408 | 0.0067 | 260 | 0.0066 | 113 |
| Extended Sources (TM2) | 0.0103 | 795 | 0.0031 | 1061 | 0.0048 | 208 |

Table 3.1: Flux ranges of relevant extraction methods for point and extended sources.

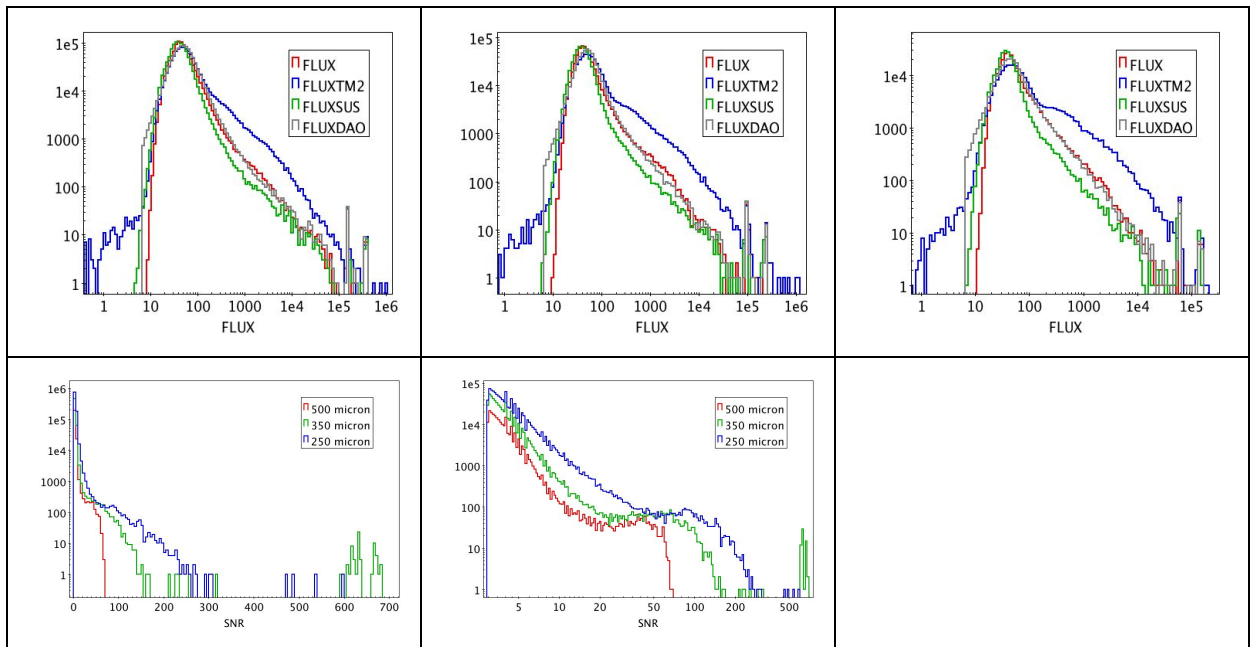

Figure 3.5: Flux distributions for the four different extraction methods. The wavelengths are 250µm, 350µm, and 500µm from left to right.
The lower row shows the SNR distributions for all three wavelengths with a linear (left) and a logarithmic (center) x-axis.

The distributions of SNRs for the three filters are shown in the lower row of the same figure with a linear and a logarithmic x-axis. The highest SNR values come from the brightest detections, in particular those of the planets that were separated out with special identifiers.

Although there are SNR cuts of 3 applied to the Sussextractor extractions and the Timeline Fitter, there is a sizeable fraction of objects (163032, 103364, 87527), where the Daophot SNR is smaller than 3. We find that in these cases there is a chance that a high energy glitch has made it through the filters, posing as an object. However, there are also cases where this is an extended source that is larger than the Daophot aperture with substantial loss in flux due to its point source optimized background annulus. In such a case the TM2 photometry could still provide a reasonable estimate.

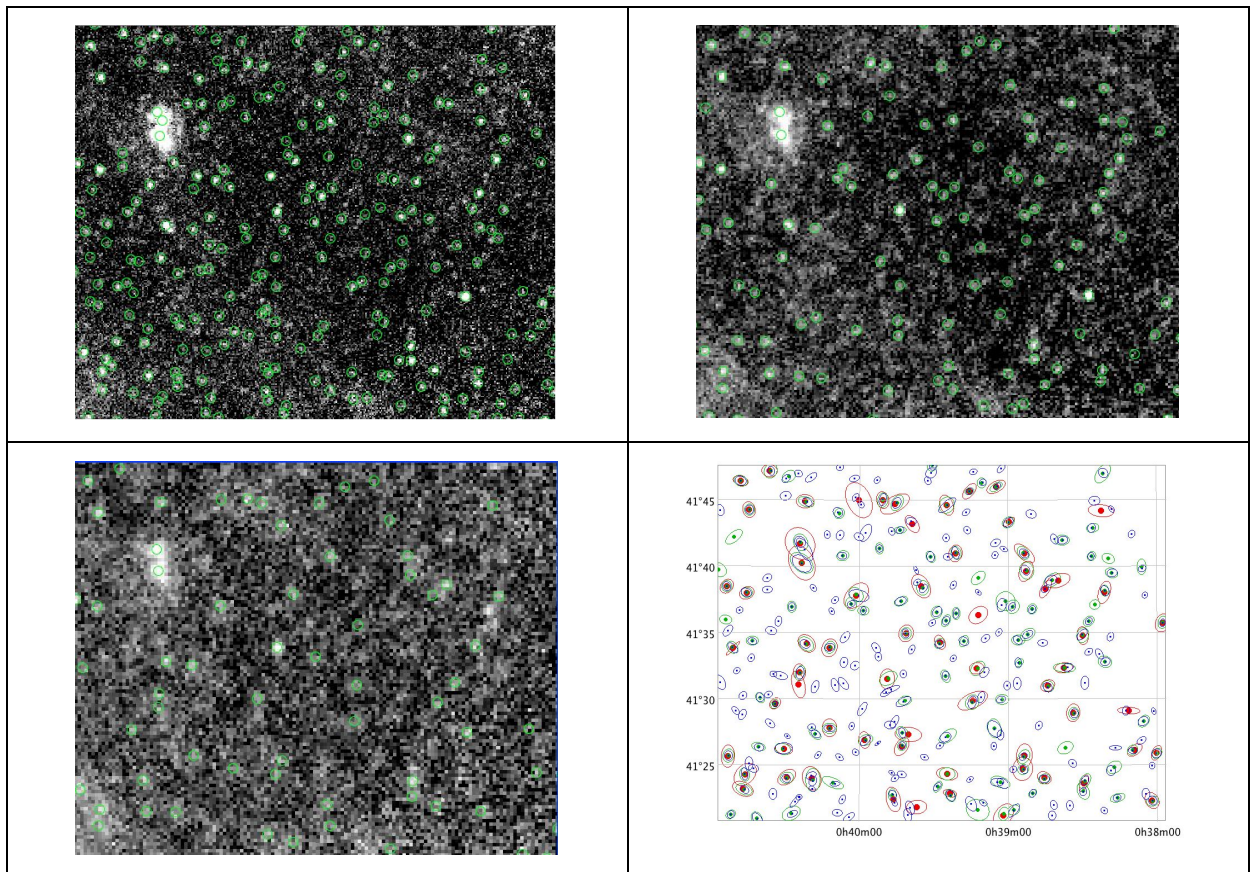

Figure 3.6: 250µm (top left), 350µm (top right), and 500µm (bottom left) maps (ObsID: 1342211319) with overplotted object positions. The plot in the lower right corner shows the ellipses derived from the shape parameters. The wavelengths are indicated blue, green, red respectively.

## Shape

The FWHM parameters and rotation are provided by the TM2 fit, giving an indication of the potential extent of an object and its orientation if extended. We did spot checks of the plotted ellipses against the maps and they show good agreement (see Figure 3.6).

Note that when plotting with Topcat, 90 degrees must be added to the rotation angle as it uses the North direction as origin.

The distributions of the FWHM parameters are shown in Figure 3.7. There is a bias arising from the requirement of $FWHM1 \geq FWHM2$ as expected also from the known small ellipticity of the SPIRE beam profiles, but also from the many weak sources that can not be classified reliably anymore.

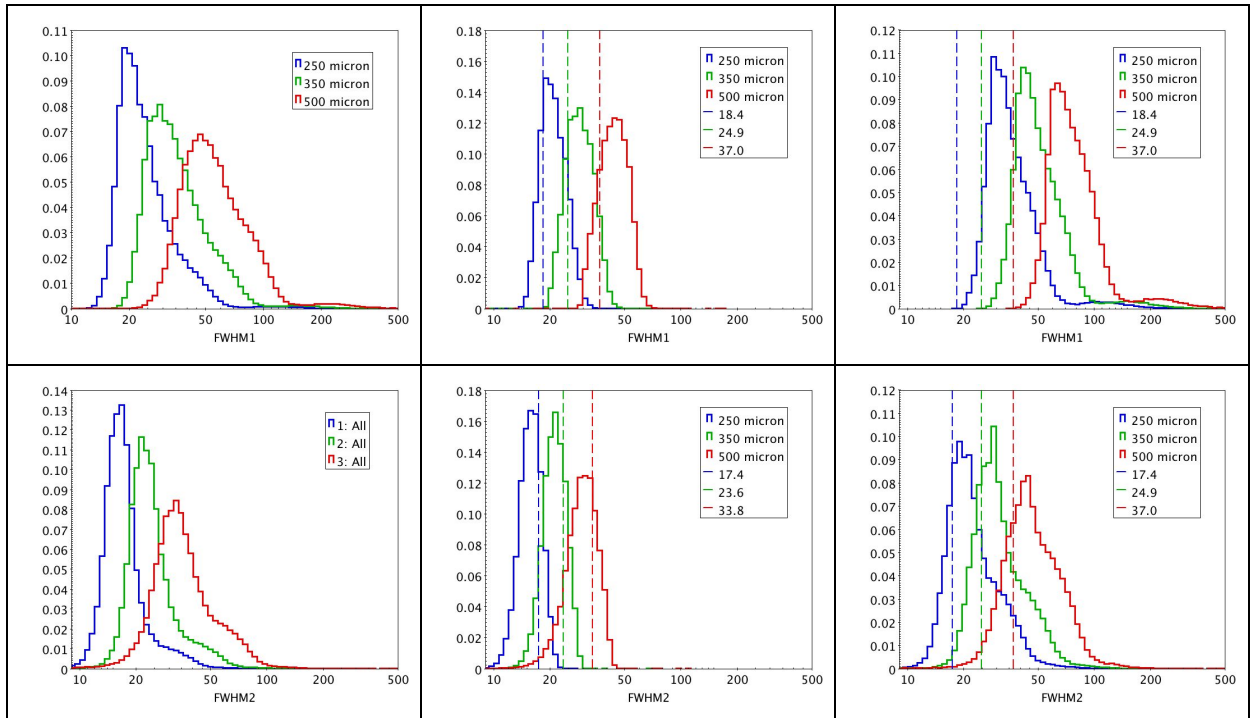

Figure 3.7: Distributions of major (upper row) and minor (lower row) FWHM. The first column shows all objects while the second and third show only objects classified as point and extended sources respectively. The filter bands are indicated by the colors red, green and blue.

A different representation shown in Figure 3.8 illustrates the dependence of the two FWHM fit parameters. The density of the data points is represented by the color scale with the lowest density being red, increasing via yellow, green, blue, violet, towards black. The measured nominal FWHM for a point source are indicated as dashed lines. Their loci are consistent with the highest density of data points.

The rotation angle is a free fit parameter that can turn over many times during the fit process. In the last stage of building the final catalog table the numbers are limited to the acceptable range of $0° \leq ROT < 180°$ ( Figure 3.9).

This is verified in Figure 3.7 left, showing the overall distribution of rotation angles that is expected to be statistically uniform. The slight bias may come from the combination of beam profile ellipticity and possible preferences of spacecraft rotation angles used throughout the mission.

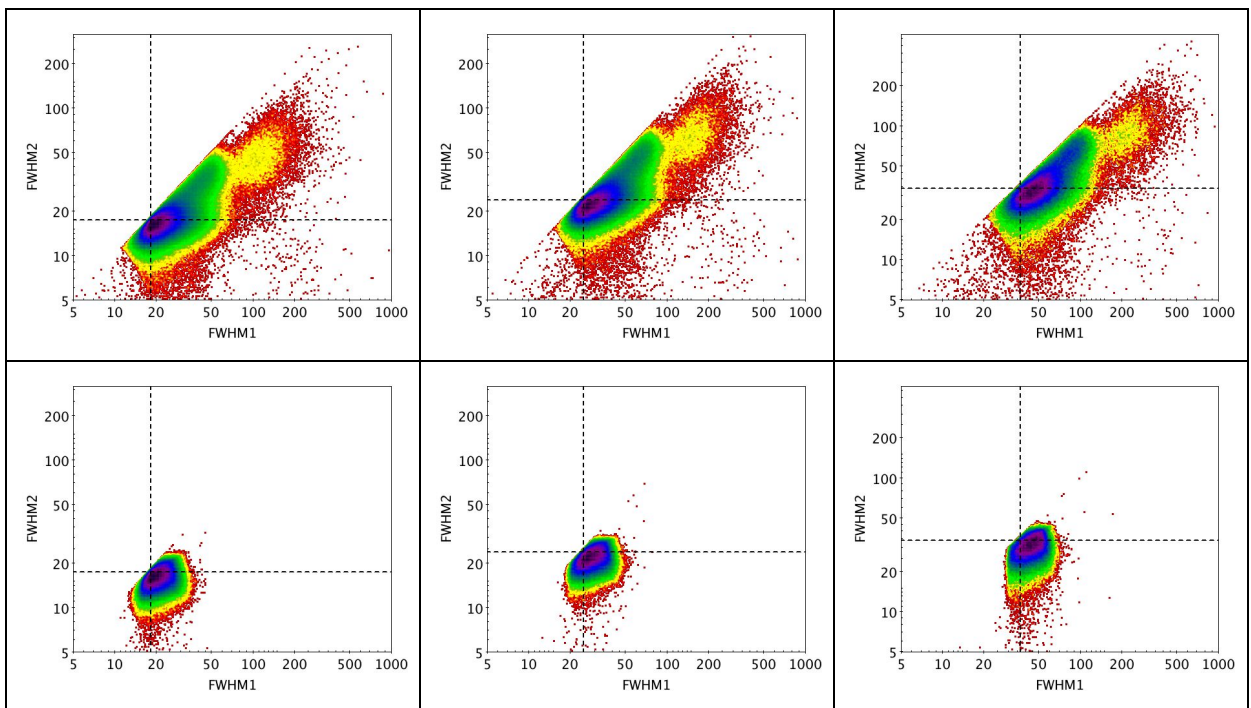

Figure 3.8: FWHM1 plotted against FWHM2 for all sources in the upper row, and point sources alone in the lower row. Data density is indicated by a rainbow color scale.

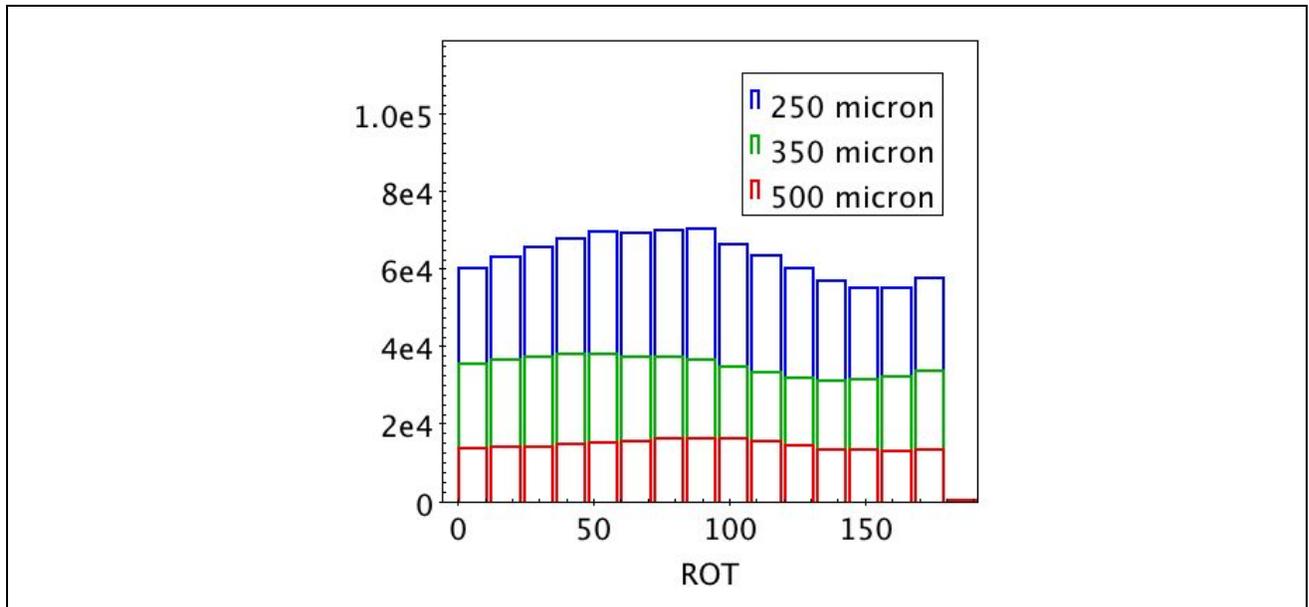

Figure 3.9: Histogram of all fitted rotation angles separated by filter band.

## Simulations

To learn our completeness and to test our photometry one way is to try to detect sources with accurate known positions and to measure their brightness, which is also very well known. This can be done if we add sources to our maps with certain flux values - we define simulations as real observations with sources artificially added to them.

The SPIRE observations cover a large variety of observations in terms of complexity. Extragalactic fields were observed for cosmological programs as well as galactic regions for star formation studies. Extragalactic windows have almost zero extended emission, but they are crowded with extragalactic objects and are confusion limited. The galactic regions are dominated by the extended emission coming from the cold dust of the interstellar medium (ISM). In this case the surface brightness of the ISM affects our detection completeness. In this work the complexity of the maps is described with the structure noise. The structure noise is closely related to the amplitude of the power spectrum at a well defined angular scale (wavenumber).

The inhomogeneities of the Herschel sky coverage means that our completeness and flux uncertainty had to be studied in all kinds of observations. Therefore we investigated the structure noise distribution of our observations in the PSW band in order to select the fields that represent our dataset in the best way. The median structure noise level of each map gives a good idea of their complexity. As the structure noise includes both the sky confusion and the instrument noise, it depends on the coverage of the map. Therefore the distribution of the median structure noise times the median coverage is plotted in Figure 3.10. The distribution shows 9 peaks, suggesting that 9 bins should be created. The boundaries of the bins are [0, 100, 120, 140, 167, 194, 235, 282, 324 and above.]

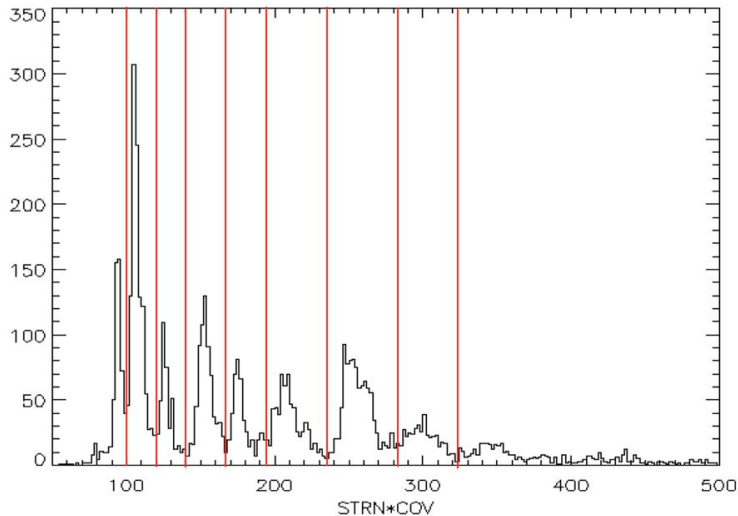

Figure 3.10: Distribution of the median structure noise times the median coverage. Red lines are the boundaries of the bins (see text).

It is also important to cover all major observing modes. From each bin a SmallScanMap, a LargeScanMap and a SpirePacsParallel map was selected. The obsid, the area size, the coverage and the median structure noise values are listed below in Table 3.2.

| obsid | nrep | targetname | maparea | obsmode | median psw strn | # sources injected |
|---|---|---|---|---|---|---|
| 1342183062 | 0 | Draco Cloud | 14400 | SpirePacsParallel | 8.64923 | 500 |
| 1342244197 | 1 | G141.25+34.37 | 400 | SpirePhotoLargeScan | 10.648829 | 100 |
| 1342234780 | 1 | 2MASX J14530794+2554327 | 16 | SpirePhotoSmallScan | 10.8349 | 10 |
| 1342213180 | 0 | Cha-II | 12500 | SpirePacsParallel | 11.5966 | 1000 |
| 1342180954 | 1 | ngc 6543 | 400 | SpirePhotoLargeScan | 9.2981 | 150 |
| 1342220637 | 1 | MRK 0304 | 16 | SpirePhotoSmallScan | 11.3171 | 10 |

| 1342195856 | 1 | COSMOS | 7225 | SpirePhotoLargeScan | 8.17359 | 1000 |
|---|---|---|---|---|---|---|
| 1342245146 | 0 | Field 209_0 | 14400 | SpirePacsParallel | 14.0428 | 500 |
| 1342234905 | 1 | NGC 5033 | 16 | SpirePhotoSmallScan | 13.245 | 10 |
| 1342215984 | 0 | OrionB-S-1 | 30000 | SpirePacsParallel | 18.5148 | 500 |
| 1342216940 | 1 | G210.90-36.55-1 | 2500 | SpirePhotoLargeScan | 10.6672 | 500 |
| 1342229600 | 2 | wise1724+3455 | 16 | SpirePhotoSmallScan | 8.62157 | 10 |
| 1342219982 | 1 | G227.95-2.98-1 | 1225 | SpirePhotoLargeScan | 11.5889 | 150 |
| 1342211604 | 0 | M31 | 12144 | SpirePacsParallel | 6.52983 | 500 |
| 1342239995 | 2 | HIP80088 | 16 | SpirePhotoSmallScan | 10.6201 | 10 |
| 1342216013 | 0 | east | 12600 | SpirePacsParallel | 21.1786 | 500 |
| 1342226626 | 1 | G155.80-14.24-1 | 2500 | SpirePhotoLargeScan | 14.6958 | 150 |
| 1342268338 | 2 | wise0426+1949 | 16 | SpirePhotoSmallScan | 13.3801 | 10 |
| 1342213455 | 1 | G89.65-7.02-1 | 1600 | SpirePhotoLargeScan | 15.5072 | 150 |
| 1342244847 | 0 | Field 90_1 | 14400 | SpirePacsParallel | 29.8449 | 500 |
| 1342226967 | 2 | IRAS 12327-6523 | 16 | SpirePhotoSmallScan | 15.4655 | 10 |
| 1342231339 | 0 | Field 63_0 | 14400 | SpirePacsParallel | 35.0556 | 500 |
| 1342228342 | 1 | G202.02+2.85-1 | 2025 | SpirePhotoLargeScan | 21.0453 | 150 |
| 1342239905 | 2 | iras 17243-4348 | 16 | SpirePhotoSmallScan | 17.2497 | 10 |
| 1342183407 | 1 | Strip Field | 4800 | SpirePhotoLargeScan | 49.8777 | 500 |
| 1342211615 | 0 | Carina Nebula Complex-1 | 19460 | SpirePacsParallel | 69.9786 | 500 |
| 1342211411 | 4 | KY Cyg | 16 | SpirePhotoSmallScan | 62.7258 | 10 |

Table 3.2: Obsid, field, area size, coverage, obsmode, median structure noise values and number of sources injected in our simulations.

As it is shown in the Table 3.2 different number of sources were injected into the observations depending on the map size. In total 7940 sources were injected per flux level per band. The total number of sources injected was 7940*34*3=809880.

The effects and artifacts coming from the different coverage of the maps and also coming from the mapmaking algorithm can be the best taken into account if we add the artificial sources to the level1 observational timeline. The sourceSubtractor task subtracts sources from SPIRE photometer timeline data using Gaussian functions with parameters specified by the user. This may be used to either remove sources so as to examine the background or, if the input source amplitude is negative, to inject artificial sources into the timeline data, which may be useful for tests with mapping, source fitting, source extraction. The task requires timelines as input and locations of sources at which to fit.

The random positions of the sources were generated once for each map and then all flux levels were injected at the same positions. We used the randomUniform() task of HIPE to randomize the positions. As the confusion for SPIRE is something we had to take into account, we did not define a minimum separation between the random positions.

The injected Gaussians were defined to have a sigma value according to the beam FWHM, eg. 17.6, 23.9 and 35.2 arcseconds. The Gaussians were scaled so they had the total flux defined for the given flux level. The flux levels we injected are the following: [5.0, 10.0, 20.0, 30.0, 40.0, 50.0, 60.0, 70.0, 80.0, 90.0, 100.0, 120.0, 140.0, 160.0, 180.0, 200.0, 225.0, 250.0, 275.0, 300.0, 350.0, 400.0, 450.0, 500.0, 600.0, 700.0, 800.0, 900.0, 1000.0, 2000.0, 3000.0, 6000.0, 10000.0, 20000.0] mJy.

After these artificial sources were added to the SPIRE observational timeline, maps were produced in the same fashion as they were produced in case of the original data. This was necessary in order to make the simulation results comparable with the real data extraction. This also involves that the source detection and source extraction was done with the same pipeline used for the catalog generation.

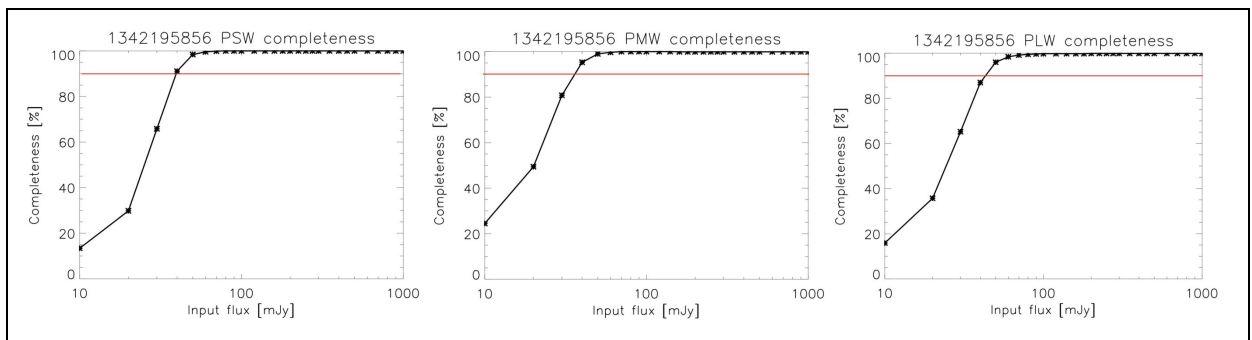

Figure 3.11.: Completeness in the COSMOS field in the PSW, PMW and PLW bands. In all bands the >90% completeness (red solid line) is reached below 50 mJy.

## Completeness

The completeness is calculated as a ratio of the number of sources injected and the number of sources detected by our pipeline. Here it is presented in two different fields. The OBSID 1342195856 stands for the COSMOS field (Figure 3.11), which is a field free of extended emission, therefore it has a low structure noise. The obsid 1342228342 (right panel) is the field G202.02+2.85-1 (Figure 3.12), close to the galactic plane. As one can see from the curves, the 90% completeness is reached at ~45 mJy, while the same completeness in the galactic field is reached at ~600 mJy.

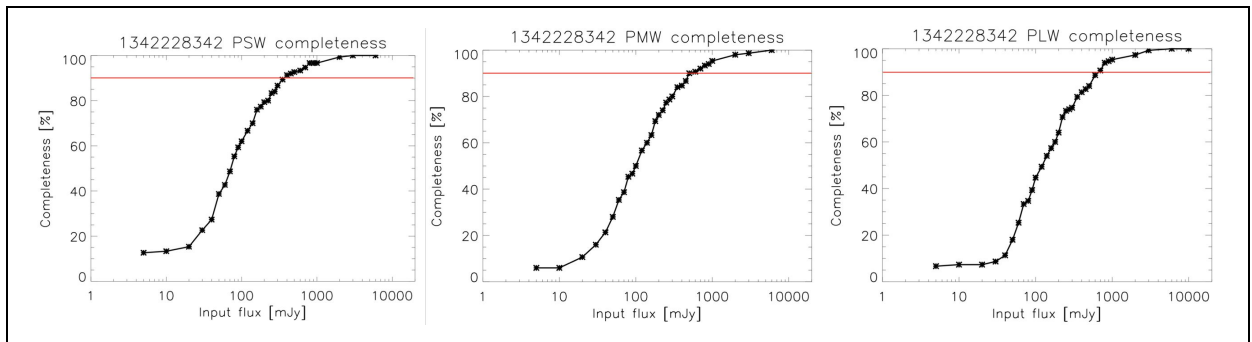

Figure 3.12.: Completeness in the Galactic region G202.02+2.85-1 in the PSW, PMW and PLW bands. In all bands the >90% completeness (red solid line) is reached above 400 mJy.

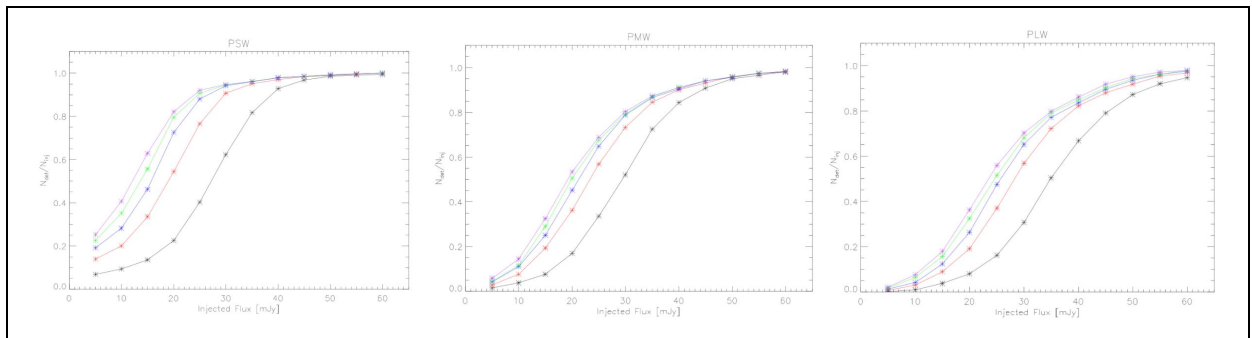

Figure 3.13: Completeness curves obtained from COSMOS observations. The curves with black, red, blue, green and purple colors correspond to maps with different coverages, where 1, 2, 3, 4 and 5 observations are combined, respectively.

An additional test we carried out shows how the completeness behaves as a function of the coverage. To simulate such a behaviour we injected 1000 sources into the COSMOS field with fluxes between 5 and 60 mJy, with 5 mJy intervals. As it was expected, the number of sources recovered from maps is increasing as the number of observations combined into a map increases. (See Figure 3.13.)

As it was written in the Structure Noise chapter, the structure noise includes noise coming from both the instrument and the sky confusion, and we just showed that the completeness highly depends on the background. Therefore it makes sense to analyse the completeness of our simulations in relation with the structure noise, or more specifically depending on the sky confusion.

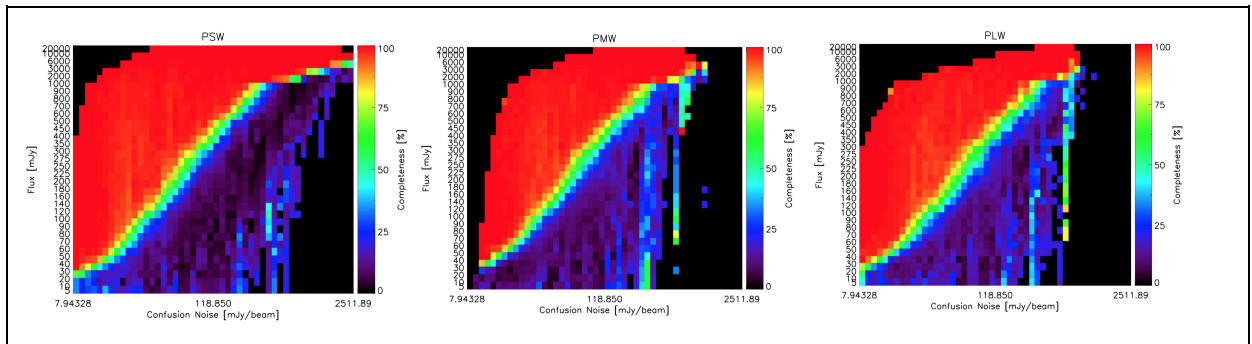

Figure 3.14.: Completeness as function of confusion error and input flux in the PSW, PMW and PLW bands (from left to right). The colours from blue to red represent completeness low and high completeness, respectively.

As it was done for the catalog objects, we separated the instrument noise portion and the sky confusion error portion of the total error. This allowed us to investigate our completeness as a function of the sky confusion, that describes the complexity of the sky background. As it was expected, faint sources are detected only at low confusion levels, while the bright sources (above ~2 Jy) are detected in most of the cases (see Figure 3.14).

## Photometric accuracy

The photometric accuracy is calculated as the ratio of the input flux and the measured flux. The photometric error is the standard deviation of the measured flux values at a given injected flux levels. For example if the injected flux is 100 mJy and the measured flux is 99.4+/-68.5, then the photometric accuracy is 99.4/100=0.994, the error is 68.5, and the S/N is 100/68.5=1.46. As it is shown in the examples below, the photometric accuracy highly depends on the environment. These tests also led us to the conclusion that areas in the Galactic plane and in star forming regions have to be excluded, as our photometric accuracy and the derived photometric errors do not fulfil our quality criteria.

The same observations were used to check the photometric accuracy that were used for the completeness analysis. In the COSMOS field the photometric error (error bars) is relatively small, the uncertainty is ~10-15 mJy at the faint flux levels. The faintest sources detected have flux values of ~30 mJy. This means that those sources injected with less than 30 mJy are detected only if they are blending with brighter sources. As the source flux increases the relative error becomes smaller, and also the deviation (red solid line) from the theoretical flux (blue dashed line) becomes small. Above 50 mJy the photometric accuracy is better than 2%.(Figure 3.15)

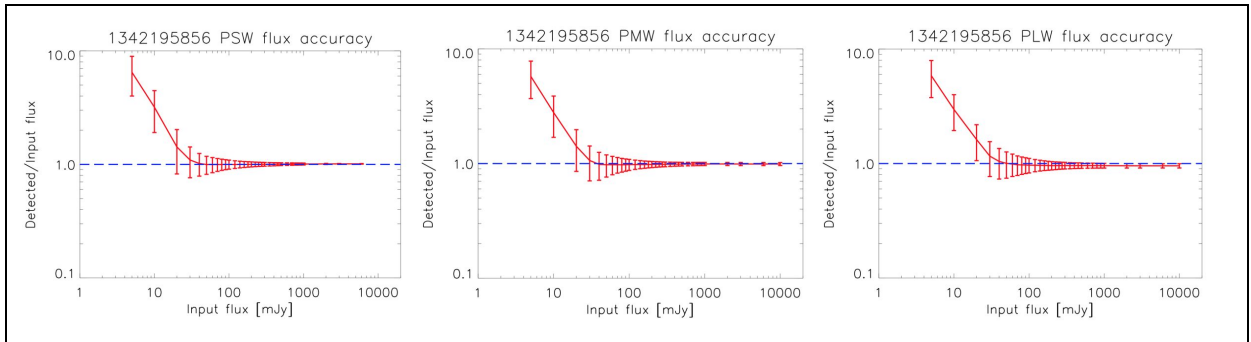

Figure 3.15.: Flux accuracy in the COSMOS field. The blue dashed line indicates the perfect photometry, where the flux ratio is 1. The red solid line and the error bars show the average measured flux and the corresponding uncertainty. Sources brighter than 100 mJy are accurate.

In the other example the the photometric accuracy was studied in a field close to the Galactic plane, same as used for the completeness analysis. The photometric errors are ~100 mJy in all cases. The photometric accuracy is below 5 percent at flux levels brighter than 80, 350 and 700 mJy in the 250µm, 350µm, and 500µm bands, respectively (Figure 3.16).

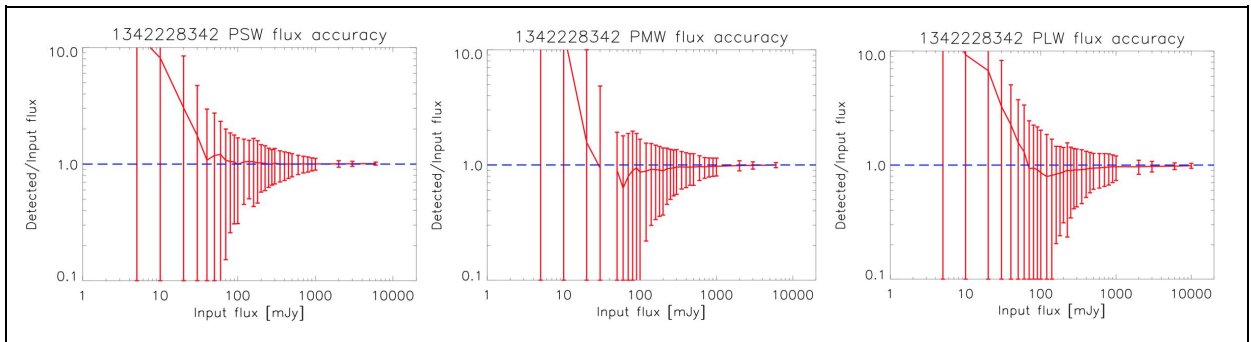

Figure 3.16.: Flux accuracy in the Galactic field G202.02+2.85-1. The blue dashed line indicates the perfect photometry, where the flux ratio is 1. The red solid line and the error bars show the average measured flux and the corresponding uncertainty. Sources below 1 Jy seem highly inaccurate.

From Figures 3.17 and 3.18 we can also conclude that the S/N=3 level is reached at 30 mJy brightness in all three bands in case of the empty sky (COSMOS), but the S/N=3 ratio is reached at 300, 350 and 500 mJy flux levels in case of the regions that is heavily affected by cirrus noise (Galactic field G202.02+2.85-1).

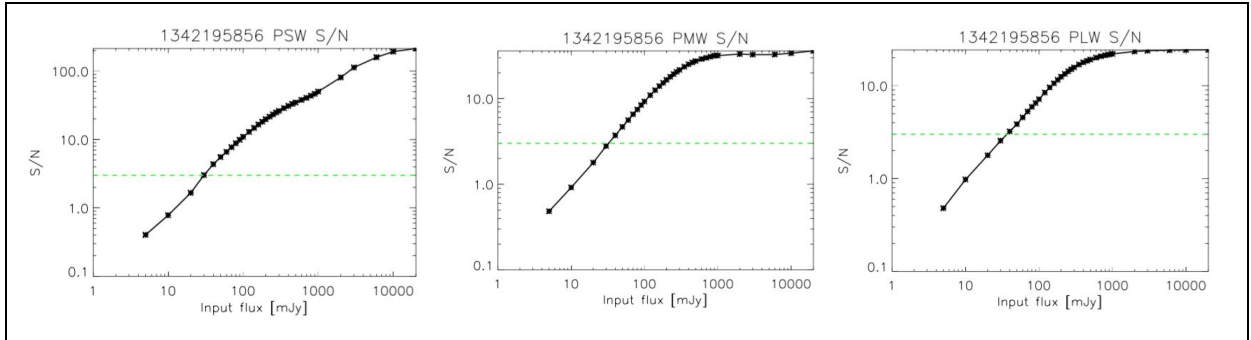

Figure 3.17.: S/N as a function of the input flux in the COSMOS field. The S/N=3 (green dashed line) is reached at ~30 mJy.

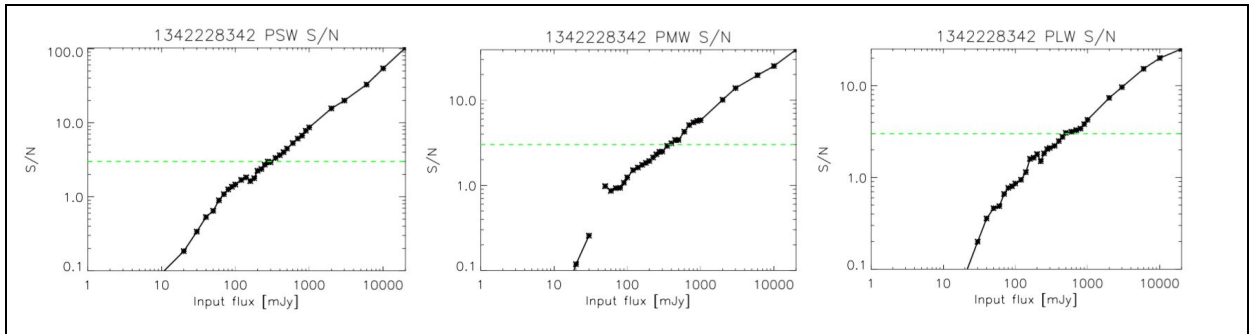

Figure 3.18.: S/N as a function of input flux in the Galactic Plane field G202.02+2.85-1. The S/N=3 is reached at several hundreds of mJy flux density.

By using the data from all of our simulations we were able to analyse the photometric accuracy (measured flux over the input flux) as a function of the confusion noise and the input flux. As it is shown on Figure 3.19 the flux of faint sources is returned accurately only if they are located in regions with low confusion error. At high confusion levels only bright sources were measured accurately.

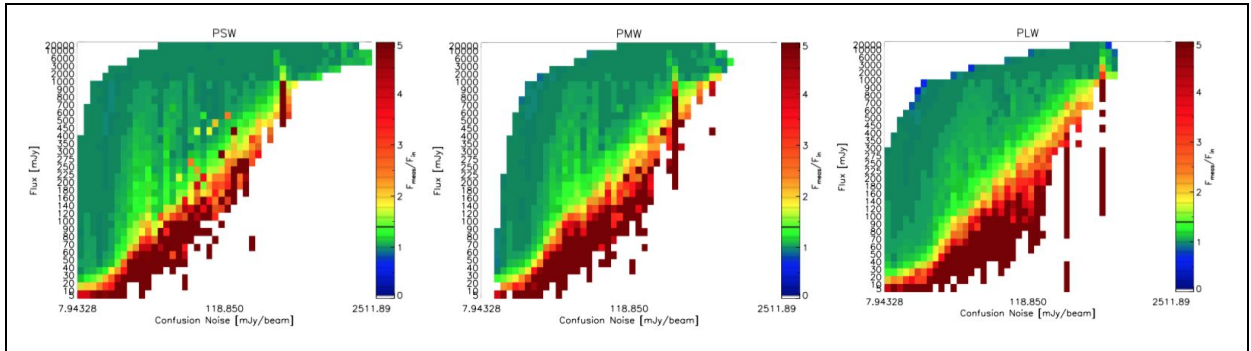

Figure 3.19.: Photometric accuracy (measured flux over the input flux) as a function of the confusion noise and input flux. Green colors mean that the photometry is accurate, red pixels indicate regions where photometry is inaccurate.

## Prime Calibrator Fluxes

The absolute SPIRE flux calibration is based solely on a radiative model of the atmosphere of Neptune as described in detail by Bendo et al. 2013. This approach was possible through a carefully determined linearization over all accessible fluxes, that formed an integral part of the SPIRE data processing pipeline.

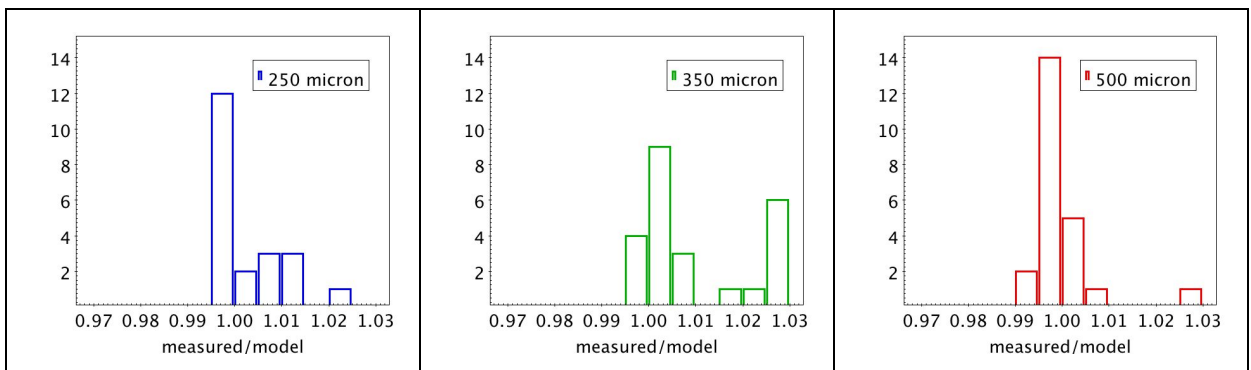

Figure 3.20: Histograms of ratios between catalog TML fluxes and calculated fluxes from an atmospheric radiation model of Neptune (ESA4).

In order to avoid the mixing up of the bright planet detections with those of much fainter background sources, the planet detections were separated out of the catalogs and given special identifiers. As a consistency check we matched the list of Neptune calibration fluxes that were calculated from the so called ESA4 model of Neptune (Moreno 1998), to the 118 catalog entries with Neptune detections. Figure 3.20 shows histograms of the ratios of measured flux and model flux of the calibrator for the three filter bands. The excellent agreement between calibrations and extracted fluxes with sub-percent standard deviations, summarized in Table 3.3, is very reassuring and proves consistency of this catalog with the official SPIRE flux calibration.

| Filter Band | Mean | Std. Deviation | Median | Number of detections |
|---|---|---|---|---|
| 250μm | 1.00284 | 0.007202 | 0.999598 | 21 |
| 350μm | 1.01048 | 0.01136 | 1.00475 | 24 |
| 500μm | 0.999771 | 0.00637 | 0.998613 | 23 |

Table 3.3: Statistics of ratios between catalog TML fluxes and calculated fluxes from an atmospheric radiation model of Neptune (ESA4).

## Serendipity Mode Slew Trails

An issue that was discovered only late during the validation effort affects completeness. It was found that maps that were crossed by the detector array prior to starting the actual map scan, sometimes had almost all sources within that scan trail missing. A particularly bad example is shown in Figure 3.21, where two trails affect a map that is a combination of five observations of a square 30' x 30' field. Almost all sources are missing within those trails and the few detected objects therein show unrealistic FWHM fits aligned with the trail direction.

The reason was traced to the inclusion of Serendipity Mode data into the Level 1 timelines, starting with SPG Version 14, a detail that was missed because the original extraction pipeline had been developed and tested on data from SPG Versions 11, 12, and 13. An exclusion of these building blocks was only implemented after the source extraction was already performed on 5657 maps. The fact that still 12.6 per cent of the sky area observed with SPIRE are not affected by this issue, is due to processing small area maps first and leaving the largest maps for extraction towards the end of the campaign. Fortunately, only a few per cent of the total number of objects were lost this way as the effect appears only under certain circumstances. However, a fix was not feasible within the schedule for this version of the catalog, as it will require a re-extraction of all 5657 maps that were source extracted before the software update.

Analysis revealed that all sources within the trails were well detected and processed by Sussextractor and Daophot, as they operate only on the maps. The Level 1 signal timelines, however, include data that is taken serendipitously on the way of the spacecraft slewing from the previous observation to the starting point of the new map. Depending on the origin, that slew can lead across the area to be mapped. After reaching the starting point of the map, the instrument electronics performs a coarse re-calibration of the offsets in order to match the dynamic range of the A/D converters to the local brightness of the sky. These offset levels are recorded and during pipeline processing added back to yield the actual bolometer voltages. The problem arises when the previous observation was performed in a celestial field with very different background flux, resulting in a set of different offset levels. In that case both Timeline Fitter runs see many additional readouts at very different flux levels that prevent conversion of the algorithm and a subsequent elimination of the respective objects from the catalog.

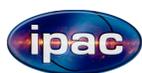 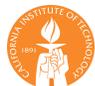 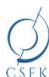 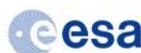 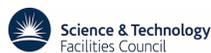 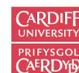

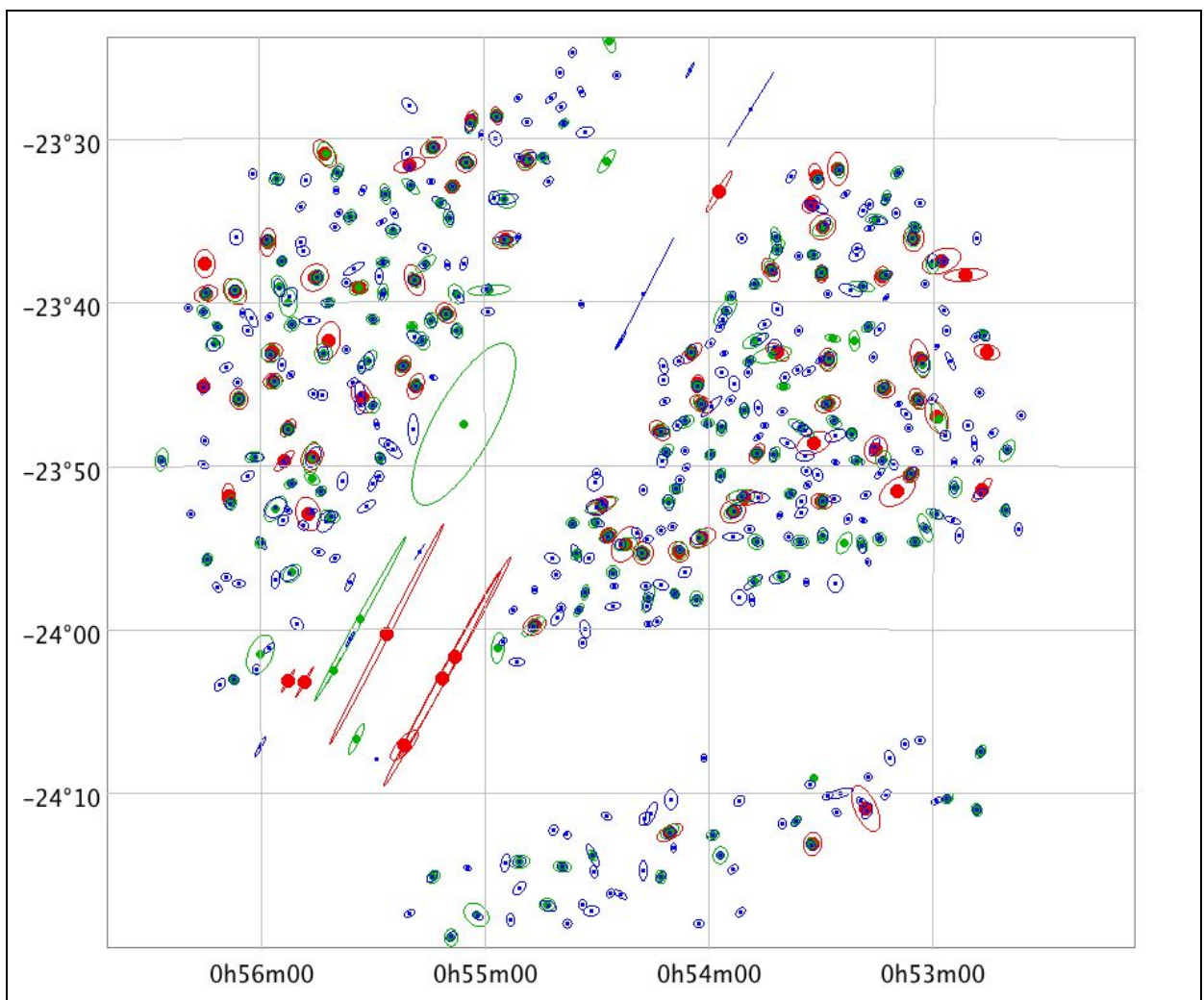

Figure 3.21: Illustration of missing sources due to Serendipity Mode trails (ObsID 1342234700). The colors blue, green, red correspond to the three filter bands and the ellipses

depict the fitted shape parameters, becoming unrealistic within the trails.

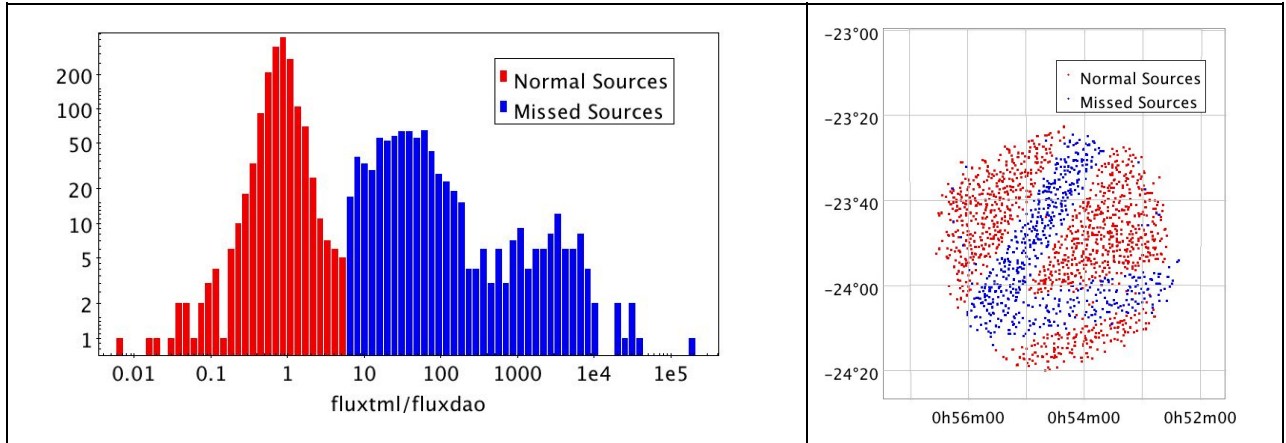

Figure 3.22: Missed sources in Serendipity Trails identified by their ratio of TML to Daophot fluxes of greater than 7. The histogram of ratios is plotted on the left and the locations of the respective sources are plotted in the same colors on the right.

The few objects that still produce converging fits and make it through the source table cleaning, are usually easy to identify by their strong discrepancy of the Timeline Fitter fluxes from Sussextractor and Daophot, as well as by an unrealistically extended FWHM1. We found that the ratio of the TM2 source peak and Daophot flux being greater than 7 is a good indicator for this condition. Figure 3.22 shows a histogram of that ratio on the left, split at a ratio of 7. The sources above that split are almost all filling the trails. Based on this criterion we estimate that about 3 per cent of all sources were lost due to this effect.

## Comparison with Extragalactic Catalogs

Even though the properties of our objects may be internally consistent and appear reasonable, comparison to other observations, preferably entirely independent, are an important step in the validation process. In this specific case we face the difficulty that Herschel was the first telescope facility to allow for large scale sky mapping in the Submm, so there are no statistically meaningful datasets to compare to at the same wavelengths and similar sensitivity. With that limitation, we can still compare to already published work based on the same data, but using different approaches, or compare to data from different wavelengths that has a high probability to correlate with the majority of sources. Table 4.1 shows the comparisons we made with other catalogs that are limited to low background confusion regions.

The MIPS 24 comparison is the only true outside validation with data entirely from different instrumentation, where we exploit the strong correlation between the 24 µm data from Spitzer and the Submm data of Herschel, which was also used for the astrometry check through stacking on 22 µm WISE catalog sources, described above.

| Field | Program | Location | | |
|---|---|---|---|---|
| | | RA (deg) | Dec (deg) | Radius [arcmin] |
| COSMOS | HerMES DR2 | 148.7935 - 151.4541 | 0.8609 - 3.5996 | - |
| COSMOS | MIPS 24 SCOSMOS | 148.7935 - 151.4541 | 0.8609 - 3.5996 | - |
| Lockman-SWIRE | HerMES DR2 | 161.753 | 58.078 | ~324 |
| CDFS | HerMES DR2 | 53.06725 | -28.26794 | ~151 |

Table 4.1: Data used for validation of the SPSC in extragalactic regions with low Cirrus background confusion.

Otherwise we compared with the HerMES Data Release 2 (Oliver at al. 2012) in three different fields, which is the largest extragalactic survey in terms of observing time, conducted with Herschel.

In the following we summarize our findings. The details can be found in the Annex below.

## Detections

At first sight the HerMES survey contains about double the number of sources or sometimes even more than the SPSC in the same area. These comparisons were performed before introducing the SNR >3 cutoff for the TML fluxes in the SPSC. Digging deeper, we found a large number of HerMES records with an SNR below 1 and sometimes zero flux. The SPSC also contained a small number of low SNR records, but not quite as many. The introduction of SNR cutoffs on both sides made the numbers more comparable.

The SNRs of both catalogs still differ. HerMES added the confusion noise limits from Nguyen et al. 2012 quadratically to the uncertainties of the extractors, which came in two flavors of source extractors, either StarFinder or Sussextractor. This is justified, as HerMES fields don't have notable Cirrus confusion. SPSC uncertainties are based on simulated source extractions taking into account local structure noise and flux. We find that these tend to be more conservative than the HerMES numbers and it appears that the SNR cutoff applicable to HerMES in order to achieve something similar to the SPSC with SNR > 3, is located somewhere around 4 or 5.

On the other hand, a very high percentage (80% - 98%) of sources that appear in the SPSC are also detected in the HerMES catalogs, indicating a high degree of reliability of the SPSC. It seems that the highest number of matching sources is achieved when the SPSC is limited to an SNR > 3, which served as another argument when deciding on the SNR cutoff of the catalog.

## Positions

The positions of matches generally agree well and the scatter is consistent with the absolute pointing uncertainty of Herschel of $\leq 2''$. We find that the agreement improves when restricting the SNR to values larger than 3. Similarly an improvement is observed when restricting the SPSC objects to point sources only.

## Fluxes

Fluxes of matching sources are generally consistent with a scatter of about 10% and better, however, we find systematically lower fluxes originating from HerMES Starfinder than those from SPSC Timeline Fitter. The same matches show no systematic offset when comparing the HerMES Sussextractor fluxes with SPSC Timeline Fitter.

There is also a sub-group of matches where the HerMES Starfinder fluxes are considerably smaller than the SPSC Timeline Fitter fluxes. It turns out that these are cases where the source is either extended or a multiplet where the constituents are closer together than the FWHM of the given wavelength and are not distinguished by the SPSC extraction algorithm. The HerMES Starfinder used for DR2 instead is able to distinguish close multiplets by successively subtracting constituents from the map and extracting new point sources from the residual. This technique has the side effect of splitting true extended sources into several point sources that are not centered on any local maxima. This conclusion is further corroborated by finding that the sum of the individual fluxes of HerMES multiplets is very close to the extended source flux of the SPSC, derived by the TM2 run.

We conclude that aside from a small systematic difference in the flux calibration of the HerMES Starfinder sources, the agreement between both catalogs is quite good. Some matches, however, don't show this agreement because extended sources and small multiplets are separated in the HerMES catalog, while the SPSC fits them as one.

## Comparison with Galactic Catalogs

Herschel observed the entire Galactic Plane and several star forming regions close to it with the goal of providing a better picture of star formation. These observations delivered many important data that helped scientist to better understand the structure of the Interstellar Medium (ISM) and the very early phases of star formation. Star forming regions (SFRs) contain large amount of extended emission and a complete zoo of stellar and pre-stellar sources. These objects are mostly deeply embedded in the ISM and their shape is not consistent with the shape of the SPIRE PSF. Therefore we need to emphasize the importance of comparison with other catalogs that targeted these regions and whose data analysis was tuned for this exercise. Our pipeline used SUSSEXtractor, which was designed to find point sources. Our photometry was also designed for point sources. This means that in confused regions we expect that we find differences when we compare to these specific catalogs that are results of specific tools.

We compared our extraction from two regions of the Galactic plane to the results provided by:

1) "Hi-GAL: the Herschel infrared Galactic Plane Survey" (KPOT_smolinar_1, PI: Sergio Molinari) The Hi-GAL team used their own tool to detect the sources and to measure their flux. The tool is called CuTEx (Molinari et al., 2011). CuTEx builds a "curvature" image from the measured image by double-differentiation in four different directions. In this way point-like and resolved, yet relatively compact, objects are easily revealed, while the slower varying fore/background is greatly diminished. Candidate sources are then identified by looking for pixels where the curvature exceeds a given threshold in absolute terms, and the methodology allows to easily pinpoint breakpoints in the source brightness profile and then derive reliable guesses for the sources' extent. Identified peaks are fit with 2D elliptical Gaussians plus an underlying planar inclined plateau, with mild constraints on size and orientation.

2) "Galactic Cold Cores: A Herschel survey of the source populations revealed by Planck" (KPOT_mjuvela_1, PI: Mika Juvela). The Cold Cores project used a different tool, called Getsources (Men'shchikov A. et al., 2012, A&A, 542, A81). It is a powerful multi-scale, multi-wavelength source extraction algorithm. Instead of the traditional approach of extracting sources in the observed images, Getsources analyzes fine spatial decompositions of original images. Sources are detected in the combined detection images by following the evolution of their segmentation masks across all spatial scales. Measurements of the source properties are done in the original background-subtracted images.

## Detections

Both CuTEx and Getsources are able to detect sources in the very confuse regions, on the highly fluctuating background. They can also well detect sources located outside, or between the filamentary structures. Our pipeline tends to detect less sources in these confused regions, but more sources are picked up in the less structured areas. In case of the Hi-GAL detections we found that less than 50% of their sources were identified with our pipeline. In case of Getsources ~60% of their sources were found in our catalogue.

### Positions

In case of the CuTEx detections the positions agree very well, the distance distribution peaks around 3-5 arcseconds in all three bands. In the comparison with *Getsources* we found that the separation between the cross-matching sources also peaks around 5 arcseconds, but with a slightly wider distribution.

### Fluxes

The primary flux our Catalogue is the output of the TimelineFitter task. It is the most useful for circular Gaussian source profiles on flat background. As most of the sources in confused regions are elongated and are located on some fluctuating background, we have compared the CuTEx and Getsources fluxes to the results of the TM2, as well.
In case of the CuTEx we found that the fluxes they provided are ~2 times higher than our point source fluxes on average. The TM2 run, however, resulted in much higher fluxes, therefore tha ratio of the CuTEx flux and our flux is ~0.8, on average.
In our other test case we found that the SPSC fluxes are always higher than the ones provided by Getsources. We have compared our Timelinefitter Flux to two different flux values coming from Getsources. The first one is the peak flux in Jy/beam units (FXPBEST), while the second one is the total flux in Jansky (FXTBEST). The ratio of our TML flux and the FXTBEST tells us that FXTBEST is always higher than the TML flux. If we divide the TML flux with the FXPBEST, we find that the TML flux is lower at 250 and 350 microns, but is systematically higher in case of the 500 micron sources. If we use the TM2 flux over the FXTBEST we find that the photometry agrees within the error bars, but the errors are high, the scatter is ~100%.

We conclude that because of these differences in both source detection and source extraction, we cannot provide reliable results from the most complex regions. Therefore we imposed the structure noise threshold of 35 mJy in the 22 bit level Q3C tiles.

# 6 Conclusions

We have constructed a point source catalog from all 6878 usable SPIRE scan map observations in a standard configuration including all applicable calibration observations. The total of 1693718 objects splits with filter band into 950688, 524734, 218296 objects for 250µm, 350µm, and 500µm respectively. These totals also contain sizeable percentages of somewhat extended sources, that, increasing with wavelength, are 29%, 38%, and 45%.

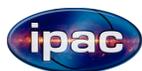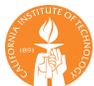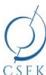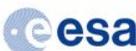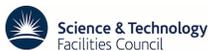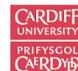

Initially the total number of objects was substantially greater, but in the interest of better reliability, we introduced an SNR threshold of 3, based on Timeline Fitter fluxes and flux and structure noise dependent uncertainties, derived from source injections into real data. A further reduction of objects resulted from excluding portions of the sky with a median structure noise of greater than 35 mJy within the respective Q3C tile, to avoid sky backgrounds where the source extraction does not work well anymore.

A comparison of the source extractions of SPIRE's prime calibrator Neptune with model data shows that the catalog fluxes are perfectly consistent with the radiation model used for the calibration. To provide good guidance to scientists, the flux values of 4 different photometric methods are listed: Two exclusively applicable to point sources (TML, Sussextractor), one to identify potentially extended sources and as sanity check (DAOPHOT), and one to derive a best guess for extended sources, assuming a Gaussian elliptical profile. These values are largely in agreement for point-like sources, and will alert the astronomer to possible source extension when they deviate from each other.

The use of structure noise dependent uncertainties based on extraction exercises of artificial source injections into real data has proven successful, yielding realistic and conservative values. Internal validation exercises as well as comparisons to other catalogs showed their validity. Additional quality indicators like the nmap and ndet values or the various flags provide additional information about individual objects and their reliability.

The validation exercise was extremely useful and revealed issues that could be fixed, but also a few issues that would have required major changes, incompatible with the schedule for this release. One of these uncorrected issues is the absolute positional shift of maps greater than 5", that was revealed using stacking on WISE 22µm catalog positions. A correction would require reprocessing of a few hundred combined Level 2.5 maps, re-running the source extraction, replacing the source lists in the database and re-running all catalog scripts. Similarly, fixing the second major uncorrected issue, caused by non-excluded serendipity mode scans, would as well require re-running the updated point source extraction procedure on the majority of all maps. Both items have limited impact on the quality of our catalog. Only 3% of the sources are estimated to have been lost due to serendipity data trails, and 110 maps of 6878, i.e. 1.6% are affected by bad pointing. Nevertheless, we are looking into the possibility of creating a second public version where these issues would be fixed. A second version would also allow to look into the design of the source extraction again with the aim of improving the performance of the different algorithms by tweaking aperture diameters, noise thresholds and other operating parameters. Additionally the software could be modified to generate rejected source lists that are useful if specific sources are not found.

Although every attempt has been made in this catalog to define point and extended sources at SPIRE wavelengths, it is important to realize that our definition of a point source is defined relative to a large beam size compared to most visible light astronomical catalogs (e. g. ~18" beam at 250 microns and larger at 350 and 500 microns). As our analyses in the Appendices have shown, our technique is insensitive to multiple sources which may be confused in the beam. Our catalog, while covering a large fraction of the SPIRE sky (excluding very complex regions), does not use priors to attempt to disentangle complex source distributions on arcsec scales. Therefore, care should be taken by users not to over-interpreting things like number counts for statistical samples based on this catalog unless likely source blending is taken into account. Furthermore, since we have not attempted to match sources in the three SPIRE bands, some sources can be defined as point sources in one band, but may be extended at others. We encourage users of the catalog to be vigilant when exploring sources, and we encourage users to inspect the SPIRE images where possible when interpreting emission from individual or larger samples of sources. Despite these caveats, we hope that the community will find the current catalog a valuable and important tool for studying emissions from the Submm sky.

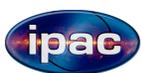

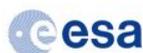
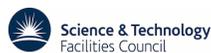

For the time being, the products explained in this document are made available through the online archives IRSA and HSA that add their own flavors and enhancing functionalities. With the next big Submillimeter telescope in space expected to come forth not before many years into the future, these datasets are expected to remain relevant for quite some time. We hope that this catalog can serve as a pathfinder for ALMA and other Submillimeter facilities to help select samples to conduct follow-up observations on, and in combination with auxiliary data from other wavelengths and instrumentations, yield new and perhaps unexpected results.

## Acknowledgements

A project of this magnitude has always many more contributors than can be found on the list of authors and also the small things can make a big difference. For that reason we would like to specifically acknowledge:
Jeff Jacobson and Rick Ebert for invaluable help with Postgres databases,
Joe Mazzarella and Ben Chan for providing the 2MASS extended galaxy ellipses,
Serge Monkewitz for essential help on coordinate averaging,
Yi Mei for providing the code to calculate map confining rectangles,
Mary Ellen McElveney for organizational help,
Marion Schmitz for helping with questions regarding catalog identifiers,
Justin Howell for help with IRSA ingestion,
Lijun Zhang for help with taming the destriper,
George Bendo for initial development of the Timeline Fitter and help with that,
Antony Smith for coding and providing Sussextractor to the community,
Natalia Larrea Brito for initial performance tests of source detection and extraction tools,
Babar Ali for generating the first representative dataset,

Göran Pilbratt, Anthony Marston, David Teyssier, and Pedro Garcia Lario for helping the project along the way,
Seb Oliver and Luisa Rebull for additional validation checks,
Jonathan Kakumasu and Wendy Burt for system support,
Luca Calzoletti, Eva Verdugo, Bruno Altieri, and Luca Conversi for help with SPIRE/PACS and HSA coordination,
George Helou for his good advice and patient support of the project until completion,
Roc Cutri and Peter Capak for good counsel on catalog building,
Roberta Paladini, Vandana Desai, Sean Carey, and Frank Masci for very useful discussions and serving on our internal review panel,
And Matt Griffin, the SPIRE Consortium, and all folks at ESA and NASA for building a phantastic instrument and observatory.
Last but not least we would like to express our appreciation to Mark Taylor and all his collaborators for writing TOPCAT that helped enormously with visualizing our data.

Herschel is an ESA space observatory with science instruments provided by European-led Principal Investigator consortia and with important participation from NASA.
SPIRE has been developed by a consortium of institutes led by Cardiff University (UK) and including Univ. Lethbridge (Canada); NAOC (China); CEA, LAM (France); IFSI, Univ. Padua (Italy); IAC (Spain); Stockholm Observatory (Sweden); Imperial College London, RAL, UCL-MSSL, UKATC, Univ. Sussex (UK); and Caltech, JPL, NHSC, Univ. Colorado (USA). This development has been supported by national funding agencies: CSA (Canada); NAOC (China); CEA, CNES, CNRS (France); ASI (Italy); MCINN (Spain); SNSB (Sweden); STFC, UKSA (UK); and NASA (USA).
HIPE is a joint development by the Herschel Science Ground Segment Consortium, consisting of ESA, the NASA Herschel Science Center, and the HIFI, PACS and SPIRE consortia.
This publication makes use of data products from the Wide-field Infrared Survey Explorer, which is a joint project of the University of California, Los Angeles, and the Jet Propulsion Laboratory/California Institute of Technology, funded by the National Aeronautics and Space Administration.
This research has made use of the NASA/ IPAC Infrared Science Archive, which is operated by the Jet Propulsion Laboratory, California Institute of Technology, under contract with the National Aeronautics and Space Administration.
This research has made use of data from the HerMES project (http://hermes.sussex.ac.uk/).
HerMES is a Herschel Key Programme utilising Guaranteed Time from the SPIRE instrument team, ESAC scientists and a mission scientist.
The HerMES data was accessed through the Herschel Database in Marseille (HeDaM-http://hedam.lam.fr) operated by CeSAM and hosted by the Laboratoire d'Astrophysique de Marseille.
This research was partially supported by OTKA grant NN-111016.

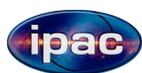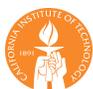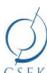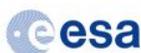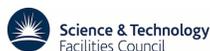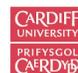

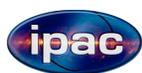

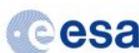
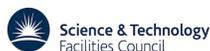

# List of Acronyms

| | |
|---|---|
| SPSC | Spire Point Source Catalog |
| SPIRE | Spectral and Photometric Imaging Receiver |
| TML | Timeline Fitter or First Timeline Fitter Run |
| TM2 | Second Timeline Fitter Run |
| HSA | Herschel Science Archive |
| ESA | European Space Agency |
| SPG | Standard Product Generation |
| FITS | Flexible Image Transport System |
| HCSS | Herschel Common Science System |
| SNR | Signal To Noise Ratio |
| SSO | Solar System Object |
| PSW | Photometer Short Wavelength detectors (250µm) |
| PMW | Photometer Medium Wavelength detectors (350µm) |
| PLW | Photometer Long Wavelength detectors (500µm) |

# List of Figures

Figure 1.1: Detection performance of the algorithms used for our test case. For flux levels >40 mJy all of them perform very well. At the faint end of the curve Sussextractor gives the best result.

Figure 1.2: Photometric accuracy of the tested algorithms. Timeline Fitter provides the best accuracy for all bands and all flux levels.

Figure 1.3: Overview flow diagram of the source extraction procedure.

Figure 1.4: Detailed flow diagram of the source extractor with its four algorithms, Sussextractor, Daophot, Timeline Fitter TML (fixed FWHM) and Timeline Fitter TM2 (free FWHM).

Figure 1.5: Flow diagram illustrating the handling of of combined (linked) and single observations in the overall source extraction procedure.

Figure 1.6: Distributions of roundness and sharpness parameters as delivered by Daophot. From left to right the diagrams correspond to 1) a range of different backgrounds, 2) an extragalactic field with predominantly weak point sources, 3) a galactic field with a mix of bright extended and point sources

Figure 1.7: Diagram showing Daophot fluxes versus the shape parameters roundness and sharpness for a mix of point and extended sources from various representative environments.

Figure 1.8: Illustration of the Timeline Fitter aperture and annulus plotted over a standard map with 10" pixel size for a 350μm map on the left, and the same map rendered with 1" pixels that better shows the actual detector timelines that are being fitted

Figure 1.9: Illustration how the beam profile models, apertures, and background annuli used by the different source extractors match with a real instrument beam profile of a point source.

Figure 1.10: llustration how the beam profile models, apertures, and background annuli used by the different source extractors match with an
extended source that still resembles a Gaussian profile.

Figure 1.11: Example of a good stack signal. The 11x11 pixel map is
composed of 250 µm map cutouts around the positions of 196 WISE 22 µm
sources. The center of the fitted circular 2D Gaussian is indicated with
a red cross, the fixed FWHM is shown as a red circle. The green star is
the stack image center.

Figure 1.12: Histogram of the number of detections of an object (NDET)
at the same position and filter band. Although the majority has only one
detection, there is a sizeable number of multiple detections where maps
overlap

Figure 1.13: Illustration of source grouping with large pointing spread comparable to search
radius

Figure 1.14: The circular configuration used for structure noise calculation. The target pixel is
shown with red square. The neighbouring pixels at the predefined angular distance are shown
with green squares

Figure 1.15.: The average structure noise value around the injected sources of 20 Jy into a
250µm map, as a function of the angular separation between the target pixel and the annulus, in
units of the beam FWHM.

Figure 1.16: The SNR values as a function of the structure noise for sources with 200 mJy
injected flux (black crosses). The SNR value is large at small structure noise values and drops
rapidly for higher . The fitted power function is presented with the solid red line

Figure 1.17: The SNR values as a function of input flux (black crosses) at a fixed structure noise
level of 200 mJy/beam. The fitted exponential function is shown as red solid line

Figure 1.18: Log-log contour plot of the SNR values as a function of structure noise and source
flux for the PSW, PMW and PLW arrays (left to right). SNR levels of 1, 3, 5 and 10 are
overplotted with white contour lines

Figure 1.19: The flux uncertainty (σ - standard deviation of the measured flux) values as a
function of the average readout number (blue squares). The uncertainty is decreasing according
to a power law. The fitted function is shown as a solid red line

Figure 1.20: Histogram, of the median STRN values of all Q3C tiles at 22 bit depth

Figure 1.22: Overall distributions of FWHM1 and FWHM2 in arcsec for the three filter bands. Note the difference in the scale of the x-axes. Note that these distributions still contain sky areas with high structure noise that were excluded for the final catalog

Figure 1.23: Plots of FWHM1 and 2 in arcsec versus TM2 flux in mJy. The distributions are fitted with upper and lower flux dependent envelopes that we use to establish a region of predominantly point sources

Figure 3.1: General source distribution across the sky.

Figure 3.2: The top row shows RA uncertainties plotted against Dec, showing the expected increase towards the celestial poles. The bottom row shows the uncertainties in Dec. Wavelengths are 250μm, 350μm, and 500μm from left to right.

Figure 3.3: Histograms of the number of detections that contributed to a given object with wavelenth increasing from left to right

Figure 3.4: ndet/nmap statistics for the three filter bands with wavelength increasing towards the right. The first row shows histograms while the second plots ndet/nmap versus Dec, indicating no special dependence on that coordinate.

Figure 3.5: Flux distributions for the four different extraction methods. The wavelengths are 250μm, 350μm, and 500μm from left to right.
The lower row shows the SNR distributions for all three wavelengths with a linear (left) and a logarithmic (center) x-axis.

Figure 3.6: 250μm (top left), 350μm (top right), and 500μm (bottom left) maps (ObsID: 1342211319) with overplotted object positions. The plot in the lower right corner shows the ellipses derived from the shape parameters. The wavelengths are indicated blue, green, red respectively.

Figure 3.7: Distributions of major (upper row) and minor (lower row) FWHM. The first column shows all objects while the second and third show only objects classified as point and extended sources respectively. The filter bands are indicated by the colors red, green and blue.

Figure 3.8: FWHM1 plotted against FWHM2 for all sources in the upper row, and point sources alone in the lower row. Data density is indicated by a rainbow color scale.

Figure 3.9: Histogram of all fitted rotation angles separated by filter band.

Figure 3.10: Distribution of the median structure noise times the median coverage. Red lines are the boundaries of the bins (see text).

Figure 3.11.: Completeness in the COSMOS field in the PSW, PMW and PLW bands. In all bands the >90% completeness (red solid line) is reached below 50 mJy.

Figure 3.12.: Completeness in the Galactic region G202.02+2.85-1 in the PSW, PMW and PLW bands. In all bands the >90% completeness (red solid line) is reached above 400 mJy.

Figure 3.13: Completeness curves obtained from COSMOS observations. The curves with black, red, blue, green and purple colors correspond to maps with different coverages, where 1, 2, 3, 4 and 5 observations are combined, respectively

Figure 3.14.: Completeness as function of confusion error and input flux in the PSW, PMW and PLW bands (from left to right). The colours from blue to red represent completeness low and high completeness, respectively.

Figure 3.15.: Flux accuracy in the COSMOS field. The blue dashed line indicates the perfect photometry, where the flux ratio is 1. The red solid line and the error bars show the average measured flux and the corresponding uncertainty. Sources brighter than 100 mJy are accurate.

Figure 3.16.: Flux accuracy in the Galactic field G202.02+2.85-1. The blue dashed line indicates the perfect photometry, where the flux ratio is 1. The red solid line and the error bars show the average measured flux and the corresponding uncertainty. Sources below 1 Jy seem highly inaccurate.

Figure 3.17.: S/N as a function of the input flux in the COSMOS field. The S/N=3 (green dashed line) is reached at ~30 mJy.

Figure 3.18.: S/N as a function of input flux in the Galactic Plane field G202.02+2.85-1. The S/N=3 is reached at several hundreds of mJy flux density.

Figure 3.19.: Photometric accuracy (measured flux over the input flux) as a function of the confusion noise and input flux. Green colors mean that the photometry is accurate, red pixels indicate regions where photometry is inaccurate.

Figure 3.20: Histograms of ratios between catalog TML fluxes and calculated fluxes from an atmospheric radiation model of Neptune (ESA4).

Figure 3.21: Illustration of missing sources due to Serendipity Mode trails (ObsID 1342234700). The colors blue, green, red correspond to the three filter bands and the ellipses depict the fitted shape parameters, becoming unrealistic within the trails.

Figure 3.22: Missed sources in Serendipity Trails identified by their ratio of TML to Daophot fluxes of greater than 7. The histogram of ratios is plotted on the left and the locations of the respective sources are plotted in the same colors on the right.

Figure 4.1: Illustration of how three exemplary catalog objects from the 250µm table are related to the SPIRE maps and their observation IDs used in the Herschel Science Archive. See text for details.

Figure 4.2: The three maps that contributed to the source cluster shown in Figure 4.1. To the left is the combined map consisting of observations 1342266649, 1342266650, 1342266651, and 1342266652. The center shows Observation 1342240055 and the right image shows Observation 1342240056.

# List of Tables